\documentclass[a4paper,reqno,oneside]{amsart}
\usepackage{etoolbox}

\newtoggle{llncs}
\makeatletter
\@ifclassloaded{llncs}{\toggletrue{llncs}}{\togglefalse{llncs}}
\makeatother

\iftoggle{llncs}{
\usepackage{amsmath}
\usepackage{amssymb}
\usepackage{bbm}
\usepackage[english]{babel}
\usepackage{color}
\usepackage{enumitem}
\usepackage{ellipsis}
\usepackage[breaklinks,colorlinks=true, citecolor=blue]{hyperref}
\usepackage[misc,geometry]{ifsym} \usepackage{dsfont}
\usepackage{mathrsfs}
\usepackage{mathtools}
\usepackage{algorithm}
\usepackage{algpseudocode}
\usepackage{stmaryrd}
\usepackage{thm-restate}
\usepackage{subcaption}
\usepackage{systeme}
\usepackage{accents}
\usepackage{tikz}
\usetikzlibrary{matrix}
\usetikzlibrary{arrows}
\usetikzlibrary{patterns, shapes, positioning}
\usetikzlibrary{decorations.pathmorphing,shapes}
\usepackage{makecell, multirow} 
\usepackage{setspace}
\usepackage[thinlines]{easytable}

\usepackage{comment}
\usepackage{bbm}
\usepackage{booktabs}

\usepackage{todonotes}
\usepackage{lipsum}

\pagestyle{plain}

 }{
\usepackage{a4wide}
\usepackage{amsmath}
\usepackage{amssymb}
\usepackage[english]{babel}
\usepackage[usenames,dvipsnames]{color}
\usepackage{enumitem}
\usepackage[breaklinks,colorlinks=true, citecolor=blue]{hyperref}
\usepackage{dsfont}
\usepackage{mathrsfs}
\usepackage{mathtools}
\usepackage{algorithm}
\usepackage[noend]{algpseudocode}
\usepackage{graphicx}
\usepackage{stmaryrd}
\usepackage{thm-restate}
\usepackage[normalem]{ulem}
\usepackage{tikz}
\usetikzlibrary{matrix}
\usetikzlibrary{arrows}
\usetikzlibrary{patterns, shapes, positioning}
\usetikzlibrary{decorations.pathmorphing,shapes}
\usepackage{makecell, multirow} 
\usepackage{setspace}
\usepackage[thinlines]{easytable}

\usepackage{comment}
\usepackage{bbm}
\usepackage{booktabs}

\usepackage[foot]{amsaddr}
\usepackage{graphicx}
\usepackage{ulem} \author{K\'evin Carrier$^{1}$}
\email{kevin.carrier@polytechnique.edu}

\author{Sana Challi$^{2}$}
\email{sana.challi@ump.ac.ma}

\author{Thomas Debris--Alazard$^{3}$}
\email{thomas.debris@inria.fr}

\address{$^{1}$ LIX, \'Ecole Polytechnique, CNRS, Inria Saclay, Institut Polytechnique de Paris, Palaiseau, France}
\address{$^{2}$ Department of Mathematics,  Mohammed First University, Morocco}
\address{$^{3}$ Inria and  Laboratoire LIX, \'Ecole Polytechnique, Palaiseau, France}

 }

\newcommand{\ie}{{\it i.e.,}}

\iftoggle{llncs}{
	\spnewtheorem{fact}{Fact}{\bfseries}{\itshape}
	\spnewtheorem{notation}{Notation}{\bfseries}{\itshape}
	}{
\newtheorem{theorem}{Theorem}[section]
	\newtheorem{proposition}[theorem]{Proposition}
	\newtheorem{lemma}[theorem]{Lemma}

\newtheorem{definition}[theorem]{Definition}
	\newtheorem{problem}[theorem]{Problem}

\newtheorem{remark}[theorem]{Remark}
	\newtheorem{notation}[theorem]{Notation}

\numberwithin{equation}{section}
	\numberwithin{figure}{section}
	\numberwithin{table}{section}
	\numberwithin{algorithm}{section}

}

\newcommand*{\eqdef}{\stackrel{\text{def}}{=}}

\renewcommand{\vec}[1]{\mathbf{#1}}

\newcommand{\negl}{\mathsf{negl}}

\newcommand{\scheme}{\textsf{MegaBight}}

\DeclareMathOperator*{\trap}{\text{\tt{Trapdoor}}}
\DeclareMathOperator*{\sampPre}{\text{\tt{InvertAlg}}}
\newcommand{\sk}{\mathsf{sk}}
\newcommand{\pk}{\mathsf{pk}}
\DeclareMathOperator*{\hash}{\text{\tt{Hash}}}
\newcommand{\Sgnsk}{\ensuremath{\mathtt{Sgn}^{{\sk}}}}
\newcommand{\Vrfypk}{\ensuremath{\mathtt{Vrfy}^{{\pk}}}}

\newcommand{\Unif}{\hookleftarrow}

\algrenewcommand\algorithmicrequire{\textbf{Input:}}
\algrenewcommand\algorithmicensure{\textbf{Output:}}
\algnewcommand\algorithmicparams{\textbf{Parameters:}}
\algnewcommand\Params{\item[\algorithmicparams]}
\algnewcommand\algorithmicparam{\textbf{Parameter:}}
\algnewcommand\Param{\item[\algorithmicparam]}

\newcommand{\signSize}{676}
\newcommand{\pkSize}{7.35}

\title{\scheme: GPV Signatures from Binary Codes}	

\begin{document}	
	
	\maketitle

	\begin{abstract}
	We present \scheme, a new family of Full-Domain-Hash (FDH) signatures based on binary codes, following the framework of Gentry, Peikert and Vaikuntanathan (\textsf{GPV}).

	The first code-based FDH scheme following the \textsf{GPV} framework, \textsf{SURF}, relied on binary~$(U,U+V)$ codes. It was broken by a hull attack which crucially exploited the structure of the binary field. A repair was given by \textsf{Wave} which moved to the ternary field. Our idea for returning to binary codes while avoiding the hull attack is rather simple: instead of increasing the alphabet size from two to three, we increase the number of underlying codes from two to three.
	
	To this aim, we introduce generalized $(U,U+V,U+V+W)$ codes built from three binary codes. We provide a decoding algorithm with an associated
	rejection sampling to prove the average preimage sampling property required by the \textsf{GPV} framework, {\it i.e.,} signatures are independent of their underlying trapdoor. Security then relies on the hardness of decoding a random binary code in the multiple-target setting and the indistinguishability of public codes from random codes.
	We propose parameters targeting $128$ bits of security, with signatures of about $\signSize$ bytes and for public keys of $\pkSize$ megabytes. The scheme enjoys very fast verification.
\end{abstract} 	\section{Introduction}

{\bf\noindent Full-Domain Hash (FDH) signatures and binary codes.}
Signing with binary codes through the FDH approach consists  in finding parity-check matrices~$\vec{H} \in \mathbb{F}_{2}^{(n-k) \times n}$ such that a solution~$\vec{e}$ of sufficiently small Hamming weight to the equation
\begin{equation}\label{eq:intro-syndrome}
	\vec{e}\vec{H}^{\top} = \vec{s} 
\end{equation}
can be found for a non-negligible proportion of all the target-vectors $\vec{s} \in \mathbb{F}_{2}^{n-k}$. The vectors~$\vec{s}$ are hashes of the messages to be signed while their associated preimages $\vec{e}$ are basically the signatures and $\vec{H}$ is the public key. The first approach to design such signature-scheme was provided by $\mathsf{CFS}$~\cite{CFS01} where it was proposed to use parity-check matrices $\vec{H}\in \mathbb{F}_{2}^{(n-k)\times n}$ whose associated binary linear codes, {\it i.e.,}~$\mathcal{C}=\left\{\vec{c}\in\mathbb{F}_2^n:\;  \vec{c}\vec{H}^{\top} = \vec{0} \right\}$, are Goppa codes. These codes admit an efficient decoding algorithm to solve~\eqref{eq:intro-syndrome} with $\vec{e}$ of Hamming weight at most $t$ whenever a solution exists. However, this decoding algorithm requires $t$ to be small and to work in an {\it injective regime of parameters}, {\it i.e.,} when Equation~\eqref{eq:intro-syndrome} admits at most one solution. This means that only a small proportion of the target-vectors~$\vec{s}$ admit a solution. The idea of $\mathsf{CFS}$ was then to choose Goppa codes of very high rate, {\it i.e.,}~$k\approx n$, because it makes this proportion 
large enough: roughly~$1/t!$ which is not too small when~$t$ is itself small. However, this hard constraint on~$t$ meant that when \textsf{CFS} was first introduced the security scaled only weakly super-polynomially in the key size for polynomial-time signature generation. The situation has recently taken a dramatic turn: \cite{BLT26} has given a polynomial-time key-recovery attack breaking $\mathsf{CFS}$ by precisely exploiting the fact that the Goppa codes used have high rate.

\medskip 
{\bf\noindent Decoding in the surjective regime.}
In order to circumvent these difficulties, it was proposed in $\mathsf{SURF}$~\cite{DST17a} to relax what we ask of the code and its associated decoding algorithm. It was no longer required to solve~\eqref{eq:intro-syndrome} in a unique decoding regime of parameters, but in a regime where it has many solutions. The weight $t$ of $\vec{e}$ is chosen large enough so that all target-vectors $\vec{s}$ admit exponentially many preimages. This is the {\em surjective regime of parameters}.

It turns out that this approach was already known in the lattice-based context through the \textsf{GPV} framework~\cite{GPV08}. In particular, it was known that choosing parameters where~\eqref{eq:intro-syndrome} admits plenty of solutions could potentially be disastrous for the security of the underlying signature scheme. Indeed, when solving~\eqref{eq:intro-syndrome}, we have to be \emph{extremely} careful: solutions of this equation are intended to be made public (these are signatures), while to compute them we precisely use a trapdoor, namely some underlying secret structure. It is therefore crucial to ensure that the distribution of solutions is independent of the trapdoor. This is the so-called design of \emph{trapdoor preimage sampleable functions}~\cite[\S5.3.1]{GPV08}. This elegant framework allows the design of secure FDH signatures, but it is generally very difficult to design an algorithm satisfying this prerequisite. In particular, very few instantiations of this framework are known.

	To instantiate this approach, $\mathsf{SURF}$~\cite{DST17a} proposed to use permuted binary~$(U,U+V)$ codes as public codes. These are obtained by permuting the following codes, which are built  from two codes $U$ and $V$,
	\begin{equation}
	(U,U+V)\eqdef
	\Big\{(\vec{u},\vec{u}+\vec{v}):
	\vec{u}\in U,\ \vec{v}\in V\Big\} \ .
	\end{equation}
It turns out that with these codes, solving~\eqref{eq:intro-syndrome} can be done by first decoding~$V$ and then using its solution to solve a decoding problem in $U$. But more importantly, by carefully choosing the dimensions of the codes $V$ and $U$, the two decoding problems which are solved (even for large decoding distances) enable to efficiently compute solutions of~\eqref{eq:intro-syndrome} which are a priori hard to find when~$\vec{H}$ is considered to be random, {\it i.e.,} without any peculiar structure. Furthermore, carefully randomizing this two steps decoder through appropriate rejection sampling even allows to obtain the distribution required by the \textsf{GPV} framework. This enabled to propose an FDH signature scheme whose trapdoor is the underlying $(U,U+V)$-structure. The security then relies on two assumptions: $(i)$ the hardness of decoding a random code, and $(ii)$ the indistinguishability of a permuted $(U,U+V)$ code from a random code.
Unfortunately, the second assumption turned out to be false precisely in the regime of parameters needed for $\mathsf{SURF}$ to work~\cite{DST17b}. To obtain a decoding advantage, the dimensions~$k_U$ and $k_V$ of~$U$ and $V$ have to satisfy~$k_U>k_V$. But, over~$\mathbb F_2$, the dual code of~$(U,U+V)$ is
\begin{equation}
(U,U+V)^{\perp}
=\Big\{(\vec{u}'+ \vec{v}',\vec{v}'):\; 
 \vec{u}'\in U^{\perp},\;  \vec{v}'\in V^{\perp}\Big\} \ .
\end{equation}
It follows that every vector $(\vec{x},\vec{x})$ with $\vec{x}\in U\cap V^{\perp}$ (which is non-trivial as $k_U> k_V$) belongs both to the code and to its dual. Indeed, its inner product with any $(\vec{u},\vec{u}+\vec{v}) \in (U,U+V)$ is 
\begin{equation}
2\langle\vec{u},\vec{x}\rangle+\langle\vec{x},\vec{v}\rangle, 
\end{equation}
and the second term is equal to $0$ as $\vec{v} \in V$ and $\vec{x} \in U \cap V^{\perp}$ by assumption while the first term vanishes over $\mathbb F_2$. This is precisely where the binary field causes a difficulty. For the parameters of $\mathsf{SURF}$, $V\cap U^{\perp}$ is typically trivial, so the hull, {\it i.e.,} the intersection of the code with its dual, is precisely
\begin{equation}
(U,U+V)\cap(U,U+V)^{\perp}
=\Big\{(\vec{x},\vec{x}):\vec{x}\in U\cap V^{\perp}\Big\} \ .
\end{equation}
Its dimension is then typically $k_U-k_V>0$, whereas a random binary code typically has a hull dimension close to zero. Computing the hull of the public code therefore gives vectors $(\vec{x},\vec{x})$ up to a secret permutation. Comparing pairs of coordinates typically allows to recover the hidden permutation breaking the $\mathsf{SURF}$ scheme.
\medskip

{\bf\noindent $\mathsf{Wave}$: increasing the alphabet size.}
Fortunately, this difficulty is closely related to the binary field and the situation changes drastically when moving to larger fields as proposed in $\mathsf{Wave}$~\cite{DST19a}. For instance, over $\mathbb F_3$ it can be considered two different ways of combining codes~$U$ and $V$, namely $(U,U+V)$ and~$(U+V,U-V)$ codes, while preserving a decoding advantage. These combinations can furthermore be chosen independently at each pair of coordinates. This gave the family of generalized $(U,U+V)$ codes. The freedom to mix these two families of codes is precisely what allowed $\mathsf{Wave}$ to avoid the hull attack against $\mathsf{SURF}$.

Another, and more surprising, ingredient of $\mathsf{Wave}$ was to introduce decoding at large distance: solutions of~\eqref{eq:intro-syndrome} over $\mathbb F_3$ have a very large Hamming weight~$t$ (typically $\approx 0.9n$) for which the trapdoor gives an advantage over generic decoding, {\it i.e.,} when $\vec{H}$ is supposed to be random. Combined with appropriate rejection sampling, this gave a secure signature scheme.
From a practical point of view, the \textsf{Wave} approach is particularly suitable for obtaining very short signatures and very fast verification. Indeed, once the message has been hashed, verification essentially amounts to checking Equation~\eqref{eq:intro-syndrome} and the weight of $\vec{e}$. For instance, $\mathsf{Wave}$-822, which targets NIST security level $I$, has signatures of at most $822$ bytes and a public key of $3.68$ MB~\cite[\S5]{BCCCDGKLNSST23}. \textsf{Wave} non-optimized reference implementation verifies a signature in about $1.23$ million cycles on an Intel Core i5-1135G7 platform at
2.40 GHz.

The main drawback of \textsf{Wave} is therefore the rather large size of its public keys. Fortunately, it has recently been shown \cite{BDS25} that useful trade-offs are possible. 
After preprocessing a \textsf{Wave} public key, a verifier can store a much smaller private verification key. For $\mathsf{Wave}$-822, construction of \cite{BDS25} replaced the storage of a $3.68$MB public key by that of a verification key of about~$172$kB with even faster verification. These gains came at the cost of using larger signatures, whose size is roughly twice that of the original signatures. This shows that the large size of the public key does not imply large storage requirements for verification.
\medskip

{\bf\noindent Returning to binary codes.} The initial question of building a secure and efficient FDH signature scheme from {\it binary} codes remains open. Obtaining one would already be of interest from an implementation point of view. Indeed, working over the ternary field as in \textsf{Wave} requires specific algorithms even for tasks as basic as hashing the message to sign. For instance, in $\mathsf{Wave}$~\cite[Appendix~\S A]{BCCCDGKLNSST23}, binary hash outputs have to be converted into ternary vectors, with an additional sampling procedure to obtain the required distribution. In the binary case, the output of a standard extendable-output hash function already belongs to $\mathbb F_2^{n-k}$ and there is no need for this conversion. More generally, working over $\mathbb F_2$ allows us to use the binary arithmetic of the machine directly.

	Unfortunately, binary decoding gives us very little structure to design trapdoors and this is well illustrated by the lattice-based setting where the situation is quite different. There, the corresponding decoding problem in its surjective regime of parameters is the so-called Inhomogeneous Short Integer Solution (\textsf{ISIS}): given a matrix $\vec{A}$ and a target $\vec{s}$ modulo $q$, the aim is to find an integer vector $\vec{x}$ with small coefficients modulo $q$ such that,\iftoggle{llncs}{$\vec{x}\vec{A}^{\top}= \vec{s}\pmod q$}{
	\begin{equation}
	\vec{x}\vec{A}^{\top}= \vec{s}\pmod q\ .
	\end{equation}}

Here $q$ can be chosen large and increased with the dimension. This \iftoggle{llncs}{}{innocent-looking} fact has provided considerable freedom to design trapdoors. This degree of freedom is completely absent in the binary case: the alphabet is limited to two elements. 
This lack of freedom partly explains why many cryptographic constructions remain \iftoggle{llncs}{}{for now} out of reach in the binary setting. For instance, adapting the \textsf{GPV} \iftoggle{llncs}{}{approach} to design an identity-based encryption (IBE) scheme in the binary setting can only lead to insecure instantiations \cite{DT18b}. Fully homomorphic encryption from standard binary decoding assumptions also remains an open problem. These difficulties persist despite the long history of binary decoding: it is the oldest problems used in post-quantum cryptography, with algorithms to solve it dating back to $1962$~\cite{P62} and cryptographic applications to $1978$~\cite{M78}. It is therefore maybe not accidental that obtaining a practical \textsf{GPV} signature with Hamming-metric codes first required increasing the alphabet size from two to three.
\medskip

{\bf\noindent Our contribution: a binary signature from three random codes.}
Motivated by this current state of affairs, we have wished to obtain a binary instantiation of\iftoggle{llncs}{}{ the} \textsf{GPV} \iftoggle{llncs}{}{framework}. Our starting point is the lesson drawn from the passage from $\mathsf{SURF}$ to $\mathsf{Wave}$. The difficulty with $\mathsf{SURF}$ was that the $(U,U+V)$--structure\iftoggle{llncs}{}{, which enabled decoding,} also appeared in the hull of the public code. $\mathsf{Wave}$ overcame this difficulty by mixing $(U,U+V)$ and $(U+V,U-V)$ codes which is possible over $\mathbb{F}_{3}$. But, over $\mathbb F_2$, the second combination $(U+V,U-V)$ becomes $(U+V,U+V)$ and it no longer offers a decoding advantage. We therefore need another way of obtaining this freedom while keeping the alphabet binary. Our idea is rather simple:
\begin{center}
	{\it instead of increasing the alphabet size, we increased the number of underlying codes.}
\end{center}

To this aim, we choose three random binary codes $U,V,W$ of length $n/3$ and introduce generalized $(U,U+V,U+V+W)$ codes. Their codewords, up to a secret permutation, have the form
\begin{equation}
\big(\vec{u}+ \vec{v},\, \vec{u}+ \vec{a} \star \vec{w},\,  \vec{u}+ \vec{b}\star \vec{v}+ \vec{w}\big),
\end{equation}
where $\vec{u}\in U$, $\vec{v}\in V$, $\mathbf w\in W$, and $\vec{a},\vec{b}\in\mathbb F_2^{n/3}$ satisfy\footnote{It ensures that $(\vec{u},\vec{v},\vec{w}) \mapsto \big(\vec{u}+ \vec{v},\, \vec{u}+ \vec{a} \star \vec{w},\,  \vec{u}+ \vec{b}\star \vec{v}+ \vec{w}\big)$ is invertible.} $\vec{a}\star\vec{b}=\vec{a}$. Here $\star$ denotes component-wise multiplication. Notice that this condition leaves three possible choices for each $(a_i,b_i)$, namely~$(0,0)$,~$(0,1)$ and~$(1,1)$. We can therefore mix three combinations across the coordinates while working over $\mathbb F_2$.

This mixing is crucial for hiding the structure. In particular, we show via experiments that the empirical distribution of the hull dimension behaves like that of random binary codes. The large hull which was at the origin of the attack against $\mathsf{SURF}$ no longer appears. However, taking three codes instead of two cannot be considered only as a way of hiding the structure: we also have to show a decoding advantage. Indeed, hiding all structure would be easy if we could simply choose the public code at random, but we would then lose the trapdoor which enables us to sign. The difficulty is therefore to obtain a more diverse family while keeping enough structure to decode efficiently. This is what we achieve with the above construction.

To decode, we proceed as in \textsf{SURF} and \textsf{Wave}: we decode successively and take into account the previous decoding outputs. In our case, we first decode in~$W$, then use the obtained solution when decoding in $V$, and finally take both solutions into account when decoding in $U$. The point is that these three decoding steps cannot be done independently: the first two solutions constrain the choices made in the last step to reach the target weight. With the right choice of parameters, their combination gives solutions of~\eqref{eq:intro-syndrome} with a weight for which the best known generic decoding algorithms remain exponential, while working in the surjective regime.

However the story is not ended: it remains to ensure that the signatures do not leak information. To achieve this task, we succeeded in designing efficient rejection sampling procedures for the three decoding steps and proving the resulting \textsf{GPV} average preimage sampling property. This crucial property enabled us to prove, as in $\mathsf{SURF}$, that the security of our \scheme-scheme, reduces to the hardness of decoding a random binary code\footnote{More precisely, we consider the multiple-target setting, namely the Decoding One-Out-of-Many problem (\textsf{DOOM}).} and the hardness of distinguishing between a public code and a random code.

It makes \scheme{} a secure signature scheme based on {\it binary} codes which scales well with the parameters: it can be shown that if one wants a
security level of $2^{\lambda}$, then signature size is of order $O(\lambda)$, the public key size is of order $O(\lambda^{2})$, signature
generation is of order $O(\lambda^{3})$, whereas signature verification is of order $O(\lambda^{2})$.
We also propose concrete parameters targeting $128$ bits of security. In particular, we obtain very short signatures of about $\signSize$B with public keys of $\pkSize$MB (see Table~\ref{tab:parameters}). This gives a binary instantiation of the \textsf{GPV} approach and answers the question left open by $\mathsf{SURF}$: designing an efficient and secure {\it binary} hash-and-sign signature scheme.

 	\section{Notation}

{\bf \noindent Basic notation.} The notation $x \eqdef y$ means that
$x$ is being defined as equal to~$y$. Let $a < b$ be integers. We let
$[a,b]$ denote the set of integers $\{a,a+1,\dots,b\}$. \iftoggle{llncs}{}{We use the convention that~$\binom{a}{b}=0$ when $b\notin[0,a]$.} Notation
$\negl(\lambda)$ refers to a function that is $O(1/\lambda^{b})$ for
every constant $b > 0$. Given a finite set $\mathcal{S}$, notation $|\mathcal{S}|$ or $\sharp \mathcal{S}$ refers to its cardinal. Vectors are in row notation and they will be
written with lowercase bold letters such as $\vec{a}$. Uppercase bold letters
such as~$\vec{A}$ are used to denote matrices. We let $\mathbb{F}_{2}$ denote the binary field\iftoggle{llncs}{.}{, {\it i.e.,} the finite field with two elements.} Given integers~$m,n$ we denote by
$\mathbb{F}_{2}^{m \times n}$ the set of $m\times n$ matrices with
entries in~$\mathbb{F}_{2}$. 
In what follows, we let $\mathcal{S}_{t}^{n}$ 
denote the sphere of radius $t$ around $\vec{0}$ in $\mathbb{F}_{2}^{n}$
for the Hamming weight~$|\cdot |$, which is defined by \iftoggle{llncs}{$\forall \vec{x} \in \mathbb{F}_{2}^{n}, \quad |\vec{x}| \eqdef \sharp \left\{ i \in [1,n]: \; x_i \neq 0  \right\}$.\medskip}{
\begin{equation}
\forall \vec{x} \in \mathbb{F}_{2}^{n}, \quad |\vec{x}| \eqdef \sharp \left\{ i \in [1,n]: \; x_i \neq 0  \right\} .
\end{equation}
}
Given two vectors $\vec{a},\vec{b}\in \mathbb{F}_{2}^{n}$, we let $\vec{a}\star\vec{b}$ denote their component-wise product, {\it i.e.,} $(a_i b_i)_{i \in [1,n]}$.

\iftoggle{llncs}{}{\medskip}
{\bf \noindent Linear codes.} A binary linear code $\mathcal{C}$ of length $n$ and dimension $k$ is a subspace
of $\mathbb{F}_{2}^{n}$ with dimension $k$. We say that it has parameters $[n,k]$ or that it is an~$[n,k]$-code. 
A parity-check matrix for $\mathcal{C}$ is a full-rank $(n-k) \times n$ matrix over $\mathbb{F}_{2}$ such that $\mathcal{C}$ is its right kernel, {\it i.e.,} \iftoggle{llncs}{$\mathcal{C} = \left\{ \vec{c} \in \mathbb{F}_{2}^{n}: \; \vec{H}\vec{x}^{\top} = \vec{0} \right\}$}{
$$
\mathcal{C} = \left\{ \vec{c} \in \mathbb{F}_{2}^{n}: \; \vec{H}\vec{x}^{\top} = \vec{0} \right\} .
$$}
The code $\mathcal{C}^{\perp}$ whose basis consists of the rows of $\vec{H}$ is called the dual code of $\mathcal{C}$. It is an $[n,n-k]$-code.
\medskip

{\bf\noindent Probabilistic notation.} For a probability
distribution $\mathcal{D}$ over some set $\mathcal{E}$, we let
$X\Unif \mathcal{D}$ denote a random variable distributed
according to $\mathcal{D}$. When~$\mathcal{E}$ is finite,~$X\Unif \mathcal{E}$ will simply denote a uniformly random
variable over $\mathcal{E}$. \iftoggle{llncs}{}{Sometimes when we compute probabilities or
expectations, we write $\mathbb{P}_{r}(A)$ or $\mathbb{E}_{r}(A)$ to denote that
they are computed over the randomness~$r$.} Throughout the paper, whenever we condition on an event, we implicitly assume that this event has positive probability.

The proximity of two random variables $X$ and $Y$ defined over a finite set~$\mathcal{E}$ can be measured with respect to their (total variation)
statistical distance $\Delta(X, Y)$ defined by\iftoggle{llncs}{ $\Delta(X, Y) \eqdef \dfrac{1}{2} \sum_{a\in \mathcal{E}} \left| \mathbb{P}(X=a) - \mathbb{P}(Y=a) \right|$.}{
\begin{equation}
\Delta(X, Y) \eqdef \dfrac{1}{2} \sum_{a\in \mathcal{E}} \left| \mathbb{P}(X=a) - \mathbb{P}(Y=a) \right| \ .
\end{equation}} 		\section{\scheme{} Trapdoor}

	\subsection{Code-based signatures via average trapdoor preimage sampleable functions}

We focus in this paper on instantiating the Full Domain Hash (FDH) paradigm~\cite{BR96} in the code--based setting. Our aim is therefore to design an FDH--scheme whose security follows from the (average) hardness of the Decoding Problem (\textsf{DP}). It is stated in its dual form \iftoggle{llncs}{}{(it also admits an equivalent {\it primal} form, see for instance~\cite[Ch. 2]{D23})} as follows. 
\begin{problem}[$\mathsf{DP}$] Let $n\geq k,t$ be integers. The parameters $n,k,t$ are functions of some security parameter $\lambda$.  Let $\vec{H} \in \mathbb{F}_{2}^{(n-k)\times n}, \vec{x} \in \mathcal{S}_{t}^{n}$ be sampled uniformly at random and $\vec{s} \eqdef \vec{x}\vec{H}^{\top}$. The~$\mathsf{DP}(n,k,t)$ problem is to find $\vec{e} \in \mathcal{S}_{t}^{n}$ such that~$\vec{s} = \vec{e}\vec{H}^{\top}$ given the pair $(\vec{H},\vec{s})$. 
\end{problem}

\iftoggle{llncs}{}{
\begin{remark}
	Notice that given $\vec{s} = \vec{x}\vec{H}^{\top}$ with $|\vec{x}|=t$, we do not require to recover $\vec{x}$. 
\end{remark}
}

This leads us to consider the following one-way function,	 
\begin{equation}
	f_{\vec{H},t}: \vec{x} \in \mathcal{S}_{t}^{n} \longmapsto \vec{x}\vec{H}^{\top} \in \mathbb{F}_{2}^{n-k}.
\end{equation}

The FDH paradigm requires to be able to invert (thanks to a trapdoor) this function on any input $\vec{s} \in \mathbb{F}_{2}^{n-k}$, or at least for a density close to one of inputs. This constrains how parameters~$n,k,t$ have to be chosen. Hopefully we know which precise constraint needs to be verified as the number of preimages can accurately be estimated (see for instance \cite[Proposition 2.3.1]{D23}). 
\begin{proposition}\label{propo:numberSolution}
	Let $n\geq k$, and $t>0$ be integers. We have, \iftoggle{llncs}{for $\vec{H} \Unif \mathbb{F}_{2}^{(n-k)\times n}$,}{}
	\begin{equation}
	\forall \vec{s} \in \mathbb{F}_{2}^{n-k}, \quad \mathbb{E}_{\vec{H}}\left( \sharp\left\{ \vec{e} \in f_{\vec{H},t}^{-1}(\vec{s}) \right\} \right) = \frac{\binom{n}{t}}{2^{n-k}} = 2^{n\Big( h(t/n) - (1-k/n)  + o(1) \Big)}
	\end{equation}
	\iftoggle{llncs}{where }{where $\vec{H} \Unif \mathbb{F}_{2}^{(n-k)\times n}$ 
and} $h: x \mapsto -x \log_2 x - (1-x)\log_2(1-x)$ denotes the binary entropy. 
\end{proposition}
It turns out that $h$ is an increasing function over $[0,1/2]$ and $h(0) = 0$ while $h(1/2) = 1$. Therefore, for a fixed rate $k/n$, the expected number of inverses will be exponential (in $n$) when~$h(t/n) > 1-k/n$ by a positive constant. The boundary of this range is  the so-called {\it Gilbert-Varshamov radius}.

\begin{definition}[Gilbert-Varshamov radius]
	Let $n\geq k$ be integers. We define the Gilbert-Varshamov radius for parameters $n,k$ as\iftoggle{llncs}{ $t_{\textup{GV}}(n,k) \eqdef \left\lceil n \cdot h^{-1}\left( 1- \tfrac{k}{n} \right) \right\rceil$}{ 
	\begin{equation}
	t_{\textup{GV}}(n,k) \eqdef \left\lceil n \cdot h^{-1}\left( 1- \tfrac{k}{n} \right) \right\rceil
	\end{equation}} 	 
	where $h^{-1}$ denotes the inverse of $h$ over $[0,1/2]$. 
\end{definition} 
Proposition \ref{propo:numberSolution} has an important consequence: in order to hope that $f_{\vec{H},t}$ is surjective, we have to choose parameters $n,k,t$ such that $t \geq t_{\textup{GV}}(n,k)$. However, our scheme will enforce a slightly more restrictive condition, namely\iftoggle{llncs}{ $t \geq (1+\varepsilon)t_{\textup{GV}}(n,k)$}{
$ 
t \geq (1+\varepsilon)t_{\textup{GV}}(n,k) 
$
}
for some constant $\varepsilon > 0$. It implies that for any input $\vec{s}$, we expect {\it an exponential number of inverses} $\vec{x} \in f_{\vec{H},t}^{-1}(\vec{s})$. This situation is reminiscent of the lattice case where FDH--schemes such as \textsf{Falcon} \cite{FHKLPPRSWZ} also rely on inverting a one-way function with an exponential number of preimages (by considering the \textsf{ISIS} problem). This innocent-looking fact has a huge consequence in terms of security: when inverting~$f_{\vec{H},t}$ with the help of the trapdoor we have to be extremely careful on how we choose the inverse, otherwise we could reveal information on the trapdoor. Some lattice--
and code--based FDH--schemes~\cite{HHPSW03,NCLKK22} were broken~\cite{NR09,DLV24} as they did not take into account this potential
flaw.

Designing a secure FDH signature scheme in this situation can fortunately be achieved by imposing additional properties \cite{GPV08} to the
one-way function. This is mostly captured by the notion of
Trapdoor Preimage Sampleable Functions (TPSF)~\cite[Definition 5.3.1]{GPV08}. However, in our case we will instead rely on a slight variation of TPSF: Average Trapdoor Preimage Sampleable Functions (ATPSF) as introduced in \cite[Definition 1]{DST19a}. This will be enough to tightly reduce the security of our scheme to a multiple-instance variant of $\mathsf{DP}$ as shown in \cite{CD20}. Roughly speaking, (A)TPSF require the distribution of the trapdoor inverse output to be independent of the trapdoor.

\begin{definition}[Average Trapdoor Preimage Sampleable Code-based Functions]\label{def:ATPSF}
	It is a pair of probabilistic
	polynomial-time algorithms $(\trap,\sampPre)$ together with parameters $n,k,t$ which are functions of some security parameter $\lambda$. 
	\begin{itemize}\setlength{\itemsep}{6pt}
		\item $\trap$, when given $\lambda$, outputs $(\vec{H},T)$ where $\vec{H}\in\mathbb{F}_{2}^{(n-k)\times n}$ and $T$ the trapdoor
		corresponding to $\vec{H}$.

		\item $\sampPre$ is a probabilistic algorithm which takes as input $\vec{s} \in \mathbb{F}_{2}^{n-k}$, $T$ and outputs~$\vec{e} \in
		\mathcal{S}_{t}^{n}$ such that $\vec{e}\vec{H}^{\top} = \vec{s}$.
	\end{itemize}
	The following average trapdoor preimage sampling property has then to hold: 
	$$
	\mathbb{E}_{(\vec{H},T)}\Big( \Delta \left( \sampPre(\vec{s},T),\vec{e} \right)\Big) \in \textup{negl}(\lambda)
	$$
	where $\vec{e} \in \mathcal{S}_{t}^{n}$ and $\vec{s} \in \mathbb{F}_{2}^{n-k}$ are uniformly distributed.
\end{definition}

Given a code-based ATPSF
$(\trap,\sampPre)$ we easily define a code-based FDH signature scheme. We first generate the public and secret keys as
$(\pk,\sk)=(\vec{H},T) \leftarrow \trap(\lambda)$.
We also select a cryptographic binary hash function $\hash: \{0,1\}^{*}
\rightarrow \mathbb{F}_{2}^{n-k}$ and a salt $\vec{r}$ of size $\lambda$.
The signing and verification algorithms \Sgnsk\ and \Vrfypk\ are then defined as follows.
\medskip 

\begin{center}
	\begin{tabular}{l@{\hspace{3mm}}|@{\hspace{3mm}}l}
		$\Sgnsk(\vec{m})\!\!: \qquad \qquad \qquad$ & $\Vrfypk(\vec{m}',(\vec{e}',\vec{r}'))\!\!:$ \\
		$\quad \vec{r} \Unif \{ 0,1 \}^{\lambda}$ &$\quad \vec{s} \leftarrow \hash(\vec{m}',\vec{r}')$ \\
		$\quad \vec{s} \leftarrow \hash(\vec{m} ,\vec{r})$ & $\quad \texttt{if } \vec{e}'\vec{H}^{\top} = \vec{s} \texttt{ and } |\vec{e}'| = t \texttt{ return } 1$\\ 
		$\quad\vec{e} \leftarrow \sampPre(\vec{s},T) $ &$\quad \texttt{else return } 0 $ \\
		$\quad \texttt{return}(\vec{e},\vec{r})$& \\			
	\end{tabular} 
\end{center}
\medskip 

\iftoggle{llncs}{
A tight security reduction (for the EUF-CMA security model) in the random oracle model for signature schemes instantiated with ATPSF  has been given in~\cite{CD20}. It reduces the security to the hardness of the Decoding One Out of Many (\textsf{DOOM}) variant of \textsf{DP} (as introduced in~\cite{S11}.
}{
A security reduction (for the EUF-CMA security model) in the random oracle model for signature schemes instantiated with TPSF  has been given in
\cite{GPV08}. However it requires collision
resistance, {\it i.e.,} the hardness of finding $\vec{x}$ and $\vec{y}$ with Hamming weight $t$ such that $\vec{x}\vec{H}^{\top} = \vec{y}\vec{H}^{\top}$, which is not a tight security assumption (and not verified by our scheme). We will instead rely on the hardness of the following Decoding One Out of Many (\textsf{DOOM}) variant of \textsf{DP} (as introduced in~\cite{S11}) which has been shown to be a tight security assumption for ATPSF in the random oracle and quantum random oracle models~\cite{CD20}.

\begin{problem}[$\mathsf{DOOM}$]
\label{prob:DOOM}
	Let $n \geq k,t$ and $N$ be integers. The parameters $n,k,t,N$ are functions of some security parameter $\lambda$. Let $\vec{H} \in \mathbb{F}_{2}^{(n-k) \times n}$ and $\vec{s}_1,\dots,\vec{s}_N \in \mathbb{F}_{2}^{n-k}$ be sampled uniformly at random. The $\mathsf{DOOM}(n,k,t,N)$ problem is to find~$\vec{e} \in \mathcal{S}_{t}^{n}$ and $i \in [1,N]$ such that $\vec{e}\vec{H}^{\top} = \vec{s}_{i}$ given~$(\vec{H},\vec{s}_1,\dots,\vec{s}_N)$. 
\end{problem}
}

\subsection{The \scheme{} family of code-based average trapdoor preimage sampleable functions}

Our trapdoor, which gives an advantage in inverting $f_{\vec{H},t}$ consists in considering parity-check matrices~$\vec{H}$ defining the following family of codes. 
\begin{definition}[Generalized $(U,U+V,U+V+W)$ codes]\label{def:UVW}
	Let $n$ be an integer,~$\vec{P} \in \mathbb{F}_{2}^{n\times n}$ be a permutation matrix and $\vec{a},\vec{b} \in \mathbb{F}_{2}^{n/3}$ satisfy \iftoggle{llncs}{}{the following equation:}
	\begin{equation}\label{eq:cdtXY}
		\vec{a}\star\vec{b} = \vec{a}\ .
	\end{equation} We define, 
\begin{equation}
		\begin{array}{lccc}
			\varphi_{\vec{a},\vec{b}, \vec{P}}: & \mathbb{F}_{2}^{n/3} \times \mathbb{F}_{2}^{n/3} \times \mathbb{F}_{2}^{n/3} & \longrightarrow &  \mathbb{F}_{2}^{n/3} \times \mathbb{F}_{2}^{n/3} \times \mathbb{F}_{2}^{n/3} \\
			& (\vec{u},\vec{v},\vec{w}) & \longmapsto &  \left( \vec{u} + \vec{v}, \vec{u}  + \vec{a}\star \vec{w}, \vec{u} + \vec{b}\star \vec{v}+ \vec{w} \right)\vec{P} \ .
		\end{array}
		\end{equation}
		Given three codes $U,V,W$ with length $n/3$, 
		\begin{equation}
		\varphi_{\vec{a},\vec{b},\vec{P}}(U,V,W) = \Big\{ \varphi_{\vec{a},\vec{b},\vec{P}}(\vec{u},\vec{v},\vec{w}): \; \vec{u} \in U, \; \vec{v} \in V \mbox{ and } \vec{w} \in W \Big\}
		\end{equation}
		defines a linear code with length $n$ called generalized $(U,U+V,U+V+W)$ code. If $k_U, k_V$ and~$k_W$ denote the dimensions of $U,V$ and $W$, then it has dimension~$k \eqdef k_U+k_V+k_W$ when~$n \geq k$. Furthermore, the dual of $\varphi_{\vec{a},\vec{b},\vec{P}}(U,V,W)$ is equal to 
\iftoggle{llncs}{\begin{multline}
		\varphi^{\perp}_{\vec{a},\vec{b},\vec{P}}(U^{\perp},V^{\perp},W^{\perp}) = \Big\{ \varphi^{\perp}_{\vec{a},\vec{b},\vec{P}}(\vec{u}^{\perp},\vec{v}^{\perp},\vec{w}^{\perp}): \\ \vec{u}^{\perp} \in U^{\perp}, \; \vec{v}^{\perp} \in V^{\perp} \mbox{ and } \vec{w}^{\perp} \in W^{\perp} \Big\}
\end{multline}}{
	\begin{equation}
	\varphi^{\perp}_{\vec{a},\vec{b},\vec{P}}(U^{\perp},V^{\perp},W^{\perp}) = \Big\{ \varphi^{\perp}_{\vec{a},\vec{b},\vec{P}}(\vec{u}^{\perp},\vec{v}^{\perp},\vec{w}^{\perp}): \; \vec{u}^{\perp} \in U^{\perp}, \; \vec{v}^{\perp} \in V^{\perp} \mbox{ and } \vec{w}^{\perp} \in W^{\perp} \Big\}
	\end{equation}
}
	\iftoggle{llncs}{}{where}
	\begin{equation}
		\begin{array}{lccc}
			\varphi^{\perp}_{\vec{a},\vec{b}, \vec{P}}: & \mathbb{F}_{2}^{n/3} \times \mathbb{F}_{2}^{n/3} \times \mathbb{F}_{2}^{n/3} & \longrightarrow &  \mathbb{F}_{2}^{n/3} \times \mathbb{F}_{2}^{n/3} \times \mathbb{F}_{2}^{n/3} \\
			& (\vec{u}^{\perp}, \vec{v}^{\perp}, \vec{w}^{\perp}) & \longmapsto &  \Big( \vec{a} \star (\vec{u}^{\perp} + \vec{v}^{\perp}) + \vec{v}^{\perp} + \vec{b}\star\vec{w}^{\perp} , \\
& & & \quad \quad \vec{u}^{\perp} + \vec{v}^{\perp} + \vec{w}^{\perp} + \vec{b}\star\vec{w}^{\perp}, \\
& & & \quad \quad \quad \quad  \vec{a}\star(\vec{u}^{\perp} + \vec{v}^{\perp}) + \vec{w}^{\perp} \Big)\vec{P} \ .
	\end{array}
	\end{equation}
	\end{definition}

	Given a parity-check matrix $\vec{H}$ of a generalized $(U,U+V,U+V+W)$ code, knowing parity-check matrices of codes $U, V$ and $W$ as well as the underlying vectors $\vec{a}, \vec{b}$ and the permutation $\vec{P}$ will provide an advantage in inverting~$f_{\vec{H},t}$ as shown in Subsection \S\ref{subsec:invertTrap}. However, it is obviously not true that any code is a generalized $(U,U+V,U+V+W)$ code. In particular random codes\iftoggle{llncs}{}{\footnote{Random codes are defined as codes whose parity-check matrix has been sampled uniformly at random.}} do not belong to this family with high probability. Therefore inverting $f_{\vec{H},t}$ 
does not amount to solving $\mathsf{DP}$. Nevertheless, as proved in \cite{CD20}, the security can still be reduced to the variant $\mathsf{DOOM}$ of~$\mathsf{DP}$, but only if we assume the hardness of distinguishing between a generalized~$(U,U+V,U+V+W)$ code and a random code, problem stated as follows. 
\begin{problem}[$\mathsf{D}^{\mathsf{UVW}}$]\label{prob:UVWD}
		Let $k_U,k_V,k_W$, $k \eqdef k_U + k_V + k_W$ and $n \geq k$ be integers. The parameters~$n, k_U, k_V$ and $k_W$ are functions of some security parameter~$\lambda$. Let~$\vec{H}_{0} \in \mathbb{F}_{2}^{(n-k) \times n}$ be sampled uniformly at random and $\vec{H}_{1}\in \mathbb{F}_{2}^{(n-k) \times n}$ be sampled uniformly at random among all parity-check matrices of generalized~$(U,U+V,U+V+W)$ codes with length $n$, dimensions $k_U,k_V$ and $k_W$ for codes $U, V$ and~$W$. Let~$b \in \{0,1\}$ be a uniform bit. The $\mathsf{D}^{\mathsf{UVW}}(n,k_U,k_V,k_W)$ problem is to find~$b$ given $\vec{H}_{b}$. 
	\end{problem}
	\iftoggle{llncs}{}{
	\begin{remark}
		It is easy to distinguish between random codes and generalized $(U,U+V,U+V+W)$ codes {\bf without} the permutation $\vec{P}$. The addition of the permutation $\vec{P}$ in our construction may seem reminiscent of the McEliece scheme \cite{M78}. However, we would like to point out that such addition is crucial in our case while it is useless in terms of security in the original McEliece instantiation: a permuted Goppa code is still a Goppa code.
\end{remark}
}
	
	Our considered family of codes may seem at first sight quite convoluted. We consider it as it gives an advantage in inverting $f_{\vec{H},t}$. However, before explaining how to exploit this structure, let us explain our first rationale for considering it: to ensure security, namely that $\mathsf{D}^{\mathsf{UVW}}$ seems indeed hard.

	Notice that the following three codes (by forgetting the permutation matrix~$\vec{P}$) are particular instantiations of generalized $(U,U+V,U+V+W)$ codes by carefully choosing vectors $\vec{a}$~and~$\vec{b}$ satisfying the condition $\vec{a}\star\vec{b} = \vec{a}$,
	\iftoggle{llncs}{
		\begin{itemize}
		\item Choosing $\vec{a} = \vec{b} = \vec{0}_{n/3}$, 
	\end{itemize}
		\begin{equation}
		(U+V,U,U+W) \eqdef \left\{ (\vec{u}+\vec{v}, \vec{u},\vec{u}+\vec{w}): \vec{u} \in U, \vec{v} \in V,\vec{w} \in W \right\}
		\end{equation}
	\begin{itemize} 
		\item Choosing $\vec{a} = \vec{0}_{n/3}$ and  $\vec{b} = \vec{1}_{n/3}$, 
		\begin{equation}\label{eq:UVW}
			(U+V,U,U+V+W) \eqdef \left\{ (\vec{u}+\vec{v}, \vec{u},\vec{u}+\vec{v}+\vec{w}): \vec{u} \in U, \vec{v} \in V, \vec{w} \in W \right\}
		\end{equation}
\end{itemize}
\begin{itemize} 
		\item Choosing $\vec{a} = \vec{b} = \vec{1}_{n/3}$, 
		\begin{equation}
		(U+V,U+W,U+V+W) \eqdef \left\{ (\vec{u}+\vec{v}, \vec{u}+\vec{w},\vec{u}+\vec{v}+\vec{w}): \vec{u} \in U, \vec{v} \in V,\vec{w} \in W \right\}
		\end{equation}
	\end{itemize} 
	}{
	\medskip 
	\begin{itemize}
		\item Choosing $\vec{a} = \vec{b} = \vec{0}_{n/3}$, 
	\begin{equation}
		(U+V,U,U+W) \eqdef \left\{ (\vec{u}+\vec{v}, \vec{u},\vec{u}+\vec{w}): \; \vec{u} \in U, \; \vec{v} \in V, \; \vec{w} \in W \right\},
	\end{equation}
	\item Choosing $\vec{a} = \vec{0}_{n/3}$ and  $\vec{b} = \vec{1}_{n/3}$, 
	\begin{equation}\label{eq:UVW}
	(U+V,U,U+V+W) \eqdef \left\{ (\vec{u}+\vec{v}, \vec{u},\vec{u}+\vec{v}+\vec{w}): \; \vec{u} \in U, \; \vec{v} \in V, \; \vec{w} \in W \right\},
	\end{equation} 
	\item Choosing $\vec{a} = \vec{b} = \vec{1}_{n/3}$, 
	\begin{equation}
	(U+V,U+W,U+V+W) \eqdef \left\{ (\vec{u}+\vec{v}, \vec{u}+\vec{w},\vec{u}+\vec{v}+\vec{w}): \; \vec{u} \in U, \; \vec{v} \in V, \; \vec{w} \in W \right\} .
	\end{equation}
\end{itemize} 
}
	
	We could have restricted ourselves to only one of these three families of codes to obtain an advantage when inverting $f_{\vec{H},t}$. However, each of these codes presents both advantages and disadvantages with respect to security. For example, the first code, $(U+V, U, U+W)$, contains codewords with smaller Hamming weights -- such as $(\vec{v}, \vec{0}, \vec{0})$ -- than those found in the~$(U+V, U+W, U+V+W)$ code. This characteristic may make it easier to distinguish a $(U+V, U, U+W)$ code from a random code rather than to distinguish a~$(U+V, U+W, U+V+W)$ code from a random code\iftoggle{llncs}{.}{\footnote{Actually, in our case, $k_U$ is much larger than $k_V$, so the shortest codewords are of the form $(\vec{u}, \vec{u}, \vec{u})$. However, the same argument remains valid for the dual code.}.}
However, though~$(U+V,U+W,U+V+W)$ codes seem a better choice, when looking at the Hamming weight distribution of their codewords, they have a security issue that $(U+V,U,U+W)$ codes do not share. It turns out that  
the hull of $(U+V,U+W,U+V+W)$ codes (and of $(U+V,U,U+V+W)$ codes) experimentally behaves slightly differently from  the hull of random codes, thus giving a way to distinguish them from random codes. 
	
	Our solution to keep only the advantages of each of the aforementioned families of codes, avoiding their flaws, is to {\it mix} them as in Definition \ref{def:UVW}. By mixing we mean that for each triplet of coordinates\iftoggle{llncs}{ $(i,i+n/3,i+2n/3)$ for~$i \in [1,n/3]$}{
	$$
	\Big(i,i+n/3,i+2n/3\Big) \quad \mbox{ for~$i \in [1,n/3]$},
	$$} 
	any codeword $\vec{c}$ of a generalized $(U,U+V,U+V+W)$ code (up to a permutation) is such that~$(\vec{c}(i), \vec{c}(i+n/3),\vec{c}(i+2n/3))$ is equal to (according to\iftoggle{llncs}{}{ the choice of}~$\vec{a}(i)$ and $\vec{b}(i)$),\iftoggle{llncs}{$\Big(\vec{u}(i)+\vec{v}(i), \vec{u}(i), \vec{u}(i)+\vec{w}(i)\Big)$, or $\Big(\vec{u}(i)+\vec{v}(i), \vec{u}(i), \vec{u}(i) + \vec{v}(i) + \vec{w}(i)\Big)$, or $\Big(\vec{u}(i)+\vec{v}(i), \vec{u}(i)+\vec{w}(i), \vec{u}(i)+\vec{v}(i)+\vec{w}(i)\Big).$}{
	$$
	\Big(\vec{u}(i)+\vec{v}(i), \vec{u}(i), \vec{u}(i)+\vec{w}(i)\Big), \mbox{ or } \; \Big(\vec{u}(i)+\vec{v}(i), \vec{u}(i), \vec{u}(i) + \vec{v}(i) + \vec{w}(i)\Big), 
	$$
	$$
	\mbox{or } \; \Big(\vec{u}(i)+\vec{v}(i), \vec{u}(i)+\vec{w}(i), \vec{u}(i)+\vec{v}(i)+\vec{w}(i)\Big)
	$$ 
}
for some $\vec{u} \in U, \vec{v} \in V$ and $\vec{w} \in W$. This ``mixing construction'' has therefore the advantage of removing codewords with Hamming weight that is too small, {\it e.g.,}~$(\vec{v},\vec{0},\vec{0})$, and at the same time we do not consider $(U+V,U+W,U+V+W)$ codes which have hull issues. Fortunately, our construction effectively erases them as we will see via experiments in Subsection \S\ref{sec:instantiation}, in particular Figure~\ref{fig:hull-distribution}.

 	\section{Inverting with and without trapdoor: the \scheme{} trapdoor functions}

\subsection{Inverting without the trapdoor}

Before showing that our trapdoor gives us an advantage in inverting $f_{\vec{H},t}$, we first need to know when this function is easy to invert (for almost all matrices~$\vec{H}$), namely for which parameters we know how to solve~$\mathsf{DP}(n,k,t)$ in polynomial time (in $n$). This is given by the condition: 
\begin{equation}
t \in \left[ t_{\textup{easy}}(n,k), k + t_{\textup{easy}}(n,k)  \right] \quad \mbox{where } \; t_{\textup{easy}}(n,k) \eqdef \frac{n-k}{2}\ .
\end{equation}
Surprisingly, despite many efforts, the range of parameters for which we can invert $f_{\vec{H},t}$ in polynomial time has not changed since $\mathsf{DP}$ was first studied \iftoggle{llncs}{}{(which dates back from more than $60$ years)}. Furthermore, Prange's algorithm \cite{P62} achieving this task is rather trivial. It basically exploits the fact that to solve $\mathsf{DP}(n,k,t)$ we have $k$ degrees of freedom to compute a solution. We describe this algorithm in Algorithm \ref{algo:Prangesdd}. We are not motivated here by overzealousness, but by the fact that we will be using some combinations of this algorithm, more precisely the so-called {\sc PrangeStep}, as a subroutine when inverting~$f_{\vec{H},t}$ with its corresponding trapdoor. 
\begin{algorithm}[htb]
	\caption{ {\sc {Prange}}{$(\vec{H},\vec{s})$}}\label{algo:Prangesdd}
	\begin{algorithmic}[1]
		\Require $\vec{H}\in\mathbb{F}_{2}^{(n-k)\times n}$, $\vec{s}\in\mathbb{F}_{2}^{n-k}$
		\Param $\mathcal{D}$ a distribution over $[0,k]$
		\Ensure $\vec{e} \in \mathbb{F}_{2}^{n}$ such that $\vec{e}\vec{H}^{\top}=\vec{s}$\medskip
		
		\State $\ell\Unif\mathcal{D}$

		\State $\mathcal{I} \Unif [1,n]$ with size $k$ such that $\vec{H}_{\overline{\mathcal{I}}}$ is non-singular\label{inst:infoSet} 
		\State $\vec{x}\Unif \mathcal{S}_{\ell}^k$
		\State $\vec{e}\gets\Call{PrangeStep}{\vec{H},\vec{s},\mathcal{I},\vec{x}}$ 
		\State \Return $\vec{e}$
		\end{algorithmic}
	\smallskip	
	\hrule
	{\bf function} \Call{PrangeStep}{$\vec{H},\vec{s},\mathcal{I},\vec{x}$} -- Prange vector completion
	\hrule
	\begin{algorithmic}[1]
	\Require $\vec{H}\in\mathbb{F}_{2}^{(n-k)\times n}$, $\vec{s}\in\mathbb{F}_{2}^{n-k}$, $\mathcal{I}\subseteq [1,n ]$ with size $k$, and $\vec{x}\in\mathbb{F}_{2}^{k}$
	\Ensure $\vec{e}\vec{H}^{\top}=\vec{s}$ and $\vec{e}_\mathcal{I}=\vec{x}$\medskip
	\State $\vec{e}_{\overline{\mathcal{I}}} \gets \left(\vec{s} - \vec{x}\vec{H}_{\mathcal{I}}^{\top}\right)\left( \vec{H}_{\overline{\mathcal{I}}}^{-1}\right)^{\top}$
	\State $\vec{e}_{\mathcal{I}} \gets \vec{x}$ 
	\State \Return $\vec{e}$
	\end{algorithmic}
\end{algorithm}

\begin{proposition}
	Let $n \geq k$ be integers, let $\vec{H} \in \mathbb{F}_{2}^{(n-k)\times n}$ have rank $n-k$ and let $\vec{s} \in \mathbb{F}_{2}^{n-k}$ be sampled uniformly at random. The algorithm {\sc{Prange}} works in polynomial time in $n$. The output~$\vec{e}$ of~{\sc{Prange}}($\vec{H},\vec{s}$) satisfies $\vec{e}\vec{H}^{\top} = \vec{s}$ and \iftoggle{llncs}{$|\vec{e}| = B+L$}{
	\begin{equation}
	|\vec{e}| = B+L
	\end{equation}}
	where $B$ has a binomial distribution with parameters $(n-k)$ and  $1/2$, and $L \Unif \mathcal{D}$ is independent of~$B$. Furthermore, for all $t\in [0,n]$,\iftoggle{llncs}{ $\mathbb{P}(|\vec{e}| = t) = \sum_{j=0}^{t} \mathcal{D}(t-j) \frac{\binom{n-k}{j}}{2^{n-k}}$ and $\mathbb{E}\left( |\vec{e}| \right) = \iftoggle{llncs}{t_{\textup{easy}}(n,k) + \mathbb{E}\left( \mathcal{D} \right)}{\frac{n-k}{2} + \mathbb{E}\left( \mathcal{D} \right) = t_{\textup{easy}}(n,k) + \mathbb{E}\left( \mathcal{D} \right)}$}{  
	\begin{equation}
	\mathbb{P}(|\vec{e}| = t) = \sum_{j=0}^{t} \mathcal{D}(t-j) \frac{\binom{n-k}{j}}{2^{n-k}} \quad \mbox{and} \quad \mathbb{E}\left( |\vec{e}| \right) = \iftoggle{llncs}{t_{\textup{easy}}(n,k) + \mathbb{E}\left( \mathcal{D} \right)}{\frac{n-k}{2} + \mathbb{E}\left( \mathcal{D} \right) = t_{\textup{easy}}(n,k) + \mathbb{E}\left( \mathcal{D} \right)} \ .
	\end{equation}
}
\end{proposition}
From this proposition, we immediately deduce that a solution of $\mathsf{DP}(n,k,t)$ with~$t \in [t_{\textup{easy}},k + t_{\textup{easy}}]$ can be computed in polynomial time with high probability. We just need to carefully choose~$\mathbb{E}(\mathcal{D}) \in [0,k]$\iftoggle{llncs}{.}{ and to repeat the {\sc{Prange}} algorithm a polynomial number of times.}

The algorithm exploiting our trapdoor to invert $f_{\vec{H},t}$ uses {\sc{PrangeStep}} as subroutine. In particular, as we will see, it will be helpful to use the (huge) degree of freedom offered by $\mathcal{D}$: it basically enables to choose $k$ coordinates (where $k$ denotes the dimension of the underlying code) of the output as we want.

\subsection{Inverting with the trapdoor}\label{subsec:invertTrap}

We have seen so far that when we want to invert~$f_{\vec{H},t}$ we necessarily have to choose parameters such that $t \geq t_{\textup{GV}}(n,k)$ to ensure the existence of preimages. However, we also need to choose~$t$ such that $t < t_{\textup{easy}}(n,k)$ otherwise anyone could invert $f_{\vec{H},t}$ even without the knowledge of the trapdoor by using the {\sc{Prange}} algorithm. We will show in this subsection how to invert $f_{\vec{H},t}$ for 
\begin{equation}
t \in \Big(t_{\textup{GV}}(n,k), t_{\textup{easy}}(n,k) \Big)
\end{equation}
when $\vec{H}$ is a parity-check matrix of a generalized $(U,U+V,U+V+W)$ code whose underlying structure is known, {\it i.e.,} codes $U, V$ and $W$ as well as vectors $\vec{a},\vec{b}$ and the permutation $\vec{P}$. 	 
One of the crucial results on which we will be relying is that $\varphi_{\vec{a},\vec{b},\vec{P}}$ used in Definition~\ref{def:UVW} is invertible thanks to the constraint verified by vectors $\vec{a},\vec{b}$ given in Equation \eqref{eq:cdtXY}.

\begin{lemma}\label{lemma:varphiInverse}
	Let $\vec{a},\vec{b},\vec{P}$ and $\varphi_{\vec{a},\vec{b},\vec{P}}$ be as in Definition \ref{def:UVW}. \iftoggle{llncs}{
	\begin{multline}
		\varphi_{\vec{a},\vec{b},\vec{P}}^{-1}: (\vec{x},\vec{y},\vec{z})\vec{P} \longmapsto \big( \vec{a} \star \vec{x} + \vec{y} + \vec{a} \star \vec{z}, (\vec{1}_{n/3}+\vec{a})\star \vec{x} + \vec{y} + \vec{a} \star \vec{z}, \\ 
		\vec{b} \star \vec{x} + (\vec{1}_{n/3}+\vec{b})\star\vec{y} + \vec{z} \big)\;\; \mbox{ where }\; \vec{x},\vec{y},\vec{z} \in \mathbb{F}_{2}^{n/3}
	\end{multline}
	}{Then $\varphi_{\vec{a},\vec{b},\vec{P}}$ is invertible and
	\begin{multline}
	\varphi_{\vec{a},\vec{b},\vec{P}}^{-1}: (\vec{x},\vec{y},\vec{z})\vec{P} \longmapsto \Big( \vec{a} \star \vec{x} + \vec{y} + \vec{a} \star \vec{z}, (\vec{1}_{n/3}+\vec{a})\star \vec{x} + \vec{y} + \vec{a} \star \vec{z},\\ 
	\vec{b} \star \vec{x} + (\vec{1}_{n/3}+\vec{b})\star\vec{y} + \vec{z} \Big)
	\end{multline}
where $\vec{x},\vec{y},\vec{z} \in \mathbb{F}_{2}^{n/3}$. 
}
	
\end{lemma}

The following notation will be extremely useful to explain how to invert $f_{\vec{H},t}$ with our trapdoor. Furthermore, we will use it many times in what follows.
\begin{notation}\label{nota:UVW}
	Let $\vec{a},\vec{b},\vec{P}$ and $\varphi_{\vec{a},\vec{b},\vec{P}}$ as in Definition \ref{def:UVW}. For any vector $\vec{e} \in \mathbb{F}_{2}^{n}$, let~$\vec{e}_U, \vec{e}_V, \vec{e}_W \in \mathbb{F}_{2}^{n/3}$ such that 
	\begin{equation}
	(\vec{e}_U, \vec{e}_V, \vec{e}_W) \eqdef \varphi_{\vec{a},\vec{b},\vec{P}}^{-1}\left( \vec{e} \right) . 
	\end{equation}
\end{notation}
\iftoggle{llncs}{}{The main reason for introducing this notation is the following proposition.} 
\begin{proposition}\label{propo:decodeUVW} 
Let $\vec{H}$  be a parity-check matrix of a generalized $(U, U+V,U+V+W)$ code. Consider the underlying permutation matrix $\vec{P} \in \mathbb{F}_{2}^{n\times n}$ and vectors $\vec{a},\vec{b} \in \mathbb{F}_{2}^{n/3}$ as well as parity-check matrices $\vec{H}_U \in \mathbb{F}_{2}^{(n/3-k_U) \times n/3}, \vec{H}_V \in \mathbb{F}_{2}^{(n/3-k_V) \times n/3}, \vec{H}_W \in \mathbb{F}_{2}^{(n/3-k_W) \times n/3}$ of associated codes~$U$,~$V$,~$W$. Let $\vec{S} \in \mathbb{F}_{2}^{(n-k) \times (n-k)}$ be non-singular where $k \eqdef k_U+k_V+k_W$, such that 
	\iftoggle{llncs}{
		\begin{equation*}
		\vec{S}\vec{H} = \vec{H}_{0} \eqdef \begin{pmatrix}
			\vec{H}_{U}\vec{Diag}(\vec{a}) & \vec{H}_{U} & \vec{H}_{U}\vec{Diag}(\vec{a}) \\
			\vec{H}_{V}\left( \vec{Id}_{n/3} + \vec{Diag}(\vec{a})\right)  & \vec{H}_V & \vec{H}_{V}\vec{Diag}(\vec{a}) \\
			\vec{H}_{W}\vec{Diag}(\vec{b}) & \vec{H}_{W}\left(\vec{Id}_{n/3} + \vec{Diag}(\vec{b})\right) & \vec{H}_{W}
		\end{pmatrix}\vec{P}
		\end{equation*}}{
	\begin{equation}\label{eq:H0}
	\vec{S}\vec{H} = \vec{H}_{0} \eqdef \begin{pmatrix}
		\vec{H}_{U}\vec{Diag}(\vec{a}) & \vec{H}_{U} & \vec{H}_{U}\vec{Diag}(\vec{a}) \\
		\vec{H}_{V}\left( \vec{Id}_{n/3} + \vec{Diag}(\vec{a})\right)  & \vec{H}_V & \vec{H}_{V}\vec{Diag}(\vec{a}) \\
		\vec{H}_{W}\vec{Diag}(\vec{b}) & \vec{H}_{W}\left(\vec{Id}_{n/3} + \vec{Diag}(\vec{b})\right) & \vec{H}_{W}
	\end{pmatrix}\vec{P}
	\end{equation}
} 
	 Then, \iftoggle{llncs}{for $\vec{s}_U\in \mathbb{F}_{2}^{n/3-k_U}, \vec{s}_V\in \mathbb{F}_{2}^{n/3-k_V}, \vec{s}_W \in \mathbb{F}_{2}^{n/3-k_W}$ with $(\vec{s}_U,\vec{s}_V,\vec{s}_W) = \vec{s}\vec{S}^{\top}$,}{}
	\begin{equation}
	\vec{e}\vec{H}^{\top} = \vec{s} \quad \Longleftrightarrow \quad \left\{ 
	\renewcommand{\arraystretch}{1.4}
	\begin{array}{c}
		\vec{e}_{U}\vec{H}_{U}^{\top} = \vec{s}_U \\
		\vec{e}_{V}\vec{H}_{V}^{\top} = \vec{s}_V \\
		\vec{e}_{W}\vec{H}_{W}^{\top} = \vec{s}_W
	\end{array}
	\right.
	\end{equation}
	\iftoggle{llncs}{}{where $\vec{s}_U\in \mathbb{F}_{2}^{n/3-k_U}$, $\vec{s}_V\in \mathbb{F}_{2}^{n/3-k_V}$ and  $\vec{s}_W \in \mathbb{F}_{2}^{n/3-k_W}$ are such that $(\vec{s}_U,\vec{s}_V,\vec{s}_W) = \vec{s}\vec{S}^{\top}$.}
\end{proposition}

\iftoggle{llncs}{}{
\begin{proof}
	First, notice that $\vec{H}_{0}$ is indeed a parity-check matrix of the considered generalized $(U,U+V,U+V+W)$ code as shown by Definition \ref{def:UVW} (in particular when exhibiting the dual structure). Therefore, for any parity-check matrix $\vec{H}$ of this code there exists a non-singular matrix $\vec{S} \in \mathbb{F}_{2}^{(n-k)\times (n-k)}$ (that we can easily compute via Gaussian elimination) such that $\vec{H}_0 = \vec{S}\vec{H}$.  
	This matrix~$\vec{S}$ is nothing more than a change-of-basis matrix. Notice that $\vec{e}\vec{H}^{\top} = \vec{s}$ is equivalent to 
	\begin{equation}\label{eq:changeBasis}  
		\vec{e}\vec{H}_{0}^{\top} = \vec{s}\vec{S}^{\top} = \left( \vec{s}_U,\vec{s}_V,\vec{s}_W\right)\ . 
		\end{equation} 
	Let us write any $\vec{e} \in \mathbb{F}_{2}^{n}$ as $\vec{e} = \left( \vec{x},\vec{y},\vec{z} \right)\vec{P}$ where $\vec{x},\vec{y},\vec{z} \in \mathbb{F}_{2}^{n/3}$. We have now the following computation (by combining Lemma~\ref{lemma:varphiInverse} with Notation~\ref{nota:UVW})
	\iftoggle{llncs}{
		\begin{align*}
		\vec{H}_0\vec{e}^{\top}
		= \vec{H}_{0}\vec{P}^{\top} (\vec{x},\vec{y},\vec{z})^{\top} = \begin{pmatrix} 
			\vec{H}_{U} \left( \vec{a} \star \vec{x} +  \vec{y} + \vec{a} \star \vec{z} \right)^{\top} \\ 
			\vec{H}_V \left( \left( \vec{1}_{n/3} + \vec{a} \right) \star \vec{x} + \vec{y} + \vec{a} \star \vec{z} \right)^{\top} \\
			\vec{H}_{W}\left( \vec{b}\star \vec{x} + \left( \vec{1}_{n/3} + \vec{b} \right) \star \vec{y} + \vec{z} \right)^{\top}
		\end{pmatrix}= \begin{pmatrix}
			\vec{H}_U\vec{e}_U^{\top} \\
			\vec{H}_{V}\vec{e}_V^{\top} \\
			\vec{H}_{W}\vec{e}_{W}^{\top}
		\end{pmatrix}
	\end{align*} 
	}{
	\begin{align*}
	\vec{H}_0\vec{e}^{\top}
	&= \vec{H}_{0}\vec{P}^{\top} (\vec{x},\vec{y},\vec{z})^{\top} \\
	&= \begin{pmatrix} 
		\vec{H}_{U} \big( \vec{a} \star \vec{x} +  \vec{y} + \vec{a} \star \vec{z} \big)^{\top} \\ 
		\vec{H}_V \big( \left( \vec{1}_{n/3} + \vec{a} \right) \star \vec{x} + \vec{y} + \vec{a} \star \vec{z} \big)^{\top} \\
		\vec{H}_{W}\big( \vec{b}\star \vec{x} + \left( \vec{1}_{n/3} + \vec{b} \right) \star \vec{y} + \vec{z} \big)^{\top}
	\end{pmatrix} \\
	&= \begin{pmatrix}
		\vec{H}_U\vec{e}_U^{\top} \\
		\vec{H}_{V}\vec{e}_V^{\top} \\
		\vec{H}_{W}\vec{e}_{W}^{\top}
	\end{pmatrix}
	\end{align*} 
}
	Combining this with Equation~\eqref{eq:changeBasis} concludes the proof. 
\end{proof}
}

This proposition shows that for inverting $f_{\vec{H},t}$ with $\vec{H}$ being a parity-check matrix of a generalized $(U,U+V,U+V+W)$ code, \iftoggle{llncs}{}{{\it i.e.,} to decode this code,} it is enough to decode its underlying~$U$,~$V$ and~$W$ codes. In our instantiation we will choose these codes as being random (to decrease as much as possible the structure of the associated generalized $(U,U+V,U+V+W)$ code) and we will use combinations of {\sc{PrangeStep}} to decode them. The \scheme{} \texttt{Trapdoor} algorithm (as described in~Algorithm \ref{algo:trap}) therefore basically consists of building $\vec{H}$ where the trapdoor will be roughly speaking its underlying structure: the random codes $U,V,W$, the chosen permutation $\vec{P}$ and vectors $\vec{a},\vec{b}$ satisfying Equation~\eqref{eq:cdtXY}. 
\begin{algorithm}[htb]
	\caption{ $\trap\left(\lambda\right)$: \scheme{} trapdoor algorithm }\label{algo:trap}
	\begin{algorithmic}[1]
		\Require a security level $\lambda$
		\Params integers $n,\; t,\; k_U,\; k_V,\; k_W$ to ensure $\lambda$ bits of security
		\Ensure  a public key $\vec{H}$ and its corresponding trapdoor $T$\medskip

\State $\vec{H}_{U} \Unif \mathbb{F}_{2}^{(n/3-k_U) \times n/3}$, $\vec{H}_{V} \Unif \mathbb{F}_{2}^{(n/3-k_V) \times n/3}$ and  $\vec{H}_{W} \Unif \mathbb{F}_{2}^{(n/3-k_W) \times n/3}$
		\State $\vec{P} \Unif \left\{ \vec{M} \in \mathbb{F}_{2}^{n \times n}: \; \vec{M} \mbox{ is a permutation matrix} \right\}$
		\State $\vec{S} \Unif \left\{ \vec{N} \in \mathbb{F}_{2}^{(n-k) \times (n-k)}: \; \vec{N} \mbox{ is a non-singular matrix} \right\}$ where $k \eqdef k_U + k_V + k_W$
		\State $(\vec{a},\vec{b}) \Unif \left\{ \vec{x},\vec{y}\in \mathbb{F}_{2}^{n/3}: \; \vec{x} \star \vec{y} = \vec{x} \right\}$ 
		\State $\vec{H} \leftarrow \vec{S}^{-1}\vec{H}_0$ where $\vec{H}_0$ is given in \iftoggle{llncs}{Proposition~\ref{propo:decodeUVW}}{Equation~\eqref{eq:H0}} 
		\State $T \leftarrow \left( \vec{a},\vec{b},\vec{P},\vec{S},\vec{H}_U,\vec{H}_V,\vec{H}_W \right)$
\State\Return $\left( \vec{H},T \right)$ 
	\end{algorithmic}
\end{algorithm}

It turns out that decoding codes~$W$, $V$ and $U$ independently only enables us to decode in polynomial time at distance $t_{\textup{easy}}$ which would make our trapdoor useless. Instead, we will decode successively codes~$W$,~$V$ and then $U$ by taking into account previous computations. In the following paragraph
we explain our decoding strategy via a particular choice of generalized~$(U,U+V,U+V+W)$ code. Our aim is to provide an intuitive understanding of how the inversion algorithm 
works. However, let us stress that this is {\it not} the algorithm we will be considering in the signing process: instead, we will use a carefully ``randomised'' version, as we will explain in detail (and rigorously) in Section \S\ref{sec:unif}. 
\newline

{\bf\noindent Our decoding strategy.} Suppose that we aim to decode the $(U+V,U,U+V+W)$ code as defined in Equation \eqref{eq:UVW} after being permuted by some permutation~$\vec{P}$. It corresponds to a generalized $(U,U+V,U+V+W)$ code where $\vec{a} = \vec{0}_{n/3}$ and~$\vec{b} = \vec{1}_{n/3}$.
Proposition \ref{propo:decodeUVW} shows that it reduces to computing some $\vec{e}_{U}$, $\vec{e}_{V}$ and $\vec{e}_{W}$, {\it i.e.,} to decoding codes $U$, $V$ and $W$. The solution is then equal to \iftoggle{llncs}{}{(according to Notation \ref{nota:UVW}),} 
\begin{equation}\label{eq:finalSolution}  
	\vec{e} = \varphi_{\vec{0}_{n/3},\vec{1}_{n/3},\vec{P}}\left( \vec{e}_U,\vec{e}_V,\vec{e}_W \right)= \left( \vec{e}_U+\vec{e}_V, \vec{e}_U, \vec{e}_{U}+\vec{e}_{V}+\vec{e}_{W} \right)\vec{P} \ . 
\end{equation} 
Our algorithm will proceed in three steps by decoding successively codes $W$,~$V$ and~$U$ using {\sc{PrangeStep}} {\it but} by using the knowledge of previous decoding outputs. Our rationale is that to minimize the Hamming weight of the solution in Equation \eqref{eq:finalSolution}, it is in our best interest for instance to choose $\vec{e}_U(i)$ as $0$ when~$\vec{e}_V(i)$ and $\vec{e}_W(i)$ are themselves equal to $0$.

In what follows we describe our algorithm with pictures. Each decoding step will use Algorithm~\ref{algo:Prangesdd}, in particular {\sc PrangeStep}. To understand how our decoder works, we will only consider {\it typical} outputs of {\sc PrangeStep}. Given a code of dimension $k$ and length $n$ to decode, if for instance we set~$\ell$ coordinates to $1$ in the $k$ coordinates given by the degrees of freedom, as the output restricted to the other $n-k$ coordinates is uniform, we {\it typically} obtain a vector with Hamming weight~$\ell + (n-k)/2$ that we represent as follows 
\iftoggle{llncs}{}{\medskip} 
\begin{center}
	\begin{tikzpicture}
			\draw (0,0) rectangle (9,0.5);
			\draw (5,0) rectangle (9,0.5);
			\draw [<->] (0,-0.25) -- node[below]{$k - \ell + \frac{n-k}{2}$} (4.95,-0.25);
			\draw [<->] (5.05,-0.25) -- node[below]{$\ell + \frac{n-k}{2}$} (9,-0.25);
			\node at (0.25,0.25) {$0$};
			\node at (4.75,0.25) {$0$};
			\node at (5.25,0.25) {$1$};
			\node at (8.75,0.25) {$1$};
			\draw[thick,-] (0.4,0.25) -- (4.6,0.25);
						\draw[thick,-] (5.4,0.25) -- (8.6,0.25);
	\end{tikzpicture}
\end{center}
Obviously all our pictures are up to a permutation (induced by the choice of $\mathcal{I}$\iftoggle{llncs}{).}{ in Instruction~\ref{inst:infoSet}).} 
\iftoggle{llncs}{}{\medskip}

\begin{enumerate}\setlength{\itemsep}{10pt}
	\item{\bf Computing $\vec{e}_W$.} First we decode $W$ which has dimension $k_W$ and length $n/3$. For this decoding we set the output on $k_W$ coordinates to $\vec{0}_{k_W}$. It gives \iftoggle{llncs}{}{the following vector $\vec{e}_W$,}
	\iftoggle{llncs}{}{\medskip} 
	\begin{center}
		 \makebox[\linewidth][c]{\hspace*{-1cm}\begin{tikzpicture}
			\draw (0,0) rectangle (11,0.5);
			\draw (6,0) rectangle (11,0.5);
			\draw [<->] (0,-0.25) -- node[below]{$k_W + \frac{n/3-k_W}{2}$} (5.95,-0.25);
			\draw [<->] (6.05,-0.25) -- node[below]{$ \frac{n/3-k_W}{2}$} (11,-0.25);
			\node at (0.25,0.25) {$0$};
			\node at (5.75,0.25) {$0$};
			\node at (6.25,0.25) {$1$};
			\node at (10.75,0.25) {$1$};
			\draw[thick,-] (0.4,0.25) -- (5.6,0.25);
			\draw[thick,-] (6.4,0.25) -- (10.6,0.25);
			\node at (-0.6,0.225) {$\vec{e}_W=$};
			\node at (-1.4,0.225) {\phantom{$\vec{e}_U+\vec{e}_V+\vec{e}_W=$}};
		\end{tikzpicture}
		\hspace*{1cm}}
	\end{center}
	\item{\bf Computing $\vec{e}_V$ knowing $\vec{e}_W$}. We now decode $V$, which has dimension $k_V$ and length~$n/3$, but by taking into account the shape of the previous decoding output $\vec{e}_W$. In particular, we will only use the degrees of freedom of {\sc Prange} in the coordinates $i$ where $\vec{e}_{W}(i) = 0$ by setting the output to $0$ on $k_V$ positions. It leads to a vector $\vec{e}_V$ with the following shape,
	\iftoggle{llncs}{}{\medskip}  
		\begin{center}
		\makebox[\linewidth][c]{\hspace*{-1cm}\begin{tikzpicture}
			\draw (0,0) rectangle (11,0.5);
			\draw (4,0) rectangle (6,0.5);
			\draw (6,0) rectangle (11,0.5);
			\draw (6,0) rectangle (8.5,0.5);
			\draw [<->] (0,-0.25) -- node[below]{$k_V + \frac{\alpha}{2}$} (3.95,-0.25);
			\draw [<->] (4.05,-0.25) -- node[below]{$\frac{\alpha}{2}$} (5.95,-0.25);
			\draw [<->] (6.05,-0.25) -- node[below]{$ \frac{n/3-k_W}{4}$} (8.45,-0.25);
			\draw [<->] (8.55,-0.25) -- node[below]{$ \frac{n/3-k_W}{4}$} (11,-0.25);
			\node at (0.25,0.25) {$0$};
			\node at (3.75,0.25) {$0$};
			\node at (4.25,0.25) {$1$};
			\node at (5.75,0.25) {$1$};
			\node at (6.25,0.25) {$0$};
			\node at (8.25,0.25) {$0$};
			\node at (8.75,0.25) {$1$};
			\node at (10.75,0.25) {$1$};
			\draw[thick,-] (0.4,0.25) -- (3.6,0.25);
			\draw[thick,-] (4.4,0.25) -- (5.6,0.25);
			\draw[thick,-] (6.4,0.25) -- (8.1,0.25);
			\draw[thick,-] (8.9,0.25) -- (10.6,0.25);
			\node at (-0.6,0.225) {$\vec{e}_V=$};
			\node at (-1.4,0.225) {\phantom{$\vec{e}_U+\vec{e}_V+\vec{e}_W=$}};
		\end{tikzpicture}
		\hspace*{1cm}}
	\end{center}
	where\iftoggle{llncs}{ $\alpha \eqdef k_W  + \frac{n/3-k_W}{2} - k_V$.}{, 
	\begin{equation}\label{eq:alpha}
		\alpha \eqdef k_W  + \frac{n/3-k_W}{2} - k_V \ .
	\end{equation}}
	\item{\bf Computing $\vec{e}_U$ knowing $\vec{e}_V$ and $\vec{e}_W$}. To end the whole decoding it remains now to  decode $U$, which has dimension $k_U$ and length $n/3$, to form the solution given in Equation~\eqref{eq:finalSolution}. 
	Notice here that when $\vec{e}_{V}(i) = \vec{e}_W(i) = 0$, then choosing $\vec{e}_U(i)$ as $0$ leads to {\it three} zeros in the solution. In other words, using {\it one} degree of freedom in $U$ leads to {\it three} zeros. A similar thing happens when $\vec{e}_V(i) = 0$ and $\vec{e}_W(i) = 1$ ({\it resp.}  $\vec{e}_V(i) = 1$ and~$\vec{e}_W(i) = 0$), choosing~$\vec{e}_U(i) = 0$ ({\it resp.}~$\vec{e}_U(i) = 1$) leads to {\it two} zeros in the solution. We will also use the degrees of freedom given by {\sc Prange} in the last set of positions, namely when~$\vec{e}_{V}(i) = \vec{e}_W(i) = 1$. In this case, choosing $\vec{e}_{U}(i) = 0$ also leads to {\it two} zeros in the solution.

	When decoding~$U$ via {\sc PrangeStep} we take into account these four areas of coordinates. This is why we split $k_U$ as $k_U^{00} + k_U^{10} + k_U^{01}+k_U^{11}$ where~$k_U^{00}$ denotes the number of coordinates where we choose~$\vec{e}_U(i) = 0$ but when~$\vec{e}_V(i) = \vec{e}_W(i) = 0$,~$k_U^{10}$ ({\it resp.} $k_U^{01}$) the number of coordinates where we choose~$\vec{e}_U(i) = 1$ ({\it resp.} $\vec{e}_U(i) = 0$) but when~$\vec{e}_V(i) = 1$ and~$\vec{e}_W(i) = 0$ ({\it resp.}~$\vec{e}_V(i) = 0$ and~$\vec{e}_W(i) = 1$) and $k_{U}^{11}$ the number of coordinates where we choose $\vec{e}_U(i) = 0$ but when~ $\vec{e}_V(i) = \vec{e}_W(i) = 1$. We therefore obtain as decoding output the following~$\vec{e}_U$,

	\iftoggle{llncs}{}{\medskip}   
			\begin{center}
		\makebox[\linewidth][c]{\hspace*{-1cm}\begin{tikzpicture}
			\draw (0,0) rectangle (11,0.5);
			\draw (3,0) rectangle (4,0.5);
			\draw (4,0) rectangle (5.5,0.5);
			\draw (4,0) rectangle (6,0.5);
			\draw (6,0) rectangle (11,0.5);
			\draw (6,0) rectangle (8.5,0.5);
			\draw (6,0) rectangle (8,0.5);
			\draw (10.5,0) rectangle (11,0.5);
			\draw [<->] (0,-0.25) -- node[below]{$k_U^{00} + \frac{\beta}{2}$} (2.95,-0.25);
			\draw [<->] (3.05,-0.25) -- node[below]{$\frac{\beta}{2}$} (3.95,-0.25);
			\draw [<->] (3.05,-0.25) -- node[below]{$\frac{\beta}{2}$} (3.95,-0.25);
			\draw [<->] (4.05,-0.25) -- node[below]{$k_U^{10} + \frac{\gamma}{2}$} (5.4,-0.25);
			\draw [<->] (5.5,-0.25) -- node[below]{$\frac{\gamma}{2}$} (5.9,-0.25);
			\draw [<->] (6.05,-0.25) -- node[below]{$k_U^{01}+ \frac{\delta}{2}$} (8,-0.25);
			\draw [<->] (8.05,-0.25) -- node[below]{$\frac{\delta}{2}$} (8.45,-0.25);
			\draw [<->] (8.55,-0.25) -- node[below]{$k_{U}^{11} + \frac{\varepsilon}{2}$} (10.45,-0.25);
			\draw [<->] (10.5,-0.25) -- node[below]{$\frac{\varepsilon}{2}$} (11,-0.25);
			\node at (0.25,0.25) {$0$};
			\node at (2.75,0.25) {$0$};
			\node at (3.25,0.25) {$1$};
			\node at (3.75,0.25) {$1$};
			\node at (4.25,0.25) {$1$};
			\node at (5.25,0.25) {$1$};
			\node at (5.75,0.25) {$0$};
			\node at (6.25,0.25) {$0$};
			\node at (7.75,0.25) {$0$};
			\node at (8.25,0.25) {$1$};
			\node at (8.75,0.25) {$0$};
			\node at (10.25,0.25) {$0$};
\node at (10.75,0.25) {$1$};
			\draw[thick,-] (0.4,0.25) -- (2.6,0.25);
			\draw[thick,-] (3.4,0.25) -- (3.6,0.25);
			\draw[thick,-] (4.4,0.25) -- (5.1,0.25);
			\draw[thick,-] (6.4,0.25) -- (7.6,0.25);
			\draw[thick,-] (8.9,0.25) -- (10.1,0.25);
\node at (-0.6,0.225) {$\vec{e}_U=$};
			\node at (-1.4,0.225) {\phantom{$\vec{e}_U+\vec{e}_V+\vec{e}_W=$}};
		\end{tikzpicture}
		\hspace*{1cm}}
	\end{center}
	where $\beta \eqdef k_V + \frac{\alpha}{2} - k_{U}^{00}$, $\gamma \eqdef \frac{\alpha}{2} - k_{U}^{10}$, $\delta \eqdef \frac{n/3 - k_W}{4} - k_{U}^{01}$,\iftoggle{llncs}{}{ and} $\varepsilon \eqdef \frac{n/3-k_W}{4} - k_{U}^{11}$. \iftoggle{llncs}{}{Notice that we have the following sanity-check (by definition $k_U + \beta + \gamma + \delta + \varepsilon = n/3$),
	\begin{align}
		\beta &+ \gamma + \delta + \varepsilon \nonumber \\
		&\quad= \left( k_V + \frac{\alpha}{2} - k_{U}^{00} \right) + \left( \frac{\alpha}{2} - k_{U}^{10} \right) + \left( \frac{n/3-k_W}{4} - k_U^{01} \right) + \left( \frac{n/3-k_W}{4} - k_{U}^{11} \right) \nonumber \\
		&\quad= k_V - k_U + \alpha + \frac{n/3-k_W}{2} \nonumber\\
		&\quad=k_V - k_U + k_W + (n/3-k_W) - k_V \quad \left(\mbox{we used Equation~\eqref{eq:alpha}}\right) \nonumber\\
		&\quad= n/3-k_U \label{eq:betagamedeltaeps} 
	\end{align}
}
\item{\bf Building the solution.} To form the solution we simply combine decoding outputs~$\vec{e}_W, \vec{e}_V$ and $\vec{e}_U$ to form the vector given in Equation~\eqref{eq:finalSolution}. It gives as solution a vector with the following three components.  
	
				\begin{center}
		\makebox[\linewidth][c]{\hspace*{-1cm}\begin{tikzpicture}
						\draw (0,0) rectangle (11,0.5);
			\draw (3,0) rectangle (4,0.5);
			\draw (4,0) rectangle (5.5,0.5);
			\draw (4,0) rectangle (6,0.5);
			\draw (6,0) rectangle (11,0.5);
			\draw (6,0) rectangle (8.5,0.5);
			\draw (6,0) rectangle (8,0.5);
			\draw (10.5,0) rectangle (11,0.5);
			\draw [<->] (0,-0.25) -- node[below]{$k_U^{00} + \frac{\beta}{2}$} (2.95,-0.25);
			\draw [<->] (3.05,-0.25) -- node[below]{$\frac{\beta}{2}$} (3.95,-0.25);
			\draw [<->] (3.05,-0.25) -- node[below]{$\frac{\beta}{2}$} (3.95,-0.25);
			\draw [<->] (4.05,-0.25) -- node[below]{$k_U^{10} + \frac{\gamma}{2}$} (5.4,-0.25);
			\draw [<->] (5.5,-0.25) -- node[below]{$\frac{\gamma}{2}$} (5.9,-0.25);
			\draw [<->] (6.05,-0.25) -- node[below]{$k_U^{01}+ \frac{\delta}{2}$} (8,-0.25);
			\draw [<->] (8.05,-0.25) -- node[below]{$\frac{\delta}{2}$} (8.45,-0.25);
			\draw [<->] (8.55,-0.25) -- node[below]{$k_{U}^{11} + \frac{\varepsilon}{2}$} (10.45,-0.25);
			\draw [<->] (10.5,-0.25) -- node[below]{$\frac{\varepsilon}{2}$} (11,-0.25);
			\node at (0.25,0.25) {$0$};
			\node at (2.75,0.25) {$0$};
			\node at (3.25,0.25) {$1$};
			\node at (3.75,0.25) {$1$};
			\node at (4.25,0.25) {$0$};
			\node at (5.25,0.25) {$0$};
			\node at (5.75,0.25) {$1$};
			\node at (6.25,0.25) {$0$};
			\node at (7.75,0.25) {$0$};
			\node at (8.25,0.25) {$1$};
			\node at (8.75,0.25) {$1$};
			\node at (10.25,0.25) {$1$};
\node at (10.75,0.25) {$0$};
			\draw[thick,-] (0.4,0.25) -- (2.6,0.25);
			\draw[thick,-] (3.4,0.25) -- (3.6,0.25);
			\draw[thick,-] (4.4,0.25) -- (5.1,0.25);
			\draw[thick,-] (6.4,0.25) -- (7.6,0.25);
			\draw[thick,-] (8.9,0.25) -- (10.1,0.25);
			\node at (-1,0.225) {$\vec{e}_U+\vec{e}_V=$};
			\node at (-1.4,0.225) {\phantom{$\vec{e}_U+\vec{e}_V+\vec{e}_W=$}};
		\end{tikzpicture}
		\hspace*{1cm}}
	\end{center}

				\begin{center}
		\makebox[\linewidth][c]{\hspace*{-1cm}\begin{tikzpicture}
						\draw (0,0) rectangle (11,0.5);
			\draw (3,0) rectangle (4,0.5);
			\draw (4,0) rectangle (5.5,0.5);
			\draw (4,0) rectangle (6,0.5);
			\draw (6,0) rectangle (11,0.5);
			\draw (6,0) rectangle (8.5,0.5);
			\draw (6,0) rectangle (8,0.5);
			\draw (10.5,0) rectangle (11,0.5);
			\draw [<->] (0,-0.25) -- node[below]{$k_U^{00} + \frac{\beta}{2}$} (2.95,-0.25);
			\draw [<->] (3.05,-0.25) -- node[below]{$\frac{\beta}{2}$} (3.95,-0.25);
			\draw [<->] (3.05,-0.25) -- node[below]{$\frac{\beta}{2}$} (3.95,-0.25);
			\draw [<->] (4.05,-0.25) -- node[below]{$k_U^{10} + \frac{\gamma}{2}$} (5.4,-0.25);
			\draw [<->] (5.5,-0.25) -- node[below]{$\frac{\gamma}{2}$} (5.9,-0.25);
			\draw [<->] (6.05,-0.25) -- node[below]{$k_U^{01}+ \frac{\delta}{2}$} (8,-0.25);
			\draw [<->] (8.05,-0.25) -- node[below]{$\frac{\delta}{2}$} (8.45,-0.25);
			\draw [<->] (8.55,-0.25) -- node[below]{$k_{U}^{11} + \frac{\varepsilon}{2}$} (10.45,-0.25);
			\draw [<->] (10.5,-0.25) -- node[below]{$\frac{\varepsilon}{2}$} (11,-0.25);
			\node at (0.25,0.25) {$0$};
			\node at (2.75,0.25) {$0$};
			\node at (3.25,0.25) {$1$};
			\node at (3.75,0.25) {$1$};
			\node at (4.25,0.25) {$1$};
			\node at (5.25,0.25) {$1$};
			\node at (5.75,0.25) {$0$};
			\node at (6.25,0.25) {$0$};
			\node at (7.75,0.25) {$0$};
			\node at (8.25,0.25) {$1$};
			\node at (8.75,0.25) {$0$};
			\node at (10.25,0.25) {$0$};
\node at (10.75,0.25) {$1$};
			\draw[thick,-] (0.4,0.25) -- (2.6,0.25);
			\draw[thick,-] (3.4,0.25) -- (3.6,0.25);
			\draw[thick,-] (4.4,0.25) -- (5.1,0.25);
			\draw[thick,-] (6.4,0.25) -- (7.6,0.25);
			\draw[thick,-] (8.9,0.25) -- (10.1,0.25);
			\node at (-0.6,0.225) {$\vec{e}_U=$};
			\node at (-1.4,0.225) {\phantom{$\vec{e}_U+\vec{e}_V+\vec{e}_W=$}};
	\end{tikzpicture}
	\hspace*{1cm}}
	\end{center}

				\begin{center}
		\makebox[\linewidth][c]{\hspace*{-1cm}\begin{tikzpicture}
						\draw (0,0) rectangle (11,0.5);
			\draw (3,0) rectangle (4,0.5);
			\draw (4,0) rectangle (5.5,0.5);
			\draw (4,0) rectangle (6,0.5);
			\draw (6,0) rectangle (11,0.5);
			\draw (6,0) rectangle (8.5,0.5);
			\draw (6,0) rectangle (8,0.5);
			\draw (10.5,0) rectangle (11,0.5);
			\draw [<->] (0,-0.25) -- node[below]{$k_U^{00} + \frac{\beta}{2}$} (2.95,-0.25);
			\draw [<->] (3.05,-0.25) -- node[below]{$\frac{\beta}{2}$} (3.95,-0.25);
			\draw [<->] (3.05,-0.25) -- node[below]{$\frac{\beta}{2}$} (3.95,-0.25);
			\draw [<->] (4.05,-0.25) -- node[below]{$k_U^{10} + \frac{\gamma}{2}$} (5.4,-0.25);
			\draw [<->] (5.5,-0.25) -- node[below]{$\frac{\gamma}{2}$} (5.9,-0.25);
			\draw [<->] (6.05,-0.25) -- node[below]{$k_U^{01}+ \frac{\delta}{2}$} (8,-0.25);
			\draw [<->] (8.05,-0.25) -- node[below]{$\frac{\delta}{2}$} (8.45,-0.25);
			\draw [<->] (8.55,-0.25) -- node[below]{$k_{U}^{11} + \frac{\varepsilon}{2}$} (10.45,-0.25);
			\draw [<->] (10.5,-0.25) -- node[below]{$\frac{\varepsilon}{2}$} (11,-0.25);
			\node at (0.25,0.25) {$0$};
			\node at (2.75,0.25) {$0$};
			\node at (3.25,0.25) {$1$};
			\node at (3.75,0.25) {$1$};
			\node at (4.25,0.25) {$0$};
			\node at (5.25,0.25) {$0$};
			\node at (5.75,0.25) {$1$};
			\node at (6.25,0.25) {$1$};
			\node at (7.75,0.25) {$1$};
			\node at (8.25,0.25) {$0$};
			\node at (8.75,0.25) {$0$};
			\node at (10.25,0.25) {$0$};
\node at (10.75,0.25) {$1$};
			\draw[thick,-] (0.4,0.25) -- (2.6,0.25);
			\draw[thick,-] (3.4,0.25) -- (3.6,0.25);
			\draw[thick,-] (4.4,0.25) -- (5.1,0.25);
			\draw[thick,-] (6.4,0.25) -- (7.6,0.25);
			\draw[thick,-] (8.9,0.25) -- (10.1,0.25);
			\node at (-1.4,0.225) {$\vec{e}_U+\vec{e}_V+\vec{e}_W=$};
		\end{tikzpicture}
		\hspace*{1cm}}
	\end{center}
	
	We therefore (typically) expect a solution with Hamming weight $t$ given by \iftoggle{llncs}{}{the following computation,}
	\begin{align*}
		t\iftoggle{llncs}{}{&= 3 \frac{\beta}{2} + \left(  k_{U}^{10} + \frac{\gamma}{2} \right) + 2\frac{\gamma}{2} + \left( k_{U}^{01} + \frac{\delta}{2}\right) + 2\frac{\delta}{2} + \left( k_{U}^{11} + \frac{\varepsilon}{2}\right) + 2\frac{\varepsilon}{2} \\ 
		&= k_U - k_U^{00} + \frac{3}{2} \left( \beta + \gamma +\delta + \varepsilon \right) \quad \left(\mbox{we used $k_U = k_{U}^{00}+k_{U}^{10} + k_{U}^{01} + k_{U}^{11}$}\right)  \iftoggle{llncs}{}{\\
		&= k_U - k_U^{00} + \frac{3}{2}\left( n/3-k_U \right) \quad \left(\mbox{we used Equation~\eqref{eq:betagamedeltaeps}}\right)} \\}
		\iftoggle{llncs}{}{&}= \frac{n-k_U}{2} - k_U^{00} \iftoggle{llncs}{}{\\}
		&= \underbrace{\frac{n-k}{2}}_{t_{\textup{easy}}(n,k)} + \frac{k_V+k_W - 2k_U^{00}}{2}
	\end{align*}
where~$k \eqdef k_U+k_V+k_W$ denotes the dimension of the underlying $(U+V,U,U+V+W)$ code. It shows that our decoder indeed gives an advantage over {\sc Prange} but under the condition that 
	\begin{equation}\label{eq:cstParam1}
	2k_{U}^{00} > k_V+k_W\ .
	\end{equation} 
\end{enumerate}

Our trapdoor algorithm uses 
exactly the same strategy 
except that it relies on a particular choice of vectors $\vec{a}$ and~$\vec{b}$. The dependence on these vectors only appears in the last decoding step. When decoding $U$ by choosing $\vec{e}_{U}(i)$ via the degrees of freedom given by {\sc PrangeStep} we only ensure that (which is always possible) \iftoggle{llncs}{$\left| \left( \vec{e}_{U}(i) + \vec{e}_V(i), \vec{e}_{U}(i) + \vec{a}(i)\vec{e}_{W}(i), \vec{e}_{U}(i)+ \vec{b}(i)\vec{e}_{V}(i) + \vec{e}_{W}(i) \right)\right| \in \{0,1\}$}{
\begin{equation}
\left| \left( \vec{e}_{U}(i) + \vec{e}_V(i), \vec{e}_{U}(i) + \vec{a}(i)\vec{e}_{W}(i), \vec{e}_{U}(i)+ \vec{b}(i)\vec{e}_{V}(i) + \vec{e}_{W}(i) \right)\right| \in \{0,1\}
\end{equation}}
according to the values of $\vec{e}_V(i)$ and $\vec{e}_{W}(i) $. 
What gives our decoder an advantage over Prange's algorithm is the essential fact that the code $U$ appears {\it three} times in a generalized $(U,U+V,U+V+W)$ code, whatever the choice\footnote{The constraint given in Equation~\eqref{eq:cdtXY} is here to ensure that $\varphi_{\vec{a},\vec{b},\vec{P}}$ is invertible.} of the $\vec{a}(i)$'s and the $\vec{b}(i)$'s.
\medskip

\iftoggle{llncs}{}{
{\bf \noindent Leakage-free signatures.} Our decoder gives an advantage over {\sc Prange} in inverting $f_{\vec{H},t}$ under the mild condition on the parameters given in Equation~\eqref{eq:cstParam1}. However, to ensure security, vectors of Hamming weight $t$ output by our algorithm {\it i.e.,} signatures, must not reveal any information on the secret key used\iftoggle{llncs}{.}{, as required by ATPFS (Definition~\ref{def:ATPSF})}. In particular, decoding solutions have to be uniformly distributed over words of Hamming weight $t$. But if we return to the above example with the $(U+V,U,U+V+W)$ code, there is no reason for this.

As we are now going to explain (inspired by~\cite{DLV24}), vectors output by the above procedure leak information on their underlying trapdoor: an attacker could exploit them to recover the secret key if the decoder is implemented as specified so far\footnote{Notice that so far our decoder is under-specified, for instance we did not specify how to choose $k_V, k_W$ and $k_U$.}. 
Suppose that an attacker collects outputs of the above procedure.
They have the following shape
\begin{equation}
\vec{e} \eqdef \left( \vec{e}_U+\vec{e}_V, \vec{e}_U, \vec{e}_U+\vec{e}_V+\vec{e}_W\right) \vec{P}
\end{equation}
where $\vec{P}$ is some {\em secret} permutation.
Let $(a,b,c)$ be the coordinates which are the images under~$\vec{P}$ of the {\em matched triplets} (up to a permutation of the three positions), 
\begin{equation}
(i,i+n/3,i+2n/3) \ . 
\end{equation}
In the typical configuration the three bits $(\vec{e}(a),\vec{e}(b),\vec{e}(c))$ are $(1,1,1)$ precisely when $\vec{e}_V(i) = \vec{e}_W(i) = 0$ and $\vec{e}_U(i) = 1$. It turns out that there are
$\beta/2$ (where $\beta = \tfrac{k_V}{2} + \tfrac{n/3 + k_W}{4} - k_{U}^{00}$) such positions among the $n/3$ matched triplets. Thus, for a uniformly chosen matched triplet $(a,b,c)$,
\begin{equation}
\mathbb{P}\left(   (\vec{e}(a),\vec{e}(b),\vec{e}(c)) = (1,1,1) \right) \approx \frac{\beta/2}{n/3} \ . 
\end{equation}

By comparison, if $\vec{e}$ were uniform over $\mathcal{S}_{t}^{n}$ (and thus independent of $\vec{P}$), any three distinct positions~$(a,b,c)$ would satisfy
\begin{equation}
\mathbb{P}\left( (\vec{e}(a),\vec{e}(b),\vec{e}(c)) = (1,1,1) \right) = \frac{\binom{n-3}{t-3}}{\binom{n}{t}} \ .
\end{equation}
There is no reason for these two quantities to agree. Any attacker could therefore collect vectors~$\vec{e}$ and estimate empirically for each triplet~$(a,b,c)$ the probability that~$(\vec{e}(a),\vec{e}(b),\vec{e}(c))$ is equal to~$(1,1,1)$ to deduce whether a triplet is matched or not, therefore gaining information about the secret~$\vec{P}$ to ultimately recover it with a bunch of signatures. }

 	\section{Preimage Sampling with Trapdoor: Achieving Uniformly Distributed Outputs}\label{sec:unif}

To instantiate our signature scheme, it is crucial to ensure that signatures, {\it i.e.,} outputs $\vec{e}^{\mathrm{out}} \in f_{\vec{H},t}^{-1}(\vec{s})$ for some $\vec{s}$ of our trapdoor algorithm, are statistically independent of the underlying trapdoor to produce them\iftoggle{llncs}{.}{, and thus do not leak any information about it.} 
This is precisely the {\it average trapdoor preimage sampling} property required in Definition \ref{def:ATPSF}. The above procedure obviously does not have this property. Fortunately,  by changing it slightly, we can achieve this no-leakage property. \iftoggle{llncs}{All the proofs of this section can be found in the {\it Auxiliary Material}.}{} 

\subsection{Rejection sampling to reach uniformly distributed output}
\label{sec:rejection}

Our algorithm proceeds as in the previous section: it decodes 
successively the codes $W$, $V$ and $U$ while taking into account the previous
decoding outputs. The idea to reach uniformity is to impose at each decoding step some ``condition''. This is given in Algorithm~\ref{algo:inv1}. 

\begin{algorithm}[htb]
	\caption{ $\mathsf{InvertAlg}(\vec{s},T)$: \scheme{} inversion algorithm }\label{algo:inv1}
	\begin{algorithmic}[1]
		\Require A vector $\vec{s} \in \mathbb{F}_2^{n-k}$ and
         an output $T \eqdef (\vec{a},\vec{b},\vec{P},\vec{S},
         \vec{H}_U,\vec{H}_V,\vec{H}_W)$
         of Algorithm~\ref{algo:trap}

\Ensure A vector $\vec{e} \in \mathcal{S}_{t}^{n}$
        satisfying $\vec{e}\vec{H}^{\top} = \vec{s}$,
        where $\vec{H}$ is the second output of
        Algorithm~\ref{algo:trap}\medskip

		\State $(\vec{s}_U,\vec{s}_V,\vec{s}_W) \leftarrow \vec{s}\vec{S}^{\top}$ 
		\Repeat
		\State\label{inst:W1} $\vec{e}_W\gets\mathsf{DecodeW}(\vec{H}_W,\vec{s}_W)$
		\Until{Condition 1 is met}
		\Repeat
		\State\label{inst:V1} $\vec{e}_V\gets\mathsf{DecodeV}(\vec{H}_V,\vec{s}_V,\vec{e}_W)$
		\Until{Condition 2 is met}
		\Repeat
		\State\label{inst:U1} $\vec{e}_U\gets\mathsf{DecodeU}(\vec{H}_U,\vec{s}_U,\vec{e}_V,\vec{e}_W,\varphi_{\vec{a},\vec{b},\vec{P}})$
		\Until{Condition 3 is met}
		\State $\vec{e}^{\mathrm{out}}\gets\varphi_{\vec{a},\vec{b},\vec{P}}(\vec{e}_U,\vec{e}_V,\vec{e}_W)$
		\State \Return $\vec{e}^{\mathrm{out}}$
	\end{algorithmic}
\end{algorithm}

\iftoggle{llncs}{}{
From now, and on numerous occasions until the end of this section, we will use the following notation. 
}
\begin{notation}\iftoggle{llncs}{$\left(\vec{e}_U^{\mathrm{unif}}, \vec{e}_V^{\mathrm{unif}}, \vec{e}_W^{\mathrm{unif}}\right) = \varphi_{\vec{a},\vec{b},\vec{P}}^{-1}\left( \vec{e}^{\mathrm{unif}} \right)$ where $\vec{e}^{\mathrm{unif}} \Unif \mathcal{S}_{t}^{n}$}{
	\begin{equation}
	 \left(\vec{e}_U^{\mathrm{unif}}, \vec{e}_V^{\mathrm{unif}}, \vec{e}_W^{\mathrm{unif}}\right) = \varphi_{\vec{a},\vec{b},\vec{P}}^{-1}\left( \vec{e}^{\mathrm{unif}} \right) \quad \mbox{ where } \; \vec{e}^{\mathrm{unif}} \Unif \mathcal{S}_{t}^{n} \ .
	 \end{equation}}
\end{notation}

Our objective is to make the output of Algorithm \ref{algo:inv1}, denoted by $\vec{e}^{\mathrm{out}}$,
 statistically close to~$\vec{e}^{\mathrm{unif}}$. A sufficient condition is to ensure via the conditions of Algorithm \ref{algo:inv1} that $\vec{e}_W$, $\vec{e}_V$ knowing~$\vec{e}_W$ and~$\vec{e}_U$ knowing $(\vec{e}_W,\vec{e}_V)$ are distributed as~$\vec{e}^{\mathrm{unif}}_W$, $\vec{e}^{\mathrm{unif}}_V$ knowing $\vec{e}^{\mathrm{unif}}_W$ and $\vec{e}^{\mathrm{unif}}_U$ knowing $(\vec{e}^{\mathrm{unif}}_W,\vec{e}^{\mathrm{unif}}_V)$. However, as these are vectors in $\mathbb{F}_{2}^{n/3}$, this requires handling an exponential number of statistics to ensure that they coincide. Fortunately, the underlying distributions of $\vec{e}^{\mathrm{unif}}$ 
are ``only'' functions of the following few statistics (as shown in Lemma~\ref{lem:reference-uniformity}) and it will be enough to ensure that $\vec{e}^{\mathrm{unif}}$ and $\vec{e}^{\mathrm{out}}$ coincide on them (as shown in Lemma~\ref{lem:uniform}) .

\begin{definition}\label{def:statistics}
	Let $\vec{x}_U, \vec{x}_V, \vec{x}_W, \vec{a}, \vec{b} \in \mathbb{F}_2^{n/3}$.
We first define
\begin{equation}
    \omega(\vec{x}_W) \eqdef |\vec{x}_W|
    \quad \text{and} \quad
    \nu(\vec{x}_V,\vec{x}_W)
    \eqdef \sharp \left\{i\in [1,n/3]:
    (\vec{x}_V(i),\vec{x}_W(i))\neq(0,0)\right\} \ .
    \label{eq:tau-sigma}
\end{equation}
For each $(\alpha,\beta)\in\{0,1\}^2$, let
\begin{equation}
    \mathcal M_{\alpha\beta}(\vec{x}_V,\vec{x}_W)
    \eqdef \left\{i\in[1,n/3]:
    (\vec{x}_V(i),\vec{x}_W(i))=(\alpha,\beta)\right\},
\end{equation}
and define
\begin{equation}
    \vec{m}(\vec{x}_V,\vec{x}_W)
    \eqdef (m_{00},m_{10},m_{01},m_{11}),
    \quad \text{where} \quad
    m_{\alpha\beta}\eqdef \sharp \mathcal{M}_{\alpha\beta} \ .
\end{equation}
These quantities satisfy
\begin{equation}
    \omega(\vec{x}_W)=m_{01}+m_{11}
    \quad\text{and}\quad
    \nu(\vec{x}_V,\vec{x}_W)=\frac n3-m_{00}\ .
\end{equation}
Next, for each $i\in[1,n/3]$, define
\begin{equation}\label{eq:Omegai}
    \Omega_i(\vec{x}_U;\vec{x}_V,\vec{x}_W)
    \eqdef \left|
    \left(
    \vec{x}_U(i)+\vec{x}_V(i),\
    \vec{x}_U(i)+\vec{a}(i)\vec{x}_W(i),\
    \vec{x}_U(i)+\vec{b}(i)\vec{x}_V(i)+\vec{x}_W(i)
    \right)\right|\ .
\end{equation}
Finally, let
\begin{equation}
    \vec{h}(\vec{x}_U;\vec{x}_V,\vec{x}_W)
    \eqdef(h_{00},h_{10},h_{01},h_{11}),
\end{equation}
where, for each $(\alpha,\beta)\in\{0,1\}^2$,
\begin{equation}
    h_{\alpha\beta}(\vec{x}_U;\vec{x}_V,\vec{x}_W)
    \eqdef\sharp \left\{
    i\in\mathcal M_{\alpha\beta}:
    \Omega_i(\vec{x}_U;\vec{x}_V,\vec{x}_W)\in\{2,3\}
    \right\} \ .
\end{equation}
When the arguments are clear from the context, we omit them
and simply write $\omega$, $\nu$, $\vec{m}$, $\vec{h}$,
and~$\Omega_i$.

\end{definition}

	Notice that $\vec{h}$ counts the positions $i$ where the choice of $\vec{x}_U(i)$ gives the larger of the two possible ``local weights'' $\Omega_i(\vec{x}_U;\vec{x}_V,\vec{x}_W)$. These weights are given by the following proposition.

	\begin{restatable}{proposition}{propoStat}\label{prop:counts}
	
Let $\vec{a}, \vec{b} \in \mathbb{F}_2^{n/3}$ satisfy
$\vec{a}\star\vec{b}=\vec{a}$, and let
$\vec{x}_U,\vec{x}_V,\vec{x}_W\in\mathbb{F}_2^{n/3}$.
For each fixed pair~$(\vec{x}_V(i),\vec{x}_W(i))$,
the bit $c$ minimizing $\Omega_i$ as defined in Equation~\eqref{eq:Omegai}, together with the
corresponding values of $\Omega_i$ for~$\vec{x}_U(i)=c$ and $\vec{x}_U(i)=1+c$, are given
in the following table:
\[
\begin{array}{c|c|c|c}
(\vec{x}_V(i),\vec{x}_W(i))
& \text{Minimizing bit }c
& \Omega_i \text{ if }\vec{x}_U(i)=c
& \Omega_i \text{ if }\vec{x}_U(i)=1+c
\\ \hline
(0,0) & 0 & 0 & 3\\
(1,0) & \vec{b}(i) & 1 & 2\\
(0,1) & \vec{a}(i) & 1 & 2\\
(1,1) & 1+\vec{a}(i)+\vec{b}(i) & 1 & 2
\end{array}
\]
Consequently, the Hamming weight of
$\varphi_{\vec{a},\vec{b},\vec{P}}
(\vec{x}_U,\vec{x}_V,\vec{x}_W)$ is given by
\begin{equation}\label{eq:weight}
    \big|\varphi_{\vec{a},\vec{b},\vec{P}}
    (\vec{x}_U,\vec{x}_V,\vec{x}_W)\big|
    =\nu+3h_{00}+h_{10}+h_{01}+h_{11}\ .
\end{equation}
Furthermore, for fixed $\vec{x}_V$ and $\vec{x}_W$,
the number of vectors $\vec{x}_U\in\mathbb{F}_2^{n/3}$
yielding a prescribed value of~$\vec{h}$ is
\begin{equation}\label{eq:Nm}
    N_{\vec{m}}(\vec{h})
    =\prod_{\alpha,\beta\in\{0,1\}}
    \binom{m_{\alpha\beta}}{h_{\alpha\beta}} \ .
\end{equation}
	\end{restatable}
	\iftoggle{llncs}{}{
	
\begin{proof}
    Fix $i\in[1,n/3]$, and write
    $a\eqdef\vec{a}(i)$, $b\eqdef\vec{b}(i)$, and
    $(\alpha,\beta)\eqdef(\vec{x}_V(i),\vec{x}_W(i))$.
    For $\vec{x}_U(i)=u$, the three-bit vector
    whose Hamming weight is $\Omega_i$ is
    \[
        (u+\alpha,\ u+a\beta,\ u+b\alpha+\beta)\ .
    \]
    Replacing $u$ by $1+u$ complements all three bits,
    so the corresponding Hamming weights sum to three.

    When $(\alpha,\beta)=(0,0)$, the choices $u=0$
    and $u=1$ yield $(0,0,0)$ and $(1,1,1)$,
    respectively. For the remaining cases, substituting
    the minimizing bit $c$ given in the table yields
    \[
    \begin{array}{c|c}
        (\alpha,\beta) & \text{Three-bit vector for }u=c\\
        \hline
        (1,0) & (1+b,b,0)\\
        (0,1) & (a,0,1+a)\\
        (1,1) & (a+b,1+b,a)
    \end{array}
    \]
    The first two vectors have Hamming weight one.
    In the last case, the condition $ab=a$ restricts
    $(a,b)$ to $(0,0)$, $(0,1)$, or $(1,1)$,
    yielding $(0,1,0)$, $(1,0,0)$, and $(0,0,1)$,
    respectively. Thus, the minimizing bit always
    yields weight one when $(\alpha,\beta)\neq(0,0)$,
    while the complementary bit yields weight two.
    This establishes the table.

    We now establish the weight identity~\eqref{eq:weight}.
    For positions in $\mathcal M_{00}$, exactly
   $h_{00}$ three-bit vectors have weight three,
    while the remaining ones have weight zero.
    For each $(\alpha,\beta)\neq(0,0)$, exactly~$h_{\alpha\beta}$ positions contribute weight two,
    while the remaining
    $m_{\alpha\beta}-h_{\alpha\beta}$ positions
    contribute weight one. Consequently,
    \begin{align*}
        \big|\varphi_{\vec{a},\vec{b},\vec{P}}
        (\vec{x}_U,\vec{x}_V,\vec{x}_W)\big|
        &=3h_{00}
        +\sum_{(\alpha,\beta)\neq(0,0)}
        \big(2h_{\alpha\beta}
        +m_{\alpha\beta}-h_{\alpha\beta}\big)\\
        &=3h_{00}
        +m_{10}+m_{01}+m_{11}
        +h_{10}+h_{01}+h_{11}\\
        &=\nu+3h_{00}+h_{10}+h_{01}+h_{11},
    \end{align*}
    where the last equality follows from
    $\nu=n/3-m_{00}=m_{10}+m_{01}+m_{11}$.

    Finally, for fixed $\vec{x}_V$ and $\vec{x}_W$,
    prescribing $\vec{h}$ amounts to choosing,
    within each $\mathcal M_{\alpha\beta}$,
    exactly~$h_{\alpha\beta}$ positions at which
    $\vec{x}_U(i)$ takes the larger-weight value.
    Since each choice uniquely determines $\vec{x}_U$,
    the number of such vectors is
    \[
        N_{\vec{m}}(\vec{h})
        =\prod_{\alpha,\beta\in\{0,1\}}
        \binom{m_{\alpha\beta}}{h_{\alpha\beta}},
    \]
    proving~\eqref{eq:Nm}.
\end{proof}
}

	We can now explain why the statistics of Definition~\ref{def:statistics} are useful. The following lemma shows that the probabilities associated to $\vec{e}^{\mathrm{unif}}$ are only functions of these statistics.
	
	\begin{restatable}{lemma}{lemmaRefUnif}\label{lem:reference-uniformity}
		For any $\vec{x}_U, \vec{x}_V, \vec{x}_W \in \mathbb{F}_{2}^{n/3}$,
		\begin{enumerate}\setlength{\itemsep}{6pt}
			\item[\textup{(1)}] $\mathbb{P}\left(\vec{e}_W^{\mathrm{unif}}=\vec{x}_W\right)$ depends only on $\omega(\vec{x}_W)$,
			$$
			\mathbb{P}\left(\vec{e}_W^{\mathrm{unif}}=\vec{x}_W \right) = \frac{A_{\omega}}{\binom{n}{t}} \mbox{ where } A_{\omega} \eqdef 2^{\omega}\sum_{j} \binom{n/3-\omega}{j}\binom{n/3}{t-\omega-2j} \ . 
			$$
			\item[\textup{(2)}] $\mathbb{P}\left(\vec{e}_V^{\mathrm{unif}}=\vec{x}_V\mid\vec{e}_W^{\mathrm{unif}}=\vec{x}_W\right)$ depends only on
			$(\omega(\vec{x}_W),\nu(\vec{x}_V,\vec{x}_W))$,
			$$
			\mathbb{P}\left(\vec{e}_V^{\mathrm{unif}}=\vec{x}_V\mid\vec{e}_W^{\mathrm{unif}}=\vec{x}_W\right) = \frac{B_{\nu}}{A_{\omega}} \mbox{ where } B_{\nu}\eqdef\sum_{j=0}^{n/3-\nu}
			\binom{n/3-\nu}{j}\binom{\nu}{t-\nu-3j} \ . 
			$$
			\item[\textup{(3)}] $\mathbb{P}\left(\vec{e}_U^{\mathrm{unif}}=\vec{x}_U\mid
			\vec{e}_V^{\mathrm{unif}}=\vec{x}_V,\vec{e}_W^{\mathrm{unif}}=\vec{x}_W\right)$ depends only on
			$\left(\vec{m}(\vec{x}_V,\vec{x}_W),\vec{h}(\vec{x}_U;\vec{x}_V,\vec{x}_W)\right)$,
			$$
			\mathbb{P}\left(\vec{e}_U^{\mathrm{unif}}=\vec{x}_U\mid
			\vec{e}_V^{\mathrm{unif}}=\vec{x}_V,\vec{e}_W^{\mathrm{unif}}=\vec{x}_W\right) = 
			\begin{cases}
				\frac{1}{B_{\nu}}
				& \text{if }\nu + 3h_{00}+h_{10}+h_{01}+h_{11}=t,\\[1.2ex]
				0 & \text{otherwise}.
			\end{cases} 
			$$
		\end{enumerate}
		
All conditional probabilities are understood to be
conditioned on events of positive probability.
	\end{restatable}
	
	\iftoggle{llncs}{}{

\begin{proof}
    Write $\vec{e}^{\mathrm{unif}}=(\vec{x},\vec{y},\vec{z})\vec{P}$.
    Since $\vec{P}$ is a permutation matrix,
    $(\vec{x},\vec{y},\vec{z})$ is uniformly distributed
    over vectors of Hamming weight $t$.

    We first prove (1). By Lemma~\ref{lemma:varphiInverse}, for $i\in[1,n/3]$,
    \[
        \vec{e}_W^{\mathrm{unif}}(i)=
        \begin{cases}
            \vec{x}(i)+\vec{z}(i)
                & \text{if }\vec{b}(i)=1,\\
            \vec{y}(i)+\vec{z}(i)
                & \text{otherwise}.
        \end{cases}
    \]
    Define $\vec{p},\vec{r}\in\mathbb{F}_2^{n/3}$ by
    \[
        \vec{p}(i)\eqdef
        \begin{cases}
            \vec{x}(i) & \text{if }\vec{b}(i)=1\\
            \vec{y}(i) & \text{otherwise}
        \end{cases}
        \qquad
        \vec{r}(i)\eqdef
        \begin{cases}
            \vec{y}(i) & \text{if }\vec{b}(i)=1,\\
            \vec{x}(i) & \text{otherwise}.
        \end{cases}
    \]
    Thus, $\vec{e}_W^{\mathrm{unif}}=\vec{p}+\vec{z}$.
    The map
    $(\vec{x},\vec{y},\vec{z})\mapsto
    (\vec{p},\vec{r},\vec{z})$
    only exchanges $\vec{x}(i)$ and $\vec{y}(i)$
    at positions where~$\vec{b}(i)=0$, and hence it keeps the distribution invariant.

    Fix $\vec{x}_W\in\mathbb{F}_2^{n/3}$ and let
    $\omega\eqdef\omega(\vec{x}_W)$.
    The condition $\vec{e}_W^{\mathrm{unif}}=\vec{x}_W$
    imposes odd parity on exactly~$\omega$ pairs
    $(\vec{p}(i),\vec{z}(i))$ and even parity on
    the remaining $n/3-\omega$ pairs.
    Each odd-parity pair has two possible values,
    both of weight one.
    Among the even-parity pairs, choose $j$ positions
    to take the value $(1,1)$, while the others
    take the value $(0,0)$.
    The vector $\vec{r}$ must then have weight
    $t-\omega-2j$.

    Consequently, the number of vectors
    $(\vec{p},\vec{r},\vec{z})$ of weight $t$
    satisfying $\vec{p}+\vec{z}=\vec{x}_W$ is
    \[
        2^\omega\sum_j
        \binom{n/3-\omega}{j}
        \binom{n/3}{t-\omega-2j}
        =A_\omega\ .
    \]
    Since there are $\binom{n}{t}$ vectors of weight $t$,
    we obtain (1).

    Now, let us prove (2).
    Fix $\vec{x}_V,\vec{x}_W\in\mathbb{F}_2^{n/3}$
    and let $\nu\eqdef\nu(\vec{x}_V,\vec{x}_W)$.
    By Proposition~\ref{prop:counts}, each position
    in $\mathcal{M}_{00}$ contributes either weight
    zero or three, while each position in
    $\mathcal{M}_{\alpha\beta}$ with~$(\alpha,\beta)\neq(0,0)$ contributes either
    weight one or two.
    Moreover, each choice of weight uniquely
    determines the corresponding bit of $\vec{x}_U$.

    Suppose that $j$ of the $n/3-\nu$ positions
    in $\mathcal{M}_{00}$ contribute weight three,
    and that $s$ of the remaining $\nu$ positions
    contribute weight two.
    The total Hamming weight is then
    $$
        3j+2s+(\nu-s)=\nu+3j+s \ .
    $$
    Hence, the condition that the total weight
    equals $t$ imposes $s=t-\nu-3j$.
    For each $j$, there are~$\binom{n/3-\nu}{j}$ ways to choose
    the positions contributing weight three,
    and~$\binom{\nu}{t-\nu-3j}$ ways to choose
    those contributing weight two.
    Since these choices uniquely determine $\vec{x}_U$,
    the number of vectors~$\vec{x}_U$ compatible
    with the fixed pair $(\vec{x}_V,\vec{x}_W)$ is
    \[
        B_\nu
        =\sum_{j=0}^{n/3-\nu}
        \binom{n/3-\nu}{j}
        \binom{\nu}{t-\nu-3j}\ .
    \]
    Therefore,
    \[
        \mathbb{P}\left(
        \vec{e}_V^{\mathrm{unif}}=\vec{x}_V,
        \vec{e}_W^{\mathrm{unif}}=\vec{x}_W
        \right)
        =\frac{B_\nu}{\binom{n}{t}}\ .
    \]
    Combining this with (1), we obtain
    \[
        \mathbb{P}\left(
        \vec{e}_V^{\mathrm{unif}}=\vec{x}_V
        \mid
        \vec{e}_W^{\mathrm{unif}}=\vec{x}_W
        \right)
        =\frac{B_\nu}{A_\omega},
    \]
    which proves (2).

    Finally, we prove (3).
    Fix $\vec{x}_V,\vec{x}_W$ such that
    $B_\nu>0$.
    Conditioned on
    $(\vec{e}_V^{\mathrm{unif}},
      \vec{e}_W^{\mathrm{unif}})
    =(\vec{x}_V,\vec{x}_W)$,
    the vector $\vec{e}_U^{\mathrm{unif}}$
    is uniformly distributed over the $B_\nu$
    compatible vectors.
    Thus, for any $\vec{x}_U$,
    \[
        \mathbb{P}\left(
        \vec{e}_U^{\mathrm{unif}}=\vec{x}_U
        \mid
        \vec{e}_V^{\mathrm{unif}}=\vec{x}_V,
        \vec{e}_W^{\mathrm{unif}}=\vec{x}_W
        \right)
        =
        \begin{cases}
            \dfrac{1}{B_\nu}
            & \text{if }
            \big|\varphi_{\vec{a},\vec{b},\vec{P}}
            (\vec{x}_U,\vec{x}_V,\vec{x}_W)\big|=t,\\[1.2ex]
            0 & \text{otherwise}.
        \end{cases}
    \]
    By Equation~\eqref{eq:weight},
    \[
        \big|\varphi_{\vec{a},\vec{b},\vec{P}}
        (\vec{x}_U,\vec{x}_V,\vec{x}_W)\big|
        =\nu+3h_{00}+h_{10}+h_{01}+h_{11}\ .
    \]
    Consequently,
    \[
        \mathbb{P}\left(
        \vec{e}_U^{\mathrm{unif}}=\vec{x}_U
        \mid
        \vec{e}_V^{\mathrm{unif}}=\vec{x}_V,
        \vec{e}_W^{\mathrm{unif}}=\vec{x}_W
        \right)
        =
        \begin{cases}
            \dfrac{1}{B_\nu}
            & \text{if }
            \nu+3h_{00}+h_{10}+h_{01}+h_{11}=t,\\[1.2ex]
            0 & \text{otherwise}.
        \end{cases}
    \]
    Since $\nu=n/3-m_{00}$, both the denominator
    and the condition depend only on
    $(\vec{m},\vec{h})$.
    This proves~(3) and completes the proof.
\end{proof}

}

Lemma \ref{lem:reference-uniformity} characterizes the successive conditional distributions of $\vec{e}_W^{\mathrm{unif}}$, $\vec{e}_V^{\mathrm{unif}}$, and $\vec{e}_U^{\mathrm{unif}}$, showing that they depend only on the corresponding statistics. Our goal is to design the decoders $\mathsf{DecodeW}$, $\mathsf{DecodeV}$, and $\mathsf{DecodeU}$ in Instructions \ref{inst:W1}, \ref{inst:V1}, and \ref{inst:U1} of Algorithm~\ref{algo:inv1} to reproduce these distributions as closely as possible, thereby ensuring that the output $\vec{e}^{\mathrm{out}}$ is close to uniformly distributed over $\mathcal{S}_t^n$. To this end, we first suppose that our decoders satisfy some ``ideal'' properties, before showing that our concrete instantiations sufficiently approximate their ideal counterparts.

	\begin{definition}[Perfect decoders]\label{def:uniformity}
		The three decoders $\mathsf{DecodeW}$, $\mathsf{DecodeV}$ and $\mathsf{DecodeU}$ are said to be
		\emph{perfect} if there exist three functions $p_W,p_V,p_U$ satisfying:
	\begin{enumerate}\setlength{\itemsep}{6pt}
		\item[\textup{(1)}]
		$$
		\mathbb{P}\big(\mathsf{DecodeW}(\vec H_W,\vec{s}_W)=\vec x_W\big)
		=p_W\big(\omega(\vec x_W)\big),
		$$
		\item[\textup{(2)}] for every fixed $\vec x_W$,
		$$
		\mathbb{P}\big(\mathsf{DecodeV}(\vec H_V,\vec{s}_V,\vec x_W)=\vec x_V\big)
		=p_V\big(\omega(\vec x_W),\nu(\vec x_V,\vec x_W)\big),
		$$
		\item[\textup{(3)}] for every fixed $(\vec x_V,\vec x_W)$,
		$$	
			\mathbb{P}\big(\mathsf{DecodeU}(\vec H_U,\vec{s}_U,\vec x_V,\vec x_W,
			\varphi_{\vec a,\vec b,\vec P})=\vec x_U\big) =p_U\big(\vec m(\vec x_V,\vec x_W), \vec h(\vec x_U;\vec x_V,\vec x_W) \big) \ .
		$$
	\end{enumerate}
	The probabilities are \iftoggle{llncs}{}{taken here} over uniform $\vec{s}_U$, $\vec{s}_V$ and $\vec{s}_W$ and the internal randomness of the decoders. In particular matrices $\vec{H}_{U}$, $\vec{H}_V$ and $\vec{H}_W$ are supposed to be fixed. 
\end{definition}

Thus, for a uniform input vector, each {\em perfect decoder} returns vectors with the same statistics with equal probability. To obtain outputs distributed as $\vec{e}^{\mathrm{unif}}$, it remains to ensure that these statistics themselves follow the corresponding statistics from $\vec{e}^{\mathrm{unif}}$ as shown by the following lemma.

\begin{restatable}{lemma}{lemmaUnifSampl}
\label{lem:uniform}
Let $\vec e_U,\vec e_V,\vec e_W\in\mathbb F_2^{n/3}$
be random vectors, and define $\vec e^{\mathrm{out}}
    \eqdef
    \varphi_{\vec a,\vec b,\vec P}
    (\vec e_U,\vec e_V,\vec e_W)$. Assume that the probability of the event
$\vec e_W=\vec x_W$ depends only on $\omega(\vec x_W)$,
that the probability of~$\vec e_V=\vec x_V$ conditioned
on $\vec e_W=\vec x_W$ depends only on
$(\omega(\vec x_W),\nu(\vec x_V,\vec x_W))$,
and that the probability of $\vec e_U=\vec x_U$
conditioned on $(\vec e_V,\vec e_W)=(\vec x_V,\vec x_W)$
depends only on
$(\vec m(\vec x_V,\vec x_W),
\vec h(\vec x_U;\vec x_V,\vec x_W))$.
Suppose furthermore that, for all admissible
values $\widetilde{\omega}$, $\widetilde{\nu}$, $\widetilde{\vec m}$, and $\widetilde{\vec h}$,
\begin{align}
    \mathbb P\big(\omega(\vec e_W)=\widetilde{\omega}\big)
    &=
    \mathbb P\big(\omega(\vec e_W^{\mathrm{unif}})=\widetilde{\omega}\big),
    \label{eq:uniform-omega}
  \end{align} 
 \begin{align} 
    \mathbb P\big(
        \nu(\vec e_V,\vec e_W)=\widetilde{\nu}
        \mid \omega(\vec e_W)=\widetilde{\omega}
    \big)
    &=
    \mathbb P\big(
        \nu(\vec e_V^{\mathrm{unif}},
            \vec e_W^{\mathrm{unif}})=\widetilde{\nu}
        \mid \omega(\vec e_W^{\mathrm{unif}})=\widetilde{\omega}
    \big),
    \label{eq:uniform-nu}
\end{align}
\iftoggle{llncs}{
\begin{align}
    &\mathbb P\big(
        \vec h(\vec e_U;\vec e_V,\vec e_W)=\widetilde{\vec h}
        \mid \vec m(\vec e_V,\vec e_W)=\widetilde{\vec m}
    \big)
    \notag\\
    &\qquad =
    \mathbb P\big(
        \vec h(\vec e_U^{\mathrm{unif}};
            \vec e_V^{\mathrm{unif}},
            \vec e_W^{\mathrm{unif}})=\widetilde{\vec h}
        \mid
        \vec m(\vec e_V^{\mathrm{unif}},
            \vec e_W^{\mathrm{unif}})=\widetilde{\vec m}
    \big).
    \label{eq:uniform-h}
\end{align}
}{\begin{multline}
    \mathbb{P}\big(
        \vec h(\vec e_U;\vec e_V,\vec e_W)=\widetilde{\vec h}
        \mid \vec m(\vec e_V,\vec e_W)=\widetilde{\vec m}
    \big)
     \\ =
    \mathbb P\big(
        \vec h(\vec e_U^{\mathrm{unif}};
            \vec e_V^{\mathrm{unif}},
            \vec e_W^{\mathrm{unif}})=\widetilde{\vec h}
        \mid
        \vec m(\vec e_V^{\mathrm{unif}},
            \vec e_W^{\mathrm{unif}})=\widetilde{\vec m}
    \big)\ .
    \label{eq:uniform-h}
\end{multline}
}

Then
\begin{equation}
    \Delta(\vec e^{\mathrm{out}},
           \vec e^{\mathrm{unif}})=0\ .
\end{equation}
\end{restatable}

	\iftoggle{llncs}{}{

\begin{proof}
    Fix $\vec{x}\in\mathcal{S}_t^n$ and write $(\vec{x}_U,\vec{x}_V,\vec{x}_W)
        =\varphi_{\vec{a},\vec{b},\vec{P}}^{-1}(\vec{x})$, with corresponding statistics $\widetilde{\omega}\eqdef\omega(\vec{x}_W)$, $\widetilde{\nu}\eqdef\nu(\vec{x}_V,\vec{x}_W)$, $\widetilde{\vec{m}}
        \eqdef\vec{m}(\vec{x}_V,\vec{x}_W)$ and $\widetilde{\vec{h}} \eqdef\vec{h}(\vec{x}_U;\vec{x}_V,\vec{x}_W)$.

There are $\binom{n/3}{\widetilde{\omega}}$ vectors $\vec{x}_W$ of weight $\widetilde{\omega}$. For a fixed $\vec{x}_W$, a vector $\vec{x}_V$ with statistic $\widetilde{\nu}$ must have exactly $\widetilde{\nu}-\widetilde{\omega}$ non-zero coordinates among the $n/3-\widetilde{\omega}$ positions where $\vec{x}_W$ is zero, while its coordinates on the support of $\vec{x}_W$ are arbitrary. There are therefore $2^{\widetilde{\omega}}
        \binom{n/3-\widetilde{\omega}}
              {\widetilde{\nu}-\widetilde{\omega}}$ such vectors. Finally, for a fixed pair $(\vec{x}_V,\vec{x}_W)$, Equation~\eqref{eq:Nm} counts $N_{\widetilde{\vec{m}}}(\widetilde{\vec{h}})$ possible vectors $\vec{x}_U$ with statistic~$\widetilde{\vec{h}}$.

By the assumptions of the lemma, all vectors sharing the same statistics have equal (conditional) probabilities. Therefore, each probability can be expressed as the probability of the corresponding statistic divided by the number of compatible vectors.

Using the chain rule for conditional probabilities, we obtain
    \begin{align*}
        \mathbb{P}(\vec{e}^{\mathrm{out}}=\vec{x})
        &=
        \mathbb{P}\big(
            \vec{e}_U=\vec{x}_U
            \mid
            \vec{e}_V=\vec{x}_V,
            \vec{e}_W=\vec{x}_W
        \big) \cdot
        \mathbb{P}\big(
            \vec{e}_V=\vec{x}_V
            \mid
            \vec{e}_W=\vec{x}_W
        \big)
        \cdot
        \mathbb{P}\big(
            \vec{e}_W=\vec{x}_W
        \big)\\[1ex]
        &=
        \frac{
            \mathbb{P}\big(
                \vec{h}(\vec{e}_U;\vec{e}_V,\vec{e}_W)
                =\widetilde{\vec{h}}
                \mid
                \vec{m}(\vec{e}_V,\vec{e}_W)
                =\widetilde{\vec{m}}
            \big)
        }{
            N_{\widetilde{\vec{m}}}
              (\widetilde{\vec{h}})
        }\\
        &\qquad\qquad \cdot
        \frac{
            \mathbb{P}\big(
                \nu(\vec{e}_V,\vec{e}_W)
                =\widetilde{\nu}
                \mid
                \omega(\vec{e}_W)
                =\widetilde{\omega}
            \big)
        }{
            2^{\widetilde{\omega}}
            \binom{n/3-\widetilde{\omega}}
                  {\widetilde{\nu}-\widetilde{\omega}}
        }\cdot
        \frac{
            \mathbb{P}\big(
                \omega(\vec{e}_W)=\widetilde{\omega}
            \big)
        }{
            \binom{n/3}{\widetilde{\omega}}
        }.
    \end{align*}

By the assumptions of the lemma, the last expression equals
		\begin{align*}
			&\frac{\mathbb{P}(\vec{h}(\vec{e}_U^{\mathrm{unif}};\vec{e}_V^{\mathrm{unif}},\vec{e}_W^{\mathrm{unif}})=\widetilde{\vec{h}}\mid
				\vec{m}(\vec{e}_V^{\mathrm{unif}},\vec{e}_W^{\mathrm{unif}})=\widetilde{\vec{m}})}{N_{\widetilde{\vec{m}}}(\widetilde{\vec{h}})}\\
				&\qquad\qquad \cdot
			\frac{\mathbb{P}(\nu(\vec{e}_V^{\mathrm{unif}},\vec{e}_W^{\mathrm{unif}})=\widetilde{\nu}\mid
				\omega(\vec{e}_W^{\mathrm{unif})}=\widetilde{\omega})}
			{2^{\widetilde{\omega}}\binom{n/3-\widetilde{\omega}}{\widetilde{\nu} - \widetilde{\omega}}}\cdot
			\frac{\mathbb{P}(\omega(\vec{e}_W^{\mathrm{unif}})=\widetilde{\omega})}{\binom{n/3}{\widetilde{\omega}}}\\
			&\qquad=\mathbb{P}(\vec{e}_U^{\mathrm{unif}}=\vec{x}_U\mid
			\vec{e}_V^{\mathrm{unif}}=\vec{x}_V,\vec{e}_W^{\mathrm{unif}}=\vec{x}_W)\cdot\mathbb{P}(\vec{e}_V^{\mathrm{unif}}=\vec{x}_V\mid\vec{e}_W^{\mathrm{unif}}=\vec{x}_W)
			\mathbb{P}(\vec{e}_W^{\mathrm{unif}}=\vec{x}_W)\notag\\
			&\qquad=\mathbb{P}(\vec{e}^{\mathrm{unif}}=\vec{x})
		\end{align*}
		where we used Lemma~\ref{lem:reference-uniformity} in the first
		equality and the bijectivity of $\varphi_{\vec{a},\vec{b},\vec{P}}$ in the last one.
		We have therefore proved that
		$\mathbb{P}(\vec{e}^{\mathrm{out}}=\vec{x})=\mathbb{P}(\vec{e}^{\mathrm{unif}}=\vec{x})$ for every
		$\vec{x}\in \mathcal{S}_t^n$ which concludes the proof. 
\end{proof}

}
	
	Suppose now that the three decoders satisfy the uniformity conditions of Definition~\ref{def:uniformity}. By Lemma~\ref{lem:uniform}, to show that $\vec{e}^{\mathrm{out}}$ is uniformly distributed over $\mathcal{S}^n_t$, it remains to control the intermediate distribution of $\omega(\vec{e}_W)$, that of $\nu(\vec{e}_V, \vec{e}_W)$ knowing~$\omega(\vec{e}_W)$, and that of $\vec{h}(\vec{e}_U;\vec{e}_V, \vec{e}_W)$ knowing~$\vec{m}(\vec{e}_V, \vec{e}_W)$. A way to a priori ensure this is to use some {\em rejection sampling}:  at each step, we accept the proposed vector with a probability depending only on these statistics. This enables to change their distributions and this gives\footnote{In this algorithm, $z \Unif [0,1]$ means that we pick a uniform real $z$ between $0$ and $1$.} Algorithm \ref{algo:inv2}. 

\begin{algorithm}[htbp]
	\caption{ $\mathsf{InvertAlg}(\vec{s},T)$: \scheme{} inversion algorithm }\label{algo:inv2}
	\begin{algorithmic}[1]
		\Require A vector $\vec{s} \in \mathbb{F}_2^{n-k}$ and
         an output $T \eqdef (\vec{a},\vec{b},\vec{P},\vec{S},
         \vec{H}_U,\vec{H}_V,\vec{H}_W)$
         of Algorithm~\ref{algo:trap}

\Ensure A vector $\vec{e} \in \mathcal{S}_{t}^{n}$
        satisfying $\vec{e}\vec{H}^{\top} = \vec{s}$,
        where $\vec{H}$ is the second output of
        Algorithm~\ref{algo:trap}\medskip

		\State $(\vec{s}_U,\vec{s}_V,\vec{s}_W) \leftarrow \vec{s}\vec{S}^{\top}$ 
			\Repeat
			\State $\vec{e}_W\gets\mathsf{DecodeW}(\vec{H}_W,\vec{s}_W)$
			\State $z\Unif [0,1]$
			\Until{$z\le \vec{r}_W(\omega(\vec{e}_W))$} 	\Comment{$\vec{r}_W$ defined in Proposition \ref{propo:rejetUnif}}
			\Repeat
			\State $\vec{e}_V\gets\mathsf{DecodeV}(\vec{H}_V,\vec{s}_V,\vec{e}_W)$
			\State $z\Unif[0,1]$
			\Until{$z\le \vec{r}_V(\omega(\vec{e}_W), \nu(\vec{e}_V,\vec{e}_W))$} \Comment{$\vec{r}_V$ defined in Proposition \ref{propo:rejetUnif}}
			\Repeat
			\State $\vec{e}_U\gets\mathsf{DecodeU}(\vec{H}_U,\vec{s}_U,\vec{e}_V,\vec{e}_W,\varphi_{\vec{a},\vec{b},\vec{P}})$
			\State $z \Unif [0,1]$
			\Until{$z\le \vec{r}_U(\vec{m}(\vec{e}_V,\vec{e}_W), \vec{h}(\vec{e}_U;\vec{e}_V,\vec{e}_W))$} \Comment{$\vec{r}_U$ defined in Proposition \ref{propo:rejetUnif}}
			\State $\vec{e}^{\mathrm{out}}\gets\varphi_{\vec{a}, \vec{b}, \vec{P}}(\vec{e}_U,\vec{e}_V,\vec{e}_W)$ \label{algomark:inv2}
			\State \Return $\vec{e}^{\mathrm{out}}$
	\end{algorithmic}
\end{algorithm}

However there is a caveat: rejections are made based on the output distributions of the decoders (which are supposed for now to be perfect as in Definition~\ref{def:uniformity}) for a uniform input~$\vec{s}_W$,~$\vec{s}_V$ and~$\vec{s}_U$. In Algorithm~\ref{algo:inv2}, these vectors are kept fixed throughout each rejection loop. Rejection therefore normalises the output distribution separately for each~$\vec{s}_W$,~$\vec{s}_V$ and $\vec{s}_U$, by its own probability of acceptance. Since this probability may depend on them, the distributions obtained by averaging a single decoding call over uniform~$\vec{s}_W$,~$\vec{s}_V$ and~$\vec{s}_U$ do not suffice to analyse the repeated calls. This difficulty concerns both the perfect decoders and their subsequent instantiations where probabilities are computed over uniform $\vec{s}_W$,~$\vec{s}_V$ and~$\vec{s}_U$.

We must therefore justify that rejection with a fixed $\vec{s}_W$, $\vec{s}_V$ and $\vec{s}_U$ gives the desired output distribution, up to a small statistical distance. We first identify the conditional output laws when they are fixed. This will allow us to bound the distance to $\vec{e}^{\mathrm{unif}}$ in terms of the deviations from uniformity of $\vec{e}_W\vec{H}_{W}^{\top}$, $\vec{e}_V\vec{H}_{V}^{\top}$ and $\vec{e}_U\vec{H}_{U}^{\top}$ via the leftover hash lemma. It is at this point that we will crucially use that the underlying codes $W$, $V$ and $U$ in our construction are random.

	\begin{restatable}{proposition}{propoRejetUnif}\label{propo:rejetUnif}
Suppose in Algorithm \ref{algo:inv2} that decoders $\mathsf{DecodeW}$, $\mathsf{DecodeV}$ and $\mathsf{DecodeU}$ are perfect, {\it i.e.,} they satisfy Definition~\ref{def:uniformity}. Fix the matrices $\vec H_U,\vec H_V,\vec H_W$ and the
vectors $\vec s_U,\vec s_V,\vec s_W$.

Using the notation of Lemma~\ref{lem:reference-uniformity} and Definition~\ref{def:uniformity}, define
	\begin{equation}
		M_W\eqdef\max_{\omega: A_\omega > 0}
		\frac{A_\omega}{\binom{n}{t}\; p_W(\omega)} \;\; \mbox{ and } \;\;
		M_V(\omega)\eqdef\max_{\nu:B_{\nu}>0}
		\frac{B_{\nu}}{A_\omega \; p_V(\omega,\nu)}
	\end{equation}
	\begin{equation}
		M_U(\vec m)\eqdef
		\max_{\substack{\vec{h}: N_{\vec{m}}(\vec{h})>0\\
				\nu + 3h_{00}+h_{10}+h_{01}+h_{11}=t\\ \nu = n/3 - m_{00}}}
		\frac{1}{B_{\nu}\; p_U(\vec{m},\vec{h})}.
	\end{equation}
	where the probabilities considered in these maximum are assumed to be strictly positive, so that these constant are finite. 
Define,
	\begin{equation}
		\vec{r}_W(\omega)\eqdef
		\frac{A_\omega}{M_W\binom{n}{t}\,p_W(\omega)},\;\;
		\vec{r}_V(\omega,\nu)\eqdef
		\frac{B_{\nu}}
		{M_V(\omega)A_\omega \; p_V(\omega,\nu)},
	\end{equation}
	\begin{equation}
		\vec{r}_U(\vec{m},\vec{h})\eqdef \dfrac{1}{M_U(\vec m)B_{\nu}\; 	p_U(\vec{m},\vec{h})}
	\end{equation}
and being equal to $0$ when the considered probabilities are zero.

Then, the outputs of the three loops in Algorithm~\ref{algo:inv2} satisfy
\begin{align}
		\mathbb P(\vec e_W=\vec x_W)
		&=\mathbb P\left(\vec e_W^{\mathrm{unif}}=\vec x_W
		\mid\vec e_W^{\mathrm{unif}}\vec H_W^{\top}=\vec{s}_W\right),\label{eq:rej_W}
	\end{align}
	\begin{align} 
		\mathbb P(\vec e_V=\vec x_V\mid\vec e_W=\vec x_W)
		&=\mathbb P\left(\vec e_V^{\mathrm{unif}}=\vec x_V
		\mid\vec e_W^{\mathrm{unif}}=\vec x_W,
		\vec e_V^{\mathrm{unif}}\vec H_V^{\top}=\vec{s}_V\right), \label{eq:rej_V}
	\end{align} 
	\begin{multline}
		\mathbb P(\vec e_U=\vec x_U\mid
		\vec e_V, =\vec x_V,\vec e_W=\vec x_W)
		\\=\mathbb P\left(\vec e_U^{\mathrm{unif}}=\vec x_U
		\mid(\vec e_V^{\mathrm{unif}},\vec e_W^{\mathrm{unif}})
		=(\vec x_V,\vec x_W),
		\vec e_U^{\mathrm{unif}}\vec H_U^{\top}=\vec{s}_U\right).\label{eq:rej_U}
	\end{multline}
	\end{restatable}

\iftoggle{llncs}{}{
\begin{proof}
Throughout the proof, the matrices $\vec{H}_W$, $\vec{H}_V$, $\vec{H}_U$ and vectors $\vec{s}_W$, $\vec{s}_V$, $\vec{s}_U$ are fixed, and~$(\vec e_U,\vec e_V,\vec e_W)$ denotes the random tuple produced by Algorithm~\ref{algo:inv2} at Step~\ref{algomark:inv2}, once all three loops have been accepted.

We use the following elementary rejection-sampling fact. Suppose that the probability of proposing and accepting a value $x$ in a single trial is
\[
    c\,\mathbb P(X=x)\,\mathbf 1_{\{x\in S\}},
\]
where $c>0$ does not depend on $x$ and
$\mathbb P(X\in S)>0$.
Then independent repetitions until acceptance produce the conditional distribution of $X$ given $X\in S$. Indeed, the acceptance probability in one trial is $c\,\mathbb P(X\in S)$, and so the probability of proposing and accepting $x$ after the trials is 
\[
\begin{array}{lcl}
\sum_{j\geq 0} \Big( 1 - c\,\mathbb P(X\in S) \Big)^j c\,\mathbb P(X=x)\,\mathbf 1_{\{x\in S\}} & = & \dfrac{c\,\mathbb P(X=x)\,\mathbf 1_{\{x\in S\}}}
         {c\,\mathbb P(X\in S)} \\
& = & \mathbb P(X=x\mid X\in S)
\end{array}
\]
where $j$ is the number of rejected trials before acceptance.

We will apply this fact to each loop, keeping the outputs
of the preceding loops fixed.

By definition of $M_W$, $M_V$, and $M_U$, $\vec{r}_W$, $\vec{r}_V$ and $\vec{r}_U$ always lie in $[0,1]$.
Thus, comparing them with an independent uniform sample from $[0,1]$ implements the corresponding acceptance decision.

\medskip
\noindent\emph{The $W$-loop.}
Fix $\vec x_W$ and, for brevity, write $\omega=\omega (\vec x_W) \eqdef |\vec x_W|$. By Definition~\ref{def:uniformity},
\[
    \mathbb{P}\big(
        \mathsf{DecodeW}(\vec H_W,\vec s_W)=\vec x_W
    \big)
    =
    \begin{cases}
        2^{n/3-k_W}p_W(\omega)
        &\text{if }\vec x_W\vec H_W^\top=\vec s_W,\\
        0&\text{otherwise}.
    \end{cases}
\]
Therefore, the probability of
outputting and accepting a valid $\vec x_W$ in one trial is given by
\begin{align*}
    2^{n/3-k_W}p_W(\omega)\,\vec r_W(\omega)
    &=
    \frac{2^{n/3-k_W}}{M_W}
    \frac{A_\omega}{\binom nt}\\
    &=
    \frac{2^{n/3-k_W}}{M_W}
    \mathbb P(\vec e_W^{\mathrm{unif}}=\vec x_W)
\end{align*}
where the last equality follows from
Lemma~\ref{lem:reference-uniformity}. On the other hand, if $\vec{x}_W\vec{H}_W^{\top} \neq \vec{s}_W$, it cannot be output and thus accepted.
Using the rejection-sampling fact proves
Equation~\eqref{eq:rej_W}.

\medskip
\noindent\emph{The $V$-loop.}
Fix an output $\vec x_W$ of the $W$-loop 
and write~$\omega=\omega(\vec x_W)$. For an output $\vec x_V$ of the decoder, we denote $\nu$ as the statistic $\nu (\vec x_V,\vec x_W)$. By Definition~\ref{def:uniformity}, the definition of $\vec r_V$, and Lemma~\ref{lem:reference-uniformity}, the probability of outputting and accepting $\vec{x}_V$ 
satisfying
$\vec x_V\vec H_V^\top=\vec s_V$ in one trial is
$$
\frac{2^{n/3-k_V}}{M_V(\omega)}
    \mathbb P\big(
        \vec e_V^{\mathrm{unif}}=\vec x_V
        \mid \vec e_W^{\mathrm{unif}}=\vec x_W
    \big) \ .
$$
Since $\omega$ is fixed throughout the loop, the prefactor does not depend on $\vec x_V$. The same rejection-sampling fact therefore proves Equation~\eqref{eq:rej_V}.

\medskip
\noindent\emph{The $U$-loop.}
Fix outputs $(\vec x_V,\vec x_W)$ of the preceding loops occurring with positive probability, and write
\[
    \vec m=\vec m(\vec x_V,\vec x_W),
    \qquad \nu=n/3 - m_{00}\ .
\]
For a decoder output $\vec x_U$, write
$\vec h=\vec h(\vec x_U;\vec x_V,\vec x_W)$. By Definition~\ref{def:uniformity},  Lemma~\ref{lem:reference-uniformity}  and the definition of
$\vec r_U$, the probability of outputting and accepting $\vec{x}_U$ satisfying $\vec{x}_U\vec{H}_U^{\top}$ if  $\nu+3h_{00}+h_{10}+h_{01}+h_{11}=t$ 
in one trial is
$$
\frac{2^{n/3-k_U}}{M_U(\vec m)}
    \mathbb P\big(
        \vec e_U^{\mathrm{unif}}=\vec x_U
        \mid
        (\vec e_V^{\mathrm{unif}},\vec e_W^{\mathrm{unif}})
        =(\vec x_V,\vec x_W)
    \big) \ .
$$
Since $\vec m$ is fixed throughout the loop, the prefactor
does not depend on $\vec x_U$.
Applying the same rejection-sampling fact proves
Equation~\eqref{eq:rej_U} and concludes the proof.
	\end{proof}
}

Proposition~\ref{propo:rejetUnif} does not ensure that signatures $\vec{e}^{\mathrm{out}}$ are not leaking information on their underlying trapdoor. For example, for a fixed vector $\vec s_W$, it identifies the distribution of $\vec e_W$ with that of $\vec e_W^{\mathrm{unif}}$ conditioned on $\vec e_W^{\mathrm{unif}}\vec H_W^\top=\vec s_W$. This is not generally a uniform distribution on the admissible vectors: their probabilities are proportional to $A_{\omega}$.

Choosing the input vector uniformly does not necessarily
recover the target distribution, so Proposition~\ref{propo:rejetUnif} alone does not establish that $\vec e_W$ is independent of the trapdoor. It might still happen that the union of the sets of admissible vectors $\vec{x}_W$ satisfying $\vec{x}_W\vec{H}_W^{\top}=\vec{s}_W$, taken over most inputs $\vec{s}_W$, is a small subset of the domain having only a small probability under $\vec{e}_W^{\mathrm{unif}}$. A uniform input
$\vec s_W$ would then cause the rejection loop to return a vector in this union with high probability. This would produce a bias that could depend on the secret matrix $\vec{H}_W$, which Proposition~\ref{propo:rejetUnif} alone does not exclude.

	Hopefully, we can show that this does not happen: the conditionings in Proposition~\ref{propo:rejetUnif} can be removed, up to a statistical error, after averaging over the uniform inputs $\vec{s}_U$, $\vec{s}_V$ and $\vec{s}_W$. For instance, fix~$\vec H_W$ and let $\vec e_W$ denote the
	output of the first rejection loop with input
	a uniform~$\vec{s}_W$. Since the conditional distributions agree for all the $2^{n/3-k_W}$ vectors $\vec{s}_W$, we deduce that
	\begin{equation} 
		\Delta(\vec e_W,\vec e_W^{\mathrm{unif}})
		=\frac{1}{2}\sum_{\vec y\in\mathbb F_2^{n/3-k_W}}
		\left|2^{-(n/3-k_W)}
		-\mathbb P(\vec{e}_W^{\mathrm{unif}}\vec{H}_W^{\top}=\vec{y})\right|
		=\Delta(\vec{e}_W^{\mathrm{unif}}\vec{H}_W^{\top},\vec{s}_W) \ . 
	\end{equation} 
	Thus, if $\vec e_W^{\mathrm{unif}}\vec H_W^{\top}$ is close to uniform,
	averaging over $\vec{s}_W$ gives an output $\vec{e}_W$ close to
	$\vec e_W^{\mathrm{unif}}$. The same argument applies at the other decoding steps.
	The following version of the leftover hash lemma will control this closeness via the statistical distance and the probability of collision.

		\begin{lemma}[Leftover hash lemma, {\cite[Lemma~2.5.1]{D23}}]\label{lemma:binaryLHL}
		Let $\vec{e} \in \mathbb F_2^{n}$ be some random variable,~$\vec{s} \Unif \mathbb{F}_{2}^{n-k}$ and $\vec H \Unif \mathbb F_2^{(n-k)\times n}$. Then,
		\begin{equation}
			\mathbb{E}_{\vec{H}}\bigl(\Delta(\vec{e}\vec{H}^{\top},\vec{s})\bigr)
			\leq\frac{1}{2}\sqrt{ 2^{n-k}
				\sum_{\vec x\in\mathbb{F}_2^{n}}\mathbb{P}(\vec e=\vec x)^2} \ .
		\end{equation}
	\end{lemma}

	We are now ready to state our main theorem showing that for perfect decoders signatures are not leaking information on their underlying trapdoor.

		\begin{restatable}{theorem}{theoremNoLeakage}\label{thm:perfect-no-leakage}
		Using the notation of Lemma~\ref{lem:reference-uniformity}, define
\begin{equation}
C_W\eqdef\frac{1}{\binom{n}{t}^2} \sum_{\omega=0}^{n/3}\binom{n/3}{\omega} A_\omega^2 
\qquad 
C_V\eqdef\frac{1}{\binom{n}{t}} \sum_{\substack{0\leq\omega\leq n/3\\A_\omega>0}}
			\frac{\binom{n/3}{\omega}}{A_\omega}
			\sum_{\nu=\omega}^{n/3} 2^\omega\binom{n/3-\omega}{\nu - \omega}B_{\nu}^2
\end{equation}
\begin{equation}
C_U\eqdef\frac{1}{\binom{n}{t}} \sum_{\substack{0\leq \nu \leq n/3\\B_{\nu}>0}} \binom{n/3}{\nu}3^{\nu} \ . 
\end{equation}

		Suppose in Algorithm \ref{algo:inv2} that decoders $\mathsf{DecodeW}$, $\mathsf{DecodeV}$ and $\mathsf{DecodeU}$ are perfect, {\it i.e.,} they satisfy Definition~\ref{def:uniformity}. Then, 
		$$
			\mathbb E_{\vec H_U,\vec H_V,\vec H_W} \Big( 
				\Delta(\vec e^{\mathrm{out}},\vec e^{\mathrm{unif}}) \Big)
				\leq \frac{1}{2}\Big(\sqrt{2^{n/3-k_W}C_W}
				+\sqrt{2^{n/3-k_V}C_V}
				+\sqrt{2^{n/3-k_U}C_U} \Big)
			$$
		where $\vec{H}_U \Unif \mathbb{F}_{2}^{(n/3-k_U)\times n/3}$, $\vec{H}_V \Unif \mathbb{F}_{2}^{(n/3 - k_V) \times n/3}$ and $\vec{H}_W \Unif \mathbb{F}_{2}^{(n/3-k_W) \times n/3}$, $\vec{e}^{\mathrm{unif}} \Unif \mathcal{S}_{t}^{n}$ and~$\vec{e}^{\mathrm{out}}$ is the output of Algorithm~\ref{algo:inv2} for $\vec{s} \Unif \mathbb{F}_{2}^{n-k}$. 
	\end{restatable}
	\iftoggle{llncs}{}{
	\begin{proof}
		Let us recall the following standard decomposition of the statistical distance: given two pairs of random variables $(X,Y)$ and $(X',Y')$,
		\begin{equation}\label{eq:statDistPairs}
			\Delta\Big( (X,Y),(X',Y') \Big) \leq \Delta(X,X') +\sum_x\mathbb P(X'=x)\,
			\Delta\bigl(Y\mid X=x,\;Y'\mid X'=x\bigr)\ .
		\end{equation}

		Fix for now a trapdoor, in particular $\vec{H}_{U}$, $\vec{H}_{V}$ and $\vec{H}_{W}$. Notice that since $\varphi_{\vec{a},\vec{b},\vec{P}}$ is a bijection, we have 
		$$
		\Delta\big(\vec{e}^{\mathrm{out}},\vec{e}^{\mathrm{unif}}\big) = \Delta\big( (\vec{e}_W,\vec{e}_V,\vec{e}_U), (\vec{e}_{W}^{\mathrm{unif}},\vec{e}_V^{\mathrm{unif}},\vec{e}_U^{\mathrm{unif}} ) \big) \ . 
		$$
Applying now Equation~\eqref{eq:statDistPairs} twice gives,
\begin{multline}\label{eq:upStatDist}
			\Delta\big(\vec e^{\mathrm{out}},\vec e^{\mathrm{unif}}\big) \leq \\ \Delta\big(\vec e_W,\vec e_W^{\mathrm{unif}}\big) +\sum_{\vec x_W}\mathbb{P}(\vec e_W^{\mathrm{unif}}=\vec x_W)\cdot 
				\Delta\bigl(\vec e_V\mid\vec e_W=\vec x_W,\;
				\vec e_V^{\mathrm{unif}}\mid\vec e_W^{\mathrm{unif}}=\vec x_W\bigr)
		\\ +\sum_{\vec x_V,\vec x_W}
				\mathbb P(\vec e_V^{\mathrm{unif}}=\vec x_V,
				\vec e_W^{\mathrm{unif}}=\vec x_W)\cdot\\
		  \Delta\bigl(
			\vec e_U\mid(\vec e_V,\vec e_W)=(\vec x_V,\vec x_W),\;
			\vec e_U^{\mathrm{unif}}\mid
			(\vec e_V^{\mathrm{unif}},\vec e_W^{\mathrm{unif}})
			=(\vec x_V,\vec x_W)\bigr) \ .
		\end{multline}

		Let us now compute the above statistical distance. First, since $\vec{s}$ is uniform and $\vec{S}$ is invertible,~$(\vec{s}_U,\vec{s}_V,\vec{s}_W)=\vec{s}\vec S^{\top}$ is a uniform. 
		Recall now that we assume that decoders $\mathsf{DecodeW}$, $\mathsf{DecodeV}$ and $\mathsf{DecodeU}$ of Algorithm \ref{algo:inv2} are perfect (see Definition~\ref{def:uniformity}). Therefore Proposition \ref{propo:rejetUnif} applies. In particular, for a fixed $\vec{H}_W$,
		\begin{align*} 
		\mathbb{P}(\vec{e}_W = \vec{x}_W) &= \mathbb{P}\big(\vec{e}_W = \vec{x}_W \mid \vec{s}_W = \vec{x}_W \vec{H}_{W}^{\top}\big)\mathbb{P}\big( \vec{s}_W = \vec{x}_W \vec{H}_{W}^{\top} \big) \\
		&= \mathbb{P}\big(\vec{e}_W = \vec{x}_W \mid \vec{s}_W = \vec{x}_W\vec{H}_{W}^{\top}\big)  \;\frac{1}{2^{n/3-k_W}} \quad \left(\mbox{$\vec{s}_W$ is uniform}\right)  \\
		&= \mathbb{P}\big(\vec{e}^{\mathrm{unif}}_W = \vec{x}_W \mid \vec{e}_W^{\mathrm{unif}}\vec{H}_W^{\top} = \vec{x}_{W}\vec{H}_{W}^{\top} \big)    \;\frac{1}{2^{n/3-k_W}} \quad \left(\mbox{By Proposition \ref{propo:rejetUnif} } \right) \\
		&= \frac{\mathbb{P}\big(\vec{e}^{\mathrm{unif}}_W = \vec{x}_W\big)}{\mathbb{P}\big( \vec{e}_W^{\mathrm{unif}}\vec{H}_W^{\top} = \vec{x}_{W}\vec{H}_{W}^{\top} \big)} \;\frac{1}{2^{n/3-k_W}}
		\end{align*} 
		Therefore, 
		\begin{align*}
			\Delta\big(\vec{e}_W,\vec{e}^{\mathrm{unif}}_W\big) &= \frac{1}{2}\sum_{\vec{x}_W} \left| \mathbb{P} \big( \vec{e}_W = \vec{x}_W \big) - \mathbb{P}\big( \vec{e}_W^{\mathrm{unif}} = \vec{x}_W \big) \right|   \\
			&= \frac{1}{2}\sum_{\vec{y}} \sum_{\vec{x}_W: \; \vec{x}_W\vec{H}_W^{\top}=\vec{y}}  \left| \frac{\mathbb{P}\big(\vec{e}^{\mathrm{unif}}_W = \vec{x}_W\big)}{\mathbb{P}\big( \vec{e}_W^{\mathrm{unif}}\vec{H}_W^{\top} = \vec{x}_{W}\vec{H}_W^{\top} \big)} \;\frac{1}{2^{n/3-k_W}}-\mathbb{P}(\vec{e}_W^{\mathrm{unif}}=\vec{x}_W) \right| \\
			&=\frac{1}{2} \sum_{\vec{y}} \sum_{\vec{x}_W: \; \vec{x}_W\vec{H}_W^{\top}=\vec{y}}  \left| \frac{1}{\mathbb{P}\big( \vec{e}_W^{\mathrm{unif}}\vec{H}_W^{\top} = \vec{y} \big)} \;\frac{1}{2^{n/3-k_W}}-1 \right| \cdot \mathbb{P}(\vec{e}_W^{\mathrm{unif}}=\vec{x}_W) \\
			 &= \frac{1}{2}\sum_{\vec{y}}  \left| \frac{1}{\mathbb{P}\big( \vec{e}_W^{\mathrm{unif}}\vec{H}_W^{\top} = \vec{y} \big)} \;\frac{1}{2^{n/3-k_W}}-1 \right| \cdot \mathbb{P}(\vec{e}_W^{\mathrm{unif}}\vec{H}_W^{\top}=\vec{y}) \\
			 &=\Delta\big(\vec{s}_W, \vec{e}_W^{\mathrm{unif}}\vec{H}_W^{\top}\big)
		\end{align*}
		Now by averaging over $\vec{H}_W$ and using Lemma \ref{lemma:binaryLHL}, we easily deduce (using the notation of Lemma~\ref{lem:reference-uniformity}) that,
		\begin{equation*}
			\mathbb{E}_{\vec{H}_W}\left( \Delta\big(\vec{e}_W,\vec{e}_W^{\mathrm{unif}}\big)\right) \leq \frac{1}{2} \sqrt{2^{n/3-k_W} C_W} \ . 
		\end{equation*}
	The same computations identify the statistical distances $\Delta\bigl(\vec{e}_V^{\mathrm{unif}}\vec{H}_V^{\top},\vec{s}_V
	\mid\vec{e}_W^{\mathrm{unif}}=\vec{x}_W\bigr)$ and $\Delta(\vec e_U^{\mathrm{unif}}\vec H_U^{\top},\vec{s}_U
	\mid\vec e_V^{\mathrm{unif}}=\vec x_V,
	\vec e_W^{\mathrm{unif}}=\vec x_W)$ from which we conclude the proof by using Lemma~\ref{lemma:binaryLHL} and plugging this bound into Equation~\eqref{eq:upStatDist}. 
	\end{proof}
}

	\subsection{Instantiation via Prange Decoder}\label{subsec:instPrange}

	We now instantiate the three decoders $\mathsf{DecodeW}$, $\mathsf{DecodeV}$ and $\mathsf{DecodeU}$ of Algorithm~\ref{algo:inv2} by using some variations of
	{\sc PrangeStep} (see Algorithm~\ref{algo:Prangesdd}). We will roughly follow the strategy described in
	Subsection~\ref{subsec:invertTrap}, but we randomise the choices made on the
	free coordinates in order to control the output statistics. We also reserve at least $d$ free coordinates which
	will be filled uniformly at random.

	We first describe our proposed three decoders and
	compute the functions $p_W,p_V,p_U$ used in Proposition~\ref{propo:rejetUnif}.
	We then bound the statistical distance introduced by this instantiation as the latter are not perfect decoders.
	Throughout this subsection, let\iftoggle{llncs}{ $k_U'\eqdef k_U-d$, $k_V'\eqdef k_V-d$, $k_W'\eqdef k_W-d$.}{
	\begin{equation}
	k_U'\eqdef k_U-d,\quad k_V'\eqdef k_V-d,\quad k_W'\eqdef k_W-d \ .
	\end{equation}
}
	
	\medskip 
	{\bf\noindent Choosing the free coordinates.} Consider an input matrix $\vec H\in\mathbb F_2^{(n-k)\times n}$ with rank $n-k$ and fixed $\vec{s}\in \mathbb{F}_{2}^{n-k}$.
	Instead of setting all the $k$ free coordinates as in {\sc PrangeStep}, we will consider a variation where we will only pick a set $\mathcal{F}$ of size at most $k-d$. This is possible whenever \iftoggle{llncs}{$\mathsf{rank}(\vec H_{\overline{\mathcal{F}}})=n-k$ where $\overline{\mathcal{F}}=[1,n]\setminus\mathcal{F}$.}{
	\begin{equation}
	\mathsf{rank}(\vec H_{\overline{\mathcal{F}}})=n-k \;\; \mbox{where} \; \overline{\mathcal{F}}=[1,n]\setminus\mathcal{F} \ . 
	\end{equation}}
	Indeed, a Gaussian elimination then gives a set $\mathcal I$ of size $k$ containing $\mathcal{F}$ such that $\vec H_{\overline{\mathcal{I}}}$ is non-singular. Without loss of generality, $\mathcal F$ consists of the first $|\mathcal F|$ indices of $\mathcal I$. The $k-|\mathcal{F}| \geq d$ coordinates in $\mathcal I\setminus\mathcal{F}$ are filled uniformly at random as $\vec{u}$ before calling {\sc PrangeStep}. There is no need to choose the extension $\mathcal{I}$ uniformly: for each fixed $\vec {s}$ and chosen~$\vec{x'} \in \mathbb{F}_2^{|\mathcal{F}|}$, this procedure samples uniformly among the $2^{k-|\mathcal{F}|}$ possible completions $\vec{u}$. The chosen part of the solution given by {\sc PrangeStep} is therefore $\vec{x} \eqdef (\vec{x'}, \vec{u})$. Equivalently, the returned vector $\vec{e}$ satisfies $\vec{e}_{\mathcal{F}} = \vec{x'}$ and $\vec{e}_{\mathcal{I} \setminus \mathcal{F}} = \vec{u}$. 
	The three decoders use the common procedure of Algorithm~\ref{algo:prange-components}, with a probability distribution~$\mathcal{D}$ on\iftoggle{llncs}{$\Omega_{n,k,d}\eqdef
		\big\{(\mathcal F,\vec{x'}):\mathcal F\subseteq[1,n],\; 
		|\mathcal{F}|\le k-d,\; \vec{x'}\in\mathbb F_2^{\mathcal F}\big\}$. }{
		\begin{equation}
		\Omega_{n,k,d}\eqdef
		\left\{(\mathcal F,\vec{x'}):\mathcal F\subseteq[1,n],\; 
		|\mathcal{F}|\le k-d,\; \vec{x'}\in\mathbb F_2^{|\mathcal F|}\right\}\ .
		\end{equation}}
The constructions below specify $\mathcal D$ for each decoder.

	\begin{algorithm}[htbp]
		\caption{Common Prange procedure for $\mathsf{DecodeW}$,
			$\mathsf{DecodeV}$ and $\mathsf{DecodeU}$.}
		\label{algo:prange-components}
		\begin{algorithmic}[1]
			\Require$\vec H\in\mathbb F_2^{(n-k)\times n}$,
				$\vec s\in\mathbb F_2^{n-k}$
			\Param $0\le d\le k$ and a probability distribution $\mathcal D$ on $\Omega_{n,k,d}$
			\Ensure $\vec e\in\mathbb F_2^{n}$ such that $\vec e\vec H^{\top}=\vec s$
			\Repeat
			\State $(\mathcal F,\vec{x'})\Unif\mathcal D$
			\Until{$\operatorname{rank}(\vec H_{\overline{\mathcal F}})=n-k$}\label{inst:rankTest}
			\State Complete $\mathcal F$ into a set $\mathcal I$ of size $k$ such that $\vec H_{\overline{\mathcal I}}$ is non-singular \Comment{Order $\mathcal I$ with $\mathcal F$ first.}
			\State $\vec{x} \gets (\vec{x'}, \vec{u})$ where $\vec{u}$ is chosen uniformly at random in $\mathbb{F}_2^{k - |\mathcal{F}|}$
			\State $\vec e\gets\Call{PrangeStep}{\vec H,\vec s,\mathcal I, \vec x}$
			\State \Return $\vec e$
		\end{algorithmic}
	\end{algorithm}

	\iftoggle{llncs}{}{
	\medskip}
	The following constructions define $\mathcal{D}$ in Algorithm~\ref{algo:prange-components}. It uses as internal distributions some~$\mathcal{D}_W$, $\mathcal{D}_V^{\omega}$ and $\mathcal{D}_U^{\vec m}$ which are fixed when setting the parameters.
		\medskip 
		
	\begin{enumerate}\setlength{\itemsep}{8pt}
		\item {\bf $\mathsf{DecodeW}$: computing $\vec{e}_W$.} Apply Algorithm~\ref{algo:prange-components} with $(n,k,\vec H,\vec s)=(n/3,k_W,\vec H_W,\vec s_W)$. To this end, define $\mathcal D$ by first sampling $\ell\Unif\mathcal D_W$, where $\mathcal D_W$ is a distribution on $[0,k_W']$ to be optimized, then choosing $\mathcal F\subseteq [1,n/3]$ of size $k_W'$ and sampling $\vec{x'}$ uniformly from~$\mathcal S_\ell^{k_W'}$.
		
		\item {\bf $\mathsf{DecodeV}$: computing $\vec{e}_V$ knowing $\vec{e}_W$.} Apply Algorithm~\ref{algo:prange-components} with $(n,k,\vec H,\vec s)=(n/3,k_V,\vec H_V,\vec s_V)$. To this end, fix $\vec e_W=\vec x_W$, write $\omega=|\vec x_W|$, and set \iftoggle{llncs}{$j_\omega\eqdef\min(k_V',n/3-\omega)$. }{
\begin{equation}
	j_\omega\eqdef\min(k_V',n/3-\omega)\ .
\end{equation}}
Then, define $\mathcal D$ by first sampling $\ell\Unif\mathcal D_V^\omega$, where $\mathcal D_V^\omega$ is a distribution on $[0,j_\omega]$ to be optimized, then choosing $\mathcal F\subseteq\{i:\;\vec{x}_W(i)=0\}$ of size $j_\omega$ and sampling $\vec{x'}$ uniformly from~$\mathcal S_\ell^{j_\omega}$. Thus, the prescribed coordinates lie entirely outside the support of $\vec x_W$, while the remaining $k_V-j_\omega\geq d$ free coordinates are filled uniformly.

		\item {\bf $\mathsf{DecodeU}$: computing $\vec e_U$ knowing $(\vec e_V,\vec e_W)$.} Apply Algorithm~\ref{algo:prange-components} with $(n,k,\vec H,\vec s)=(n/3,k_U,\vec H_U,\vec s_U)$. To this end, fix $(\vec e_V,\vec e_W)=(\vec x_V,\vec x_W)$, and let $\mathcal M_{\alpha\beta}$ and $\vec m$ be as in Definition~\ref{def:statistics}. For each position $i$, let $c_i$ be the bit minimizing the local weight, as given in Proposition~\ref{prop:counts}. Then, define $\mathcal D$ by first sampling $(\vec j,\boldsymbol\ell)\Unif\mathcal D_U^{\vec m}$, where $\mathcal D_U^{\vec m}$ is a distribution to be optimized over pairs of integer vectors $\vec j=(j_{\alpha\beta})_{\alpha,\beta\in\{0,1\}}$ and $\boldsymbol\ell=(\ell_{\alpha\beta})_{\alpha,\beta\in\{0,1\}}$ satisfying
\begin{equation}
    0\leq\ell_{\alpha\beta}\leq j_{\alpha\beta}\leq m_{\alpha\beta}
    \qquad\text{and}\qquad
    \sum_{(\alpha,\beta)\in\{0,1\}^2}j_{\alpha\beta}=k_U'\ .
\end{equation}
Then, for each $(\alpha,\beta)\in\{0,1\}^2$, choose $\mathcal F_{\alpha\beta}\subseteq\mathcal M_{\alpha\beta}$ of size $j_{\alpha\beta}$ and set $\mathcal F=\bigcup_{\alpha,\beta}\mathcal F_{\alpha\beta}$. Construct the prescribed vector $\vec{x'}$, indexed by $\mathcal F$, by setting $\vec{x}'(i)=1+c_i$ at $\ell_{\alpha\beta}$ uniformly chosen positions of each $\mathcal F_{\alpha\beta}$ and $\vec{x}'(i)=c_i$ at the others. The remaining $k_U-k_U'$ free coordinates are filled uniformly. In particular, choosing $\boldsymbol\ell=\vec 0$ minimizes the local weight at every prescribed position.
\end{enumerate}

	\medskip 
	{\bf\noindent Output Distribution of the Decoders.} Our instantiated decoders are not perfect\iftoggle{llncs}{.}{, {\it i.e.,} they don't satisfy the requirements of Definition~\ref{def:uniformity}, because positions where coordinates are chosen are not picked uniformly at random.} However\iftoggle{llncs}{}{, as we will show,} decoders satisfy the equalities of Definition~\ref{def:uniformity} up to a small statistical distance (thanks to the parameter $d$) on average over the underlying codes $U$, $V$ and $W$. To understand why let us first compute their distributions obtained by setting the entries as above and completing the other entries uniformly but without imposing the rank test of Instruction~\ref{inst:rankTest} in Algorithm~\ref{algo:prange-components}. \iftoggle{llncs}{}{We then compare these distributions with the actual decoder outputs.}

\begin{restatable}{proposition}{propoPrangeProb}\label{prop:prange-probabilities}
	Consider the three sampling rules specified above. At each step, set the positions and their entries according to the corresponding rule, then complete all the remaining entries uniformly at random, without the rank test or the constraint $\vec{e}\vec{H}^{\top}=\vec{s}$ of Algorithm~\ref{algo:prange-components}. Denote the resulting vectors by $\vec{e}_W^{(0)}$, $\vec{e}_V^{(0)}$ and
	$\vec{e}_U^{(0)}$.
	Their distribution depend only on the statistics of Definition~\ref{def:statistics}. Let,
	\begin{equation}
		p_W\big(\omega(\vec x_W)\big) \eqdef\mathbb P\big(\vec e_W^{(0)}=\vec x_W\big),\;\; p_V\big(\nu(\vec x_V,\vec x_W),\omega(\vec x_W)\big) \eqdef\mathbb P\big(\vec e_V^{(0)}=\vec x_V\big),
	\end{equation}
	\begin{equation}
		p_U\big(\vec h(\vec x_U;\vec x_V,\vec x_W),\vec m(\vec x_V,\vec x_W)\big) \eqdef\mathbb P\big(\vec e_U^{(0)}=\vec x_U\big)\ .
	\end{equation}
	The last two probabilities are computed for every fixed $\vec x_W$ and $(\vec x_V,\vec x_W)$\iftoggle{llncs}{.}{, respectively.
We have,}
\begin{align}
	p_W(\omega)
	&=\frac{1}{2^{n/3-k_W'}\binom{n/3}{\omega}}
	\sum_{\ell=0}^{k_W'}\mathcal{D}_W(\ell)
	\binom{n/3-k_W'}{\omega-\ell},
\end{align}
\begin{multline} 
	p_V(\nu,\omega)
	=\frac{1}{2^{n/3-\min(k_V',n/3-\omega)}\binom{n/3-\omega}{\nu - \omega}}\\
	\sum_{\ell=0}^{\min(k_V',n/3-\omega)}\mathcal{D}_V^{\omega}(\ell)
	\binom{n/3-\omega-\min(k_V',n/3-\omega)}{\nu - \omega -\ell},
\end{multline} 
\begin{align}
	p_U(\vec{h},\vec{m})
	&=\frac{1}{2^{n/3-k_U'}N_{\vec{m}}(\vec h)}
	\sum_{\vec j,\boldsymbol\ell}
	\mathcal D_U^{\vec{m}}(\vec j,\boldsymbol{\ell})
	\prod_{\alpha,\beta\in\{0,1\}}
	\binom{m_{\alpha\beta}-j_{\alpha\beta}}
	{h_{\alpha\beta}-\ell_{\alpha\beta}} \ .
\end{align}
\end{restatable}
\iftoggle{llncs}{}{
\begin{proof}
At the first step, the weight is the sum of $\ell$ and the number of ones among $n/3-k_W'$ uniform bits. Thus, its probability of being $\omega$ is $2^{-(n/3-k_W')}\binom{n/3-k_W'}{\omega-\ell}$ for a fixed $\ell$. All vectors of weight $\omega$ have the same probability. Dividing by $\binom{n/3}{\omega}$ and averaging over $\ell$ proves the first equality.

At the second step, the number of positions in $\mathcal{M}_{10}$ (that is where $\vec x_V$ and $\vec x_W$ are $1$ and~$0$, respectively) is $\ell$ plus the number of ones among the remaining $n/3-\omega-\min(k_V',n/3-\omega)$ independent uniform bits at positions where $\vec x_W$ is zero. Hence, its probability of being $\nu - \omega$ is~$2^{-(n/3-\omega-\min(k_V',n/3-\omega))}\binom{n/3-\omega-\min(k_V',n/3-\omega)}{\nu - \omega-\ell}$ for a fixed $\ell$. The other $\omega$ bits where $\vec x_W$ is nonzero are uniform and independent of these choices. Therefore, there are $2^\omega\binom{n/3-\omega}{\nu - \omega}$ vectors with this statistic, all with the same probability. This proves the second equality.

Finally, fix $(\vec j,\boldsymbol\ell)$. For each $(\alpha,\beta)\in\{0,1\}^2$, the statistic $h_{\alpha\beta}$ is $\ell_{\alpha\beta}$ plus the number of larger local weights among~$m_{\alpha\beta}-j_{\alpha\beta}$ uniform bits. Thus,
\[
    h_{\alpha\beta}-\ell_{\alpha\beta}
    \Unif \operatorname{Binomial}(m_{\alpha\beta}-j_{\alpha\beta},1/2)\ .
\]
These four random variables are independent, so
\[
    \mathbb P(\vec h\mid\vec j,\boldsymbol\ell)
    =
    2^{-(n/3-k_U')}
    \prod_{(\alpha,\beta)\in\{0,1\}^2}
    \binom{m_{\alpha\beta}-j_{\alpha\beta}}
          {h_{\alpha\beta}-\ell_{\alpha\beta}},
\]
where we used $\sum_{\alpha,\beta}(m_{\alpha\beta}-j_{\alpha\beta}) =n/3-k_U'$. By symmetry under permutations within each~$\mathcal{M}_{\alpha\beta}$, the $N_{\vec m}(\vec h)$ vectors counted in Equation~\eqref{eq:Nm} are equally likely. Dividing by $N_{\vec m}(\vec h)$ and averaging over $(\vec j,\boldsymbol\ell)$ concludes the proof.
\end{proof}
}

The following lemma shows that the instantiated decoders satisfy up to a small statistical distance, on average over the codes $U$, $V$ and $W$, the equalities of Definition~\ref{def:uniformity} with the functions of Proposition~\ref{prop:prange-probabilities}. 

\begin{restatable}{lemma}{lemmaPrangeAlmostPerfect}\label{lem:prange-almost-perfect}
	Let $\vec{e}_W^{(0)},\vec e_V^{(0)},\vec e_U^{(0)}$ be the random variables defined in
	Proposition~\ref{prop:prange-probabilities}. For uniform matrices $\vec{H}_W$, $\vec{H}_V$, $\vec{H}_U$ and uniform vectors $\vec{s}_W$, $\vec{s}_V$, $\vec{s}_U$, we have
	\begin{equation}
		\mathbb{E}_{\vec H_W}\Big(\Delta\big(
		\mathsf{DecodeW}(\vec H_W,\vec{s}_W),\vec e_W^{(0)}\big)\Big)
		\le 2^{-d},
	\end{equation}
	\begin{equation}
		\mathbb E_{\vec H_V}\Big(\Delta\big(
		\mathsf{DecodeV}(\vec H_V,\vec{s}_V,\vec x_W),\vec e_V^{(0)}\big)\Big)
		\le 2^{-d},
	\end{equation}
	\begin{equation}
		\mathbb E_{\vec H_U}\Big(\Delta\big(
		\mathsf{DecodeU}(\vec H_U,\vec{s}_U,\vec x_V,\vec x_W,
		\varphi_{\vec a,\vec b,\vec P}),\vec e_U^{(0)}\big)\Big)
		\le 2^{-d}\ .
	\end{equation}
	The last two bounds hold for every fixed $\vec x_W$ and
	$(\vec x_V,\vec x_W)$, respectively.
	Furthermore, for each decoding step $X\in\{W,V,U\}$, let, \iftoggle{llncs}{$g_X(\vec x_X)\eqdef\mathbb P\big(\mathsf{rank}(\vec H_{X,\overline{\mathcal F}})=n/3-k_X\mid\vec{e}_X^{(0)}=\vec{x}_X\big)$,}{
	\begin{equation}
	g_X(\vec x_X)\eqdef\mathbb P\big(\mathsf{rank}(\vec H_{X,\overline{\mathcal F}})=n/3-k_X\mid\vec{e}_X^{(0)}=\vec{x}_X\big),
	\end{equation}}
	where the probability is computed over the sampling rules as in
	Proposition~\ref{prop:prange-probabilities}, with $\vec H_X$ being fixed.
	Then,
	\begin{equation}\label{eq:prange-pointwise-rank}
		\mathbb E_{\vec H_X}\big(1-g_X(\vec x_X)\big)<2^{-d}\ .
	\end{equation}
\end{restatable}
\iftoggle{llncs}{}{
\begin{proof}
For any fixed set $\mathcal F$ selected at step $X\in\{W,V,U\}$, we have $|\mathcal{F}|\le k_X-d$.
	If $\vec H_{X,\overline{\mathcal F}}$ is not of full rank, then there exists some non-zero
	$\vec{y}\in\mathbb{F}_2^{n/3-k_X}$ satisfying
	$\vec{y}\vec{H}_{X,\overline{\mathcal{F}}}=\vec{0}$.
	For a uniform~$\vec{H}_X$, this event, for each fixed non-zero $\vec{y}$, has probability
	$2^{-(n/3-|\mathcal{F}|)}$. A union bound therefore gives
	\begin{equation}\label{eq:prange-rank-bound}
		\mathbb P_{\vec H_X}\big(\mathsf{rank}(\vec H_{X,\overline{\mathcal F}})<n/3-k_X\mid\mathcal{F}\big)
		\le(2^{n/3-k_X}-1)\; 2^{-(n/3-|\mathcal{F}|)}<2^{-d}\ .
	\end{equation}
	Notice now that $\vec e_X^{(0)}$ is independent of
	$\vec H_X$. Thus, even after fixing $\vec e_X^{(0)}=\vec x_X$, the
	conditional choice of $\mathcal{F}$ is independent of $\vec H_{X}$.
	Averaging~\eqref{eq:prange-rank-bound} over this conditional choice proves
	\eqref{eq:prange-pointwise-rank}.
	
	Now fix $\vec H_X$ and a $(\mathcal{F},\vec{x'})$ passing the rank test of Instruction \ref{inst:rankTest} in Algorithm \ref{algo:prange-components}. A uniform input $\vec s_X$, together with the $k_X-|\mathcal F|$ uniform free entries, gives uniform entries on all of $\overline{\mathcal{F}}$: the map between these two descriptions is bijective
, since $\vec H_{X,\overline{\mathcal J}}$ is invertible. Algorithm~\ref{algo:prange-components} redraws $(\mathcal{F},\vec{x'})$ until $\vec H_{X,\overline{\mathcal F}}$ has rank $n/3-k_X$. Its output for a uniform input therefore has the distribution of~$\vec e_X^{(0)}$ conditioned on that rank condition. Let $b_X\eqdef\mathbb{P}\big( \mathsf{rank}(\vec H_{X,\overline{\mathcal F}})<n/3-k_X \big)$ for a fixed $\vec H_X$. We deduce that, 
	$$
	\Delta\big(\vec e_X^{(0)}\mid \mathsf{rank}(\vec H_{X,\overline{\mathcal F}})=n/3-k_X,\;\vec e_X^{(0)}\big)\le b_X \ .
	$$
	Finally, using Equation~\eqref{eq:prange-rank-bound} gives $\mathbb E_{\vec H_X}(b_X)<2^{-d}$ which concludes the proof.
\end{proof}
}

We can now instantiate Theorem~\ref{thm:perfect-no-leakage} with our particular instantiations of non-perfect decoders $\mathsf{DecodeW}$, $\mathsf{DecodeV}$ and $\mathsf{DecodeU}$ via {\sc PrangeStep}.

\begin{restatable}{theorem}{theoremPrangeNoLeakge}\label{thm:prange-no-leakage}
	Instantiate $\mathsf{DecodeW}$, $\mathsf{DecodeV}$ and $\mathsf{DecodeU}$
	in Algorithm~\ref{algo:inv2} by Algorithm~\ref{algo:prange-components}
	and the three sampling rules described in this subsection. Furthermore, we choose the rejection probabilities~$\vec{r}_W$,~$\vec{r}_V$,~$\vec{r}_U$ according to Proposition~\ref{propo:rejetUnif}, with
	$p_W,p_V,p_U$ given by Proposition~\ref{prop:prange-probabilities}. Let $\vec{e}^{\mathrm{out}}$ be the associated output of Algorithm~\ref{algo:inv2} when $\vec{s} \Unif \mathbb{F}_{2}^{n-k}$. 
We have,
	\begin{multline}
	\mathbb{E}_{\vec{H}_U,\vec{H}_V,\vec{H}_W}
	\Big(\Delta(\vec{e}^{\mathrm{out}},\vec{e}^{\mathrm{unif}})\Big)
	\\ \leq \frac{1}{2} \left( 
	\sqrt{2^{n/3-k_W}C_W}
	+\sqrt{2^{n/3-k_V}C_V}
	+\sqrt{2^{n/3-k_U}C_U}\right) +3\cdot2^{-d} \ .
	\end{multline}
	Here $\vec{H}_U \Unif \mathbb{F}_{2}^{(n/3-k_U)\times n/3}$, $\vec{H}_V \Unif \mathbb{F}_{2}^{(n/3 - k_V) \times n/3}$, $\vec{H}_W \Unif \mathbb{F}_{2}^{(n/3-k_W) \times n/3}$, $\vec{e}^{\mathrm{unif}} \Unif \mathcal{S}_{t}^{n}$ and $C_W$, $C_V$, $C_U$ are defined as in Theorem~\ref{thm:perfect-no-leakage}. 
\end{restatable}
\iftoggle{llncs}{}{
\begin{proof}
	We first consider the $W$ decoding step, {\it i.e.,} $X=W$. Let us fix $\vec{H}_W$. Use $g_W$ from Lemma~\ref{lem:prange-almost-perfect} and $b_W \eqdef \sum_{\vec{x}_W} p_{W}(\omega(\vec{x}_W))\left( 1-g_W(\vec{x}_W) \right)$. First, suppose $b_W<1$. For a uniform input, the probability of returning $\vec x_W$ in one call is $p_W(\omega(\vec{x}_W))g_W(\vec{x}_W)/(1-b_W)$. Consequently, for a fixed input $\vec{s}_W$ and $\vec x_W\vec H_W^{\top}=\vec{s}_W$, we obtain
	$$
	\mathbb{P}\big(\mathsf{DecodeW}(\vec H_W,\vec{s}_W)=\vec{x}_W\big)
	=\frac{2^{n/3-k_W}p_W(\omega(\vec{x}_W))g_W(\vec{x}_W)}{1-b_W} \ .
	$$
	Multiplying by $r_W(\omega(\vec{x}_W))$ from Proposition~\ref{propo:rejetUnif} removes $p_W(\omega(\vec{x}_W))$. After repeating the rejection with this same input, the output probabilities are therefore proportional to
	$$
	\mathbb{P}\big(\vec{e}_W^{\mathrm{unif}}=\vec{x}_W\big)g_W(\vec{x}_W)
	\quad\hbox{on}\quad
	\left\{\vec{x}_W:\vec{x}_W\vec{H}_W^{\top}=\vec{s}_W\right\} \ .
	$$
	Therefore, for a fixed $\vec{y}$, the output is obtained from $\vec{e}_W^{\mathrm{unif}}$ conditioned on $\vec{e}_W^{\mathrm{unif}}\vec{H}_W^{\top}=\vec y$ by additionally accepting $\vec{x}_W$ with probability $g_W(\vec{x}_W)$. Hence
	\begin{equation}\label{eq:statCdtse}
		\Delta\Big(\vec{e}_W\mid\vec{s}_W=\vec{y},\;
		\vec{e}_W^{\mathrm{unif}}\mid
		\vec{e}_W^{\mathrm{unif}}\vec{H}_W^{\top}=\vec{y}\Big)
		\leq
		\frac{\mathop{\sum}\limits_{\vec x_W:\vec x_W\vec H_W^{\top}=\vec y}
			\mathbb P(\vec{e}_W^{\mathrm{unif}}=\vec x_W)
			\big(1-g_W(\vec{x}_W)\big)}
		{\mathbb{P}(\vec{e}_W^{\mathrm{unif}}\vec{H}_W^{\top}=\vec{y})} \ .
	\end{equation} 
	If no acceptance is possible, the right-hand side is one and the inequality still holds. The case~$b_W=1$ is covered in the same way, since then $g_W=0$ on the support of $p_W$.
	
	We now take $\vec{s}_W$ uniform. Apply Equation~\eqref{eq:statDistPairs} to $(\vec{e}_W, \vec{s}_W)$ and $(\vec{e}_W^{\mathrm{unif}}, \vec{e}_W^{\mathrm{unif}}\vec{H}_W^{\top})$, using the input vectors as the first components in that equation. The conditional distances are weighted by~$\mathbb{P}(\vec{e}_W^{\mathrm{unif}}\vec{H}_W^{\top}=\vec y)$, which cancels the denominator above. We obtain
	\begin{align*}
		\Delta\Big(\vec e_W,\vec e_W^{\mathrm{unif}}\Big) &= \Delta\Big((\vec{e}_W, \vec{s}_W), (\vec{e}_W^{\mathrm{unif}}, \vec{e}_W^{\mathrm{unif}}\vec{H}_W^{\top}) \Big)\\ 
		&\leq\Delta(\vec{s}_W,\vec e_W^{\mathrm{unif}}\vec H_W^{\top})+ \sum_{\vec{y}} \mathbb{P}(\vec{e}_W^{\mathrm{unif}}\vec{H}_W^{\top} = \vec{y})\Delta\Big(\vec{e}_W\mid\vec{s}_W=\vec{y},\;
		\vec{e}_W^{\mathrm{unif}}\mid
		\vec{e}_W^{\mathrm{unif}}\vec{H}_W^{\top}=\vec{y}\Big)  \\
		&\leq\Delta(\vec{s}_W,\vec e_W^{\mathrm{unif}}\vec H_W^{\top})+\sum_{\vec x_W}
		\mathbb{P}(\vec{e}_W^{\mathrm{unif}}=\vec x_W)
		\big(1-g_W(\vec{x}_W)\big)
	\end{align*}
	where in the first inequality we used Equation~\eqref{eq:statDistPairs} and in the second one Equation~\eqref{eq:statCdtse}.

	Averaging over $\vec{H}_W$, using Equation~\eqref{eq:prange-pointwise-rank} and Lemma~\ref{lemma:binaryLHL} (as we did in the proof of Theorem~\ref{thm:perfect-no-leakage}) therefore gives
	\begin{equation}\label{eq:Wcase} 
	\mathbb{E}_{\vec H_W}\big(\Delta(\vec{e}_W,\vec{e}_W^{\mathrm{unif}})\big)
	\le\frac{1}{2}\sqrt{2^{n/3-k_W}C_W}+2^{-d}\ .
	\end{equation} 
	
	For the case of the $V$ decoding step, {\it i.e.,}\, $X=V$, fix $\vec{x}_W$ and use the target $\vec{e}_V^{\mathrm{unif}}$ conditioned on $\vec{e}_W^{\mathrm{unif}}=\vec x_W$. Considering now the $U$ decoding step, we fix this time $(\vec{x}_V,\vec{x}_W)$ and use the corresponding conditional target.
	The same computation and reasoning apply with $p_V,\vec r_V,g_V$ and $p_U,\vec r_U,g_U$, respectively. The next input remains uniform after fixing the preceding outputs, and each conditional target is independent of the matrix being averaged. To conclude the proof we proceed as in the proof of Theorem~\ref{thm:perfect-no-leakage} by considering Equation~\eqref{eq:upStatDist} and substituting the three bounds for $W$, $V$ and $U$ (as in Equation~\eqref{eq:Wcase}). 
\end{proof}
}

	So far we have instantiated our decoders $\mathsf{DecodeW}$, $\mathsf{DecodeV}$ and $\mathsf{DecodeU}$. However, there is a degree of freedom that we did not instantiate: the choice of their internal distributions $\mathcal D_W$,~$\mathcal D_V^\omega$ and $\mathcal D_U^{\vec m}$. The next subsection explains how to choose them so as to reduce the number of rejections (in order to ensure an efficient signing algorithm).

\subsection{Instantiating the internal distributions}\label{subsec:prange-distributions}

\iftoggle{llncs}{After choosing the different parameters, we will use the following estimation of the average total number of rejects (the different quantities are defined in Proposition~\ref{propo:rejetUnif} and Lemma~\ref{lem:reference-uniformity}),
	\begin{equation}
	M_W-1 +\sum_{\omega}
	\frac{\binom{n/3}{\omega}A_\omega}{\binom nt}
	\bigl(M_V(\omega)-1\bigr) +\sum_{\vec m}
	\frac{\binom{n/3}{m_{00},m_{10},m_{01},m_{11}}
		B_{n/3-m_{00}}}{\binom nt}
	\bigl(M_U(\vec m)-1\bigr) \ .
	\end{equation}}{
\begin{lemma}[Number of rejects for fixed inputs]\label{lem:prange-rejections}
	Suppose that $\mathsf{DecodeW}$, $\mathsf{DecodeV}$ and $\mathsf{DecodeU}$ are perfect. Fix $\vec{H}_W$, $\vec{H}_V$, $\vec{H}_U$ and $\vec{s}_U,\vec{s}_V,\vec{s}_W$ of Algorithm~\ref{algo:inv2}. 
	
Let $N_W,N_V,N_U$ denote the numbers of rejections in its three loops. For previously accepted vectors $\vec x_W,\vec x_V$, write $\omega=\omega(\vec x_W)$ and $\vec m=\vec m(\vec x_V,\vec x_W)$. Then
	\begin{align}
		\mathbb{E}\big(N_W\big)
		&=\frac{M_W}{2^{n/3-k_W}
			\mathbb{P}\big(\vec{e}_W^{\textup{unif}}\vec{H}_W^\top=\vec{s}_W\big)}-1,\\[5pt]
		\mathbb{E}\big(N_V\mid\vec{e}_W=\vec x_W\big)
		&=\frac{M_V(\omega)}{2^{n/3-k_V}
			\mathbb{P}\big(\vec{e}_V^{\textup{unif}}\vec{H}_V^\top=\vec{s}_V
			\mid\vec{e}_W^{\textup{unif}}=\vec{x}_W\big)}-1,\\[5pt]
		\mathbb{E}\big(N_U\mid\vec{e}_V=\vec{x}_V,\vec{e}_W=\vec{x}_W\big)
		&=\frac{M_U(\vec m)}{2^{n/3-k_U}
			\mathbb{P}\big(\vec{e}_U^{\textup{unif}}\vec{H}_U^\top=\vec{s}_U
			\mid\vec{e}_V^{\textup{unif}}=\vec{x}_V,\vec{e}_W^{\textup{unif}}=\vec{x}_W\big)}-1, 
	\end{align}
	where these quantities are given in Proposition~\ref{propo:rejetUnif}.

	The expectations computed are over the internal randomness of the algorithm.
	The total average number of rejections is obtained by adding these
	conditional expectations and averaging over the uniform inputs and the
	previously accepted vectors.
\end{lemma}

\begin{proof}
	The proof of Proposition~\ref{propo:rejetUnif} gives the acceptance probability in each loop. For fixed inputs and previously accepted vectors, the decoding steps are independent. With acceptance probability~$a$, their expected number of rejects is $(1-a)/a$, which gives the three expressions. The last assertion follows by taking expectations which concludes the proof. 
\end{proof}

\begin{remark}After choosing the different parameters, we will use the following estimation
	of the average total number of rejects (the different quantities are defined in Proposition~\ref{propo:rejetUnif} and Lemma~\ref{lem:reference-uniformity}),
	\begin{equation}
	\label{eq:rejection}
		M_W-1 +\sum_{\omega}
		\frac{\binom{n/3}{\omega}A_\omega}{\binom nt}
		\bigl(M_V(\omega)-1\bigr) +\sum_{\vec m}
		\frac{\binom{n/3}{m_{00},m_{10},m_{01},m_{11}}
			B_{n/3-m_{00}}}{\binom nt}
		\bigl(M_U(\vec m)-1\bigr) \ .
	\end{equation}
	It  is obtained by replacing each probability in Lemma~\ref{lem:prange-rejections} by $2^{-(n/3-k_X)}$ for $X \in \{U,V,W\}$. This approximation is justified by the fact with our parameter choices $\Delta\left( \vec{e}_X^{\textup{unif}}\vec{H}_X^\top,\vec{s}_X\right)$ where~$\vec{s}_{X}$ is chosen uniformly will be negligible (by using the bound given by the leftover hash lemma from Lemma~\ref{lemma:binaryLHL}). Furthermore, it can be shown that as in Lemma~\ref{lem:prange-almost-perfect}, that with our instantiated decoders which are not perfect, the above average number of rejects is also accurate up to exponentially small factors.
\end{remark}
}

	The three sampling rules described in the previous subsection to instantiate $\mathsf{DecodeW}$, $\mathsf{DecodeV}$ and $\mathsf{DecodeU}$ depend on distributions $\mathcal D_W,\mathcal D_V^{\omega}$ and $\mathcal D_U^{\vec m}$ which are chosen by the designer. They allow us to adjust the statistics before rejection.
First, we propose to choose their means so that the distributions before rejection match those induced by $\vec e^{\mathrm{unif}}$ as closely as possible in mean. This choice is not required for correctness, but aims to improve acceptance rates by bringing $M_W$, $M_V(\omega)$, and $M_U(\vec m)$ closer to $1$.
\iftoggle{llncs}{}{
The average-matching conditions are given by
\begin{equation}\label{mean:W} 
	\mathbb{E}_{\mathcal D_W}(\ell)+\frac{n/3-k_W'}2 =\mathbb{E}\big(\omega(\vec e_W^{\mathrm{unif}})\big) \approx \tfrac{2t}{3} (1-t/n),
\end{equation} 
\begin{multline}\label{mean:V} 
	\mathbb{E}_{\mathcal{D}_V^\omega}(\ell)+\frac{n/3-\omega-\min(k_V',n/3-\omega)}{2} =\mathbb{E}\left( \nu(\vec{e}_V^{\mathrm{unif}},\vec{e}_W^{\mathrm{unif}}) - \omega(\vec{e}_W^{\mathrm{unif}}) \mid \omega(\vec{e}_W^{\mathrm{unif}})=\omega \right)\\ 
	\approx (n/3-\omega)\frac{t/n(1-t/n)}{(t/n)^2+(1-t/n)^2}
\end{multline} 
and, for $(\alpha,\beta)\in\{0,1\}^2$,
\begin{multline}\label{mean:U} 
	\mathbb{E}_{\mathcal{D}_U^{\vec m}}(\ell_{\alpha\beta})
				+\frac{m_{\alpha\beta}-\mathbb{E}_{\mathcal D_U^{\vec m}}(j_{\alpha\beta})}2 
	=\mathbb{E}\left(
		h_{\alpha\beta}(\vec e_U^{\mathrm{unif}}; \vec{e}_V^{\mathrm{unif}},\vec{e}_W^{\mathrm{unif}}) \mid \vec{m}(\vec{e}_V^{\mathrm{unif}},\vec{e}_W^{\mathrm{unif}})=\vec{m} \right)\\  
	\approx
	\begin{cases}
		\displaystyle m_{00}\cdot \frac{(t/n)^3}{(t/n)^3+(1-t/n)^3}
		& (\alpha,\beta)=(0,0) \\[6pt]
		m_{\alpha\beta}\cdot t/n &(\alpha,\beta)\ne(0,0)
	\end{cases} \ . 
\end{multline}

For instance, one may use discrete Laplace distributions for $\mathcal D_W$, $\mathcal D_V^\omega$, and $\mathcal D_U^{\vec m}$, restricted to their respective admissible domains and renormalized. Their parameters can be chosen to approximately match the target means described above, taking into account that truncation may shift the expectations away from the corresponding Laplace location parameters and alter the variances. For fixed code parameters, empirical tuning can then be used to bring $M_W$, $M_V(\omega)$, and $M_U(\vec m)$ as close to $1$ as possible.

\medskip

The constants $M_W$ and $M_V(\omega)$ can be computed directly with
reasonable computational complexity. However, to evaluate $M_U(\vec m)$ efficiently, we exploit the factorization of the unnormalized sampling weights.
More precisely, write
\begin{equation}
	\label{eq:D_U_Laplace}
    \mathcal D_U^{\vec m}(\vec j,\boldsymbol \ell)
    \eqdef 
    \frac{1}{Z_{\vec m}}
    \prod_{\alpha,\beta\in\{0,1\}}
    \mathcal{A}_{\alpha\beta}(j_{\alpha\beta})
    \mathcal{B}_{\alpha\beta}(\ell_{\alpha\beta})
\end{equation}
on the admissible domain $0\leq \ell_{\alpha\beta}\leq j_{\alpha\beta}\leq m_{\alpha\beta}$
and $\sum_{\alpha,\beta}j_{\alpha\beta}=k'_U$.
Here, $\mathcal{A}_{\alpha\beta}$ and $\mathcal{B}_{\alpha\beta}$ are the local sampling weights for the sampling rule considered here and in our case they are unnormalized Laplace distributions. The normalizing constant $Z_{\vec m}$ makes $D_U^{\vec m}(\vec j,\boldsymbol \ell)$ a probability distribution over its admissible domain.

For each block, define
\begin{equation}
    F_{\alpha\beta}(z;h)
   \eqdef
    \sum_{s=0}^{\min(m_{\alpha\beta},n/3-k'_U)}
    \mathcal{A}_{\alpha\beta}(m_{\alpha\beta}-s)
    \left(
        \sum_{\ell=0}^{m_{\alpha\beta}-s}
        \mathcal{B}_{\alpha\beta}(\ell)
        \binom{s}{h-\ell}
    \right)z^s,
\end{equation}
and
\begin{equation}
    G_{\alpha\beta}(z)
    \eqdef
    \sum_{s=0}^{\min(m_{\alpha\beta},n/3-k'_U)}
    \mathcal{A}_{\alpha\beta}(m_{\alpha\beta}-s)
    \left(
        \sum_{\ell=0}^{m_{\alpha\beta}-s}
        \mathcal{B}_{\alpha\beta}(\ell)
    \right)z^s \ .
\end{equation}
The change of variables $s_{\alpha\beta}=m_{\alpha\beta}-j_{\alpha\beta}$
transforms the constraint on $\vec j$ into
$\sum_{\alpha,\beta}s_{\alpha\beta}=n/3-k'_U$. Consequently,
\begin{equation}
    C_{\vec m}(\vec h)
    \eqdef \mathrm{Coeff} \left( \prod_{\alpha,\beta}
    F_{\alpha\beta}(z;h_{\alpha\beta}) \, ; \, z^{n/3-k'_U} \right),
    \qquad
    Z_{\vec m}
    = \mathrm{Coeff} \left( \prod_{\alpha,\beta}G_{\alpha\beta}(z) \, ; \, z^{n/3-k'_U} \right), 
\end{equation}
and
\begin{equation}
    p_U(\vec m,\vec h)
    =
    \frac{C_{\vec m}(\vec h)}
         {2^{n/3-k'_U} Z_{\vec m}N_{\vec m}(\vec h)}\ .
\end{equation}
It follows that, for $\nu = n/3 - m_{00}$,
\begin{equation}
    M_U(\vec m)
    =
    \frac{2^{n/3-k'_U} Z_{\vec m}}{B_\nu}
    \max_{\substack{\vec{h}: N_{\vec{m}}(\vec{h})>0\\
				\nu + 3h_{00}+h_{10}+h_{01}+h_{11}=t}}
    \frac{N_{\vec m}(\vec h)}{C_{\vec m}(\vec h)}\ .
\end{equation}
Each coefficient extraction can be performed using two pairwise convolutions followed by a scalar product, avoiding explicit enumeration of the quadruplets $(j_{\alpha\beta})_{\alpha, \beta \in \{0,1\}}$ and $(\ell_{\alpha\beta})_{\alpha, \beta \in \{0,1\}}$. Moreover, $Z_{\vec m}$ is independent of $\vec h$ and is computed only once. Then, the maximization over $\vec h$ is performed separately using a descent method.
}
		
\subsection{Choosing the parameters}\label{subsec:prange-parameters}
	For a given length $n$, we first fix the code rate $R$, {\it i.e.,}  $k \eqdef  \lfloor Rn \rfloor $, and the parameter~$d$. These choices will determine the relative signature weight~$\tau$,{\it i.e.,}~$t \eqdef \lfloor \tau n \rfloor$, and the underlying dimensions $k_U,k_V,k_W$ of the random codes $U$, $V$ and $W$ under consideration (where $k = k_U + k_V + k_W$).
			
	We also introduce three new parameters $1/2 < \alpha_W, \alpha_V, \alpha_U \le 1$. The parameters $\alpha_W$ and~$\alpha_V$ specify the average proportion of prescribed positions filled with zero at the $W$ and $V$ steps. At the $U$ step, $\alpha_U$ specifies the average proportion filled with the minimizing bit~$c_i$. However, at the last two steps these proportions must be allowed to depend on the preceding outputs. We therefore write $\alpha_V(\omega)$ and $\alpha_U(\vec m,\omega)$. Since $\ell$ and~$\sum_{\alpha,\beta}\ell_{\alpha\beta}$ enumerate the chosen positions at these respective decoding steps, we set\iftoggle{llncs}{ $\mathbb{E}_{\mathcal D_W}(\ell)=(1-\alpha_W)k_W'$, $\mathbb E_{\mathcal D_V^\omega}(\ell)
				=(1-\alpha_V(\omega))\min(k_V',n/3-\omega)$ and $\mathbb{E}_{\mathcal D_U^{\vec m}}\left(\sum_{\alpha,\beta}\ell_{\alpha\beta}\right)
				=(1-\alpha_U(\vec{m},\omega))k_U'$}{
			\begin{equation}
			\mathbb{E}_{\mathcal D_W}(\ell)=(1-\alpha_W)k_W',\;\;
			\mathbb E_{\mathcal D_V^\omega}(\ell)
			=(1-\alpha_V(\omega))\min(k_V',n/3-\omega),
			\end{equation}
			\begin{equation}
			\mathbb{E}_{\mathcal D_U^{\vec m}}\left(\sum_{\alpha,\beta}\ell_{\alpha\beta}\right)
			=(1-\alpha_U(\vec{m},\omega))k_U'\ . 
			\end{equation}	
		}

			\iftoggle{llncs}{}{
	Notice that $\omega=m_{01}+m_{11}$ is already determined by $\vec{m}$. Recall now to ensure a small number of reject we need to verify Equations~\eqref{mean:U},~\eqref{mean:V} and~\eqref{mean:W}.} \iftoggle{llncs}{Then,}{Therefore,} in the typical scenario where~$\omega^{\mathrm{typ}}\approx 2n\tau(1-\tau)/3$,
			$$
			\vec m^{\mathrm{typ}}\approx \frac n3
			\big((1-\tau)^3+\tau^3,\,
			\tau(1-\tau),\,\tau(1-\tau),\,\tau(1-\tau)\big),
			$$ 
			these relations fix the dimensions $k_U'$, $k_V'$ and $k_W'$ as functions of $n$ and $\tau$:
\iftoggle{llncs}{$k_W'=\frac{n}{3} \cdot \frac{(1-2\tau)^2}{2\alpha_W-1}$, $k_V'=\frac{n}{3} \cdot \frac{(1-2\tau)^2}{2\alpha_V(\omega^{\mathrm{typ}})-1}$,
$k_U'=\frac{n}{3} \cdot \frac{1-6\tau^2+4\tau^3}{2\alpha_U(\vec{m}^{\mathrm{typ}},\omega^{\mathrm{typ}})-1}$ together with the constraints $k_W',\  k_V',\  k_U' \le \frac{n}{3} - d$ and $\alpha_V(\omega^{\mathrm{typ}}) \ge  1 - \frac{\tau (1 - \tau)}{1 - 2\tau (1 - \tau)}$.
}{
\begin{equation}
			k_W'=\frac{n}{3} \cdot \frac{(1-2\tau)^2}{2\alpha_W-1},\qquad
			k_V'=\frac{n}{3} \cdot \frac{(1-2\tau)^2}{2\alpha_V(\omega^{\mathrm{typ}})-1},\qquad
			k_U'=\frac{n}{3} \cdot \frac{1-6\tau^2+4\tau^3}{2\alpha_U(\vec{m}^{\mathrm{typ}},\omega^{\mathrm{typ}})-1}\ ,
\end{equation}
together with the constraints
\begin{equation}
k_W',\  k_V',\  k_U' \le \frac{n}{3} - d \qquad \text{ and } \qquad 
\alpha_V(\omega^{\mathrm{typ}}) \ge  1 - \frac{\tau (1 - \tau)}{1 - 2\tau (1 - \tau)}\ .
\end{equation} 
		}

			\iftoggle{llncs}{}{
			Let us now take a closer look at the distribution $\mathcal{D}_{U}^{\vec{m}}$. For now we only fixed the average value of $\sum_{\alpha,\beta}\ell_{\alpha,\beta}$. Recall that $\mathcal{D}_{U}^{\vec{m}}$ outputs $(\vec j,\boldsymbol\ell)$. Let $j_{\alpha\beta}^{\mathrm{typ}}$ and $\ell_{\alpha\beta}^{\mathrm{typ}}$ for the means of $j_{\alpha\beta}$ and $\ell_{\alpha\beta}$ under~$\mathcal D_U^{\vec m^{\mathrm{typ}}}$.
			Let us abbreviate $\alpha_U=\alpha_U(\vec m^{\mathrm{typ}},\omega^{\mathrm{typ}})$. By symmetry and the average conditions given above,
			\begin{align}
				j_{10}^{\mathrm{typ}}=j_{01}^{\mathrm{typ}}=j_{11}^{\mathrm{typ}}
				&=\frac{k_U'-j_{00}^{\mathrm{typ}}}{3},\\[3pt]
				j_{00}^{\mathrm{typ}}
				&\approx\frac n3\big((1-\tau)^3-\tau^3\big)
				+2\ell_{00}^{\mathrm{typ}}\ .
			\end{align}
			
			If $\alpha_U=1$, non-negativity forces all four counts
			$\ell_{\alpha\beta}$ to be zero. Otherwise, we choose
			\begin{equation}
			\ell_{10}^{\mathrm{typ}}=\ell_{01}^{\mathrm{typ}}=\ell_{11}^{\mathrm{typ}},
			\qquad
			\frac{\ell_{00}^{\mathrm{typ}}}{j_{00}^{\mathrm{typ}}}
			=\frac{\ell_{01}^{\mathrm{typ}}}{j_{01}^{\mathrm{typ}}}
			=1-\alpha_U,
			\end{equation}
			where the common fraction follows from the given total mean.
			This fixes all the typical means:
		\begin{equation}
				j_{00}^{\mathrm{typ}} \approx\frac n3\frac{(1-\tau)^3-\tau^3}{2\alpha_U-1}\quad , \quad
				j_{10}^{\mathrm{typ}}=j_{01}^{\mathrm{typ}}=j_{11}^{\mathrm{typ}}
				\approx\frac n3\frac{\tau(1-\tau)(1-2\tau)}{2\alpha_U-1},
		\end{equation} 
		\begin{equation} \mbox{and} \quad
				\ell_{\alpha\beta}^{\mathrm{typ}}
				=(1-\alpha_U)j_{\alpha\beta}^{\mathrm{typ}} \ .
		\end{equation}
		
}

\begin{figure}[H]
				\centering
				\includegraphics[width=0.7\linewidth]{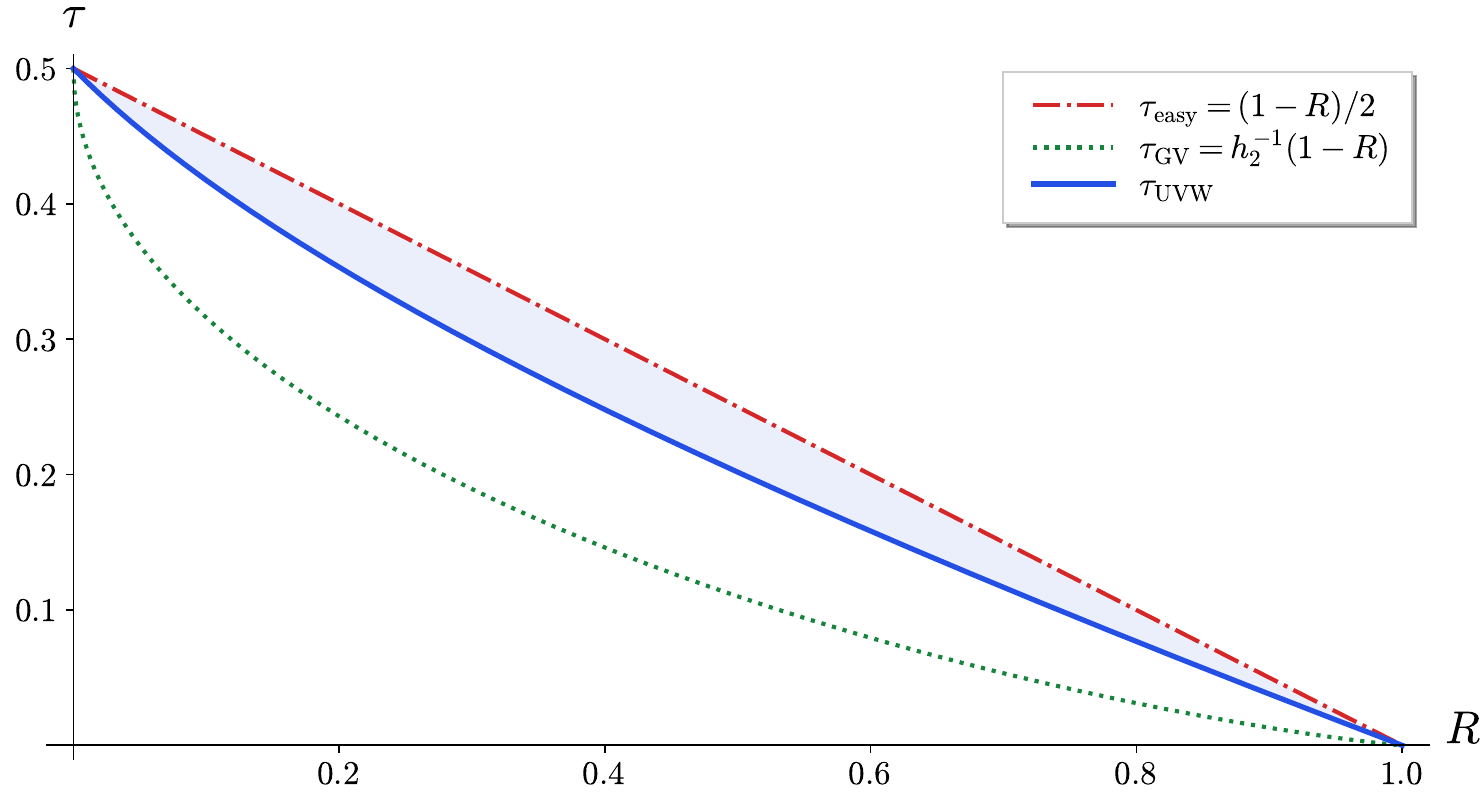}
				\caption{Relative decoding weights as functions of the rate $R=k/n$. The curve $\tau_{\mathrm{UVW}}$ is defined by the unique solution $\tau \in [0, 1/2]$ of $3R=4\tau^3+2\tau^2-8\tau+3$, obtained by neglecting $d/n$ and setting~$\alpha_U=\alpha_V=\alpha_W=1$ in the parameter choice.}
				\label{fig:decoding-advantage}
			\end{figure}

	Now \iftoggle{llncs}{}{since} $k=k_U'+k_V'+k_W'+3d$, we necessarily have to choose~$\tau\in(h^{-1}(1-R),1/2)$ \iftoggle{llncs}{to satisfy}{ by solving}
			\begin{equation}
			3R-\frac{9d}{n}
			=\frac{(1-2\tau)^2}{2\alpha_W-1}
			+\frac{(1-2\tau)^2}{2\alpha_V(\omega^{\mathrm{typ}})-1}
			+\frac{1-6\tau^2+4\tau^3}{2\alpha_U(\vec{m}^{\mathrm{typ}},\omega^{\mathrm{typ}})-1}\ .
			\end{equation}
			\iftoggle{llncs}{}{
			In particular, for $\alpha_W=\alpha_V=\alpha_U=1$, this becomes
			\begin{equation}
			3R-\frac{9d}{n}=4\tau^3+2\tau^2-8\tau+3\ .
			\end{equation}}
			Figure~\ref{fig:decoding-advantage} illustrates this case, which is close to our final parameter choice. The decoding weight exceeds the Gilbert-Varshamov bound, as required in our application of the ATPSF framework, while remaining in a regime where decoding without the trapdoor is believed to be hard. Knowledge of the private key thus provides a decoding advantage.

 	\section{Proposed Secure Instantiation}
\label{sec:instantiation}

\subsection{Concrete Parameters and Security Estimates}
\label{sec:params}

In Table~\ref{tab:parameters}, we propose several candidate parameter sets $(n,k,t)$ for \scheme{} that we claim provide $\lambda$ bits of computational security against both forgery and structural attacks. These claims are based on estimates of the attacks considered below. We propose these parameter sets as concrete targets for cryptanalysis, leaving a detailed security assessment to future work.

The public key is a parity-check matrix $\vec H$ in systematic form and can therefore be represented using $k(n-k)$ bits. A signature consists of a salt $\vec{r}\in\{0,1\}^{\lambda}$ and a vector $\vec{e}\in\mathcal{S}_t^n$. Since the verifier can derive the decoding input from the message and the salt, the systematic form of $\vec H$ allows the remaining $n-k$ coordinates of $\vec e$ to be reconstructed from the $k$ coordinates outside the identity block. Using a Bernoulli entropy model with parameter $t/n$ to compress these $k$ coordinates gives the estimated average signature size $|\sigma| \approx k h_2(t/n) + \lambda$ bits.

\begin{table}[H]
    \centering
    \begin{tabular*}{0.65\textwidth}{| @{\extracolsep\fill} c|ccc|cc|}
        \hline
        \begin{tabular}{c}
            \ $\lambda$
        \end{tabular}
        &  $n$ & $k$ & $t$
        & $|\mathrm{pk}|$ & $|\sigma|$\\
\hline
        \hline

        \multirow{1}{*}{ 128}
        & 15498 & 6617 & 3739 & 7.3457\,MB & 676\,B \\

        \hline

        \multirow{1}{*}{ 192}
        &21741 & 9208 & 5244 & 14.4255\,MB & 942\,B\\

        \hline

        \multirow{1}{*}{ 256}
        &27948 & 11876 & 6697 & 23.8589\,MB & 1212\,B\\

        \hline
    \end{tabular*}
\caption{Parameter sets and estimated public-key and signature sizes for \scheme{}.}
\label{tab:parameters}
\end{table}

The component dimensions $k_U$, $k_V$, and $k_W$ are derived from $(n,k,t)$ as described in Subsection~\ref{subsec:prange-parameters}, using the design parameters \iftoggle{llncs}{
$d=64$, $\alpha_V=\alpha_W=1$, and $\alpha_U=0.999$.
}{
\begin{equation}
    d=64,
    \qquad \alpha_V=\alpha_W=1,
    \qquad \alpha_U=0.999.
\end{equation}
}
The parameter $d$ is chosen according to the $2^{64}$ signatures which are supposed to be available to an attacker and is held fixed across the security levels considered here. It gives a gap comparable to \textsf{Wave} choice $d=40$ (\cite[p. $7$]{BCCCDGKLNSST23}) over $\mathbb F_3$, since $2^{64}\approx 3^{40}$.

\medskip

Our parameter selection considers two classes of attacks:
\begin{itemize}
    \item[$(i)$] forgery attacks, which essentially amounts to solving the Decoding One-Out-of-Many problem for the public code (see~Problem~\ref{prob:DOOM}),
    \item[$(ii)$] structural attacks, which aim to recover the hidden $(U,U+V,U+V+W)$ structure from the public code and are modeled by the $\mathsf{D}^{\mathsf{UVW}}$ problem (see~Problem~\ref{prob:UVWD}).
\end{itemize}

For each target security level $\lambda\in\{128,192,256\}$, our selection criterion is that the estimated work factor of every attack considered in either class be at least $2^\lambda$.

We do not provide here the full analysis of the attacks underlying our parameter selection. Such an analysis requires a careful study and optimization of several decoding and structural attacks, which falls outside the scope of this paper. Instead, we restrict ourselves to reporting the resulting security estimates. \iftoggle{llncs}{}{We nevertheless briefly describe below the strategies underlying the attacks considered in our analysis.

First, to estimate the complexity of~$(i)$, we use Information Set Decoding (ISD) framework \cite[Lecture $3$]{D23} combined with Wagner's generalized birthday algorithm \cite{D91,W02}. More precisely, we use a Dumer-style reduction to a higher-rate decoding instance, followed by a recursive collision search. The internal parameters of the attack are optimized for each candidate parameter set.

For~$(ii)$, we exploit the fact that the generalized $(U,U+V,U+V+W)$ structure induces special families of codewords in both the public code and its dual. Indeed, setting one or more of the components $\vec{u}$, $\vec{v}$, or $\vec{w}$ to zero gives rise to subfamilies whose weight distributions differ from that of a uniformly random code. Recovering codewords belonging to one of these families may therefore reveal information about the hidden structure. Moreover, these particular families often contain codewords of unusually low Hamming weight, which can make their recovery significantly easier. We conservatively model such an attack by assuming that the adversary can determine, at no additional computational cost, whether a recovered codeword belongs to one of these structural families. The attacker then applies ISD to both the public code and its dual in order to search for low-weight codewords of the relevant types. We estimate the corresponding work factors using Dumer's collision-based decoding algorithm~\cite{D91}, optimizing its parameters independently for each family of structural codewords and for both the primal and dual codes.}

The values of $\alpha_U$, $\alpha_V$, and $\alpha_W$ are chosen primarily for simplicity and are not claimed to minimize the expected number of rejections during signing. In particular, setting $\alpha_W=1$ and~$\alpha_V(\omega)=1$ for every admissible $\omega$ forces the corresponding nonnegative variables $\ell$ to be zero almost surely. Thus, $\mathcal D_W$ and $\mathcal D_V^\omega$ are point masses at zero.

Choosing $\alpha_U=0.999$ allows greater flexibility in $\mathcal D_U^{\vec m}$, since the variables $(\ell_{\alpha\beta})_{\alpha,\beta\in\{0,1\}}$ need not all vanish almost surely. We use this flexibility to tune the distribution numerically and reduce~$M_U(\vec m^{\mathrm{typ}})$. Specifically, we consider the unnormalized discrete Laplace weights
\begin{equation}
    L_{\mu,q}(x)\eqdef q^{|x-\mu|},
    \qquad x\in\mathbb Z,\quad \mu\in\mathbb R,\quad 0<q<1,
\end{equation}
with the additional convention $L_{0,0}(x)\eqdef\mathbf{1}_{\{x=0\}}$ for a point mass at zero. We then define \iftoggle{llncs}{
\begin{equation}
    \mathcal D_U^{\vec m}(\vec j,\boldsymbol\ell)
    \eqdef
    \frac{1}{Z_{\vec m}}
    \prod_{\alpha,\beta\in\{0,1\}}
    \mathcal A_{\alpha\beta}(j_{\alpha\beta})
    \mathcal B_{\alpha\beta}(\ell_{\alpha\beta})
\end{equation}
on the admissible domain $0\leq\ell_{\alpha\beta}\leq j_{\alpha\beta}\leq m_{\alpha\beta}$ and $\sum_{\alpha,\beta}j_{\alpha\beta}=k'_U$, and set it to zero elsewhere. Here, $Z_{\vec m}$ is the normalizing constant and we take
}{ $\mathcal D_U^{\vec m}$ as in Equation~\eqref{eq:D_U_Laplace}, with}
\begin{equation}
    \mathcal A_{\alpha\beta}(x)
    = L_{\mu_{\mathcal A_{\alpha\beta}},
          q_{\mathcal A_{\alpha\beta}}}(x),
    \qquad
    \mathcal B_{\alpha\beta}(x)
    = L_{\mu_{\mathcal B_{\alpha\beta}},
          q_{\mathcal B_{\alpha\beta}}}(x).
\end{equation}
The location parameters $\mu_{\mathcal A_{\alpha\beta}}$ and $\mu_{\mathcal B_{\alpha\beta}}$ are not, in general, the expectations of $j_{\alpha\beta}$ and $\ell_{\alpha\beta}$ under~$\mathcal D_U^{\vec m}$. These expectations, and hence the effective value of $\alpha_U$, must be computed under the joint distribution after restriction to its admissible domain and normalization.

We performed a numerical search for parameters yielding small values of $M_U(\vec m^{\mathrm{typ}})$. Table~\ref{tab:Laplace_params} reports one such candidate for the first parameter set. This search does not provide a certificate of global optimality.

\begin{table}[H]
    \centering
    \begin{tabular}{|c|rrrr|}
        \hline
        $(\alpha,\beta)$
        & $\mu_{\mathcal A_{\alpha\beta}}$
        & $q_{\mathcal A_{\alpha\beta}}$
        & $\mu_{\mathcal B_{\alpha\beta}}$
        & $q_{\mathcal B_{\alpha\beta}}$ \\
        \hline
        \hline
        $(0,0)$ & 2192.5649 & 0.9278 & 3.5696 & 0.0097 \\
        $(0,1)$ &  492.5517 & 0.9336 & 0.0000 & 0.0000 \\
        $(1,0)$ &  492.5517 & 0.9336 & 0.0000 & 0.0000 \\
        $(1,1)$ &  492.5517 & 0.9336 & 0.0000 & 0.0000 \\
        \hline
    \end{tabular}
    \caption{Discrete Laplace parameters defining $\mathcal D_U^{\vec m^{\mathrm{typ}}}$ for $(n,k,t)=(15498,6617,3739)$, $(k_U,k_V,k_W)=(3723,1447,1447)$, $d=64$, and $\alpha_U=0.999$. Here, $\vec m^{\mathrm{typ}}=(2328,946,946,946)$, and the pair $(\mu,q)=(0,0)$ denotes a point mass at zero.} \label{tab:Laplace_params}
\end{table}

Our implementation, available \href{https://github.com/kevin-carrier/MegaBight}{here}, reports the preliminary numerical estimates
\begin{equation}
M_W \approx 1.0479 \quad , \quad M_V(\omega^{\mathrm{typ}}) \approx 1.1155 \quad \text{ and } \quad M_U(\vec m^{\mathrm{typ}})\approx 78.1759\ .
\end{equation} 
This value could likely be further reduced by allowing more general values of $\alpha_U$, $\alpha_V$, and $\alpha_W$, and by jointly optimizing the distributions $\mathcal{D}_U$, $\mathcal{D}_V^{\omega^{\mathrm{typ}}}$, and $\mathcal D_U^{\vec m^{\mathrm{typ}}}$ including distributions outside the discrete Laplace family. A more thorough optimization of these internal decoding distributions is left for a future specification of the scheme.

Our choice $\alpha_V = 1$ implies $\mathcal{D}_V^{\omega} = \mathcal{D}_V^{\omega^{\mathrm{typ}}}$ for all $\omega$. however, the parameters in Table~\ref{tab:Laplace_params} specify the distribution $\mathcal D_U$ at $\vec m^{\mathrm{typ}}$ only. Applying the same weights unchanged to every admissible $\vec m$ can make some target outputs unreachable, leading to $M_U(\vec m)=+\infty$. A complete specification must therefore define $\mathcal D_U^{\vec m}$ for every admissible $\vec m$ and ensure that every output with positive target probability also has positive proposal probability.

We expect the rejection constants evaluated at typical statistics to provide a good approximation to the overall expected number of rejections. For the first parameter set, this gives the heuristic estimate
\begin{equation}
    (M_W-1)
    +\bigl(M_V(\omega^{\mathrm{typ}})-1\bigr)
    +\bigl(M_U(\vec m^{\mathrm{typ}})-1\bigr)
    \approx 77.34
\end{equation}
rejections per signature. This approximation assumes that atypical statistics contribute little to the average in Equation~\eqref{eq:rejection}.

\medskip

Finally, Theorem~\ref{thm:prange-no-leakage} allows us to bound the statistical distance between the output distribution of our algorithm and the uniform distribution on $\mathcal S_t^n$. For our first parameter set, averaging over the random choice of the component parity-check matrices gives
\begin{equation}
    \mathbb{E}_{\vec{H}_U,\vec{H}_V,\vec{H}_W}
    \left[
        \Delta\bigl(\vec{e}^{\mathrm{out}},\vec{e}^{\mathrm{unif}}\bigr)
    \right]
    \leq 2^{-62}\ .
\end{equation}

\subsection{The hull as a Distinguisher}
\label{sec:hull}

Beyond decoding-based attacks, the hull can provide a distinguisher between structured and random codes, as for instance illustrated by the attack against \textsf{SURF}, which uses binary $(U,U+V)$ codes~\cite{DST17b}. For a linear $[n,k]$ code $\mathcal C$, the hull is defined as
$$
    \mathrm{hull}(\mathcal C)\eqdef\mathcal C\cap\mathcal C^\perp.
$$
\iftoggle{llncs}{}{Its dimension can be efficiently computed from either a generator matrix $\vec{G}$ or a parity-check matrix $\vec{H}$ of $\mathcal{C}$ as
\begin{equation}
    \dim\left(\mathrm{hull}(\mathcal{C})\right)
    =
    k-\mathrm{rank}(\vec{G}\vec{G}^{\top})
    =
    n-k-\mathrm{rank}(\vec{H}\vec{H}^{\top})\ .
\end{equation}
}

The hull is a classical tool for detecting hidden structure in code-based cryptosystems. While random binary linear codes typically have a small hull~\cite{S97a}, some structured code families exhibit a significantly larger one. This property has notably been exploited to distinguish $(U,U+V)$-type constructions from random linear codes \cite{DST17b}. We therefore investigate whether the generalized $(U,U+V,U+V+W)$ codes used in \scheme{} exhibit a similar behaviour.
We compare the hull dimension of our generalized construction with that of a uniformly random binary $[n,k]$ code. Let\iftoggle{llncs}{$  X_{\mathrm{rand}}
	\eqdef
	\dim\left(\mathrm{hull}(\mathcal{C}_{\mathrm{rand}})\right)$ and $X_{\mathrm{UVW}}
	\eqdef
	\dim\left(\mathrm{hull}(\mathcal{C}_{\mathrm{UVW}})\right)$.}{
\begin{equation}
    X_{\mathrm{rand}}
    \eqdef
    \dim\left(\mathrm{hull}(\mathcal{C}_{\mathrm{rand}})\right),
    \qquad
    X_{\mathrm{UVW}}
    \eqdef
    \dim\left(\mathrm{hull}(\mathcal{C}_{\mathrm{UVW}})\right).
\end{equation}
}
For comparison, we also consider the three unmixed constructions obtained from the extreme choices of $\vec{a}$ and $\vec{b}$:\iftoggle{llncs}{
$$
X_{1} \eqdef \dim \left( \mathrm{hull}\left( (U+V, U, U+W) \right)\right),  X_{2} \eqdef \dim \left( \mathrm{hull}\left( (U+V, U, U+V+W) \right)\right)
$$
$$
X_{3} \eqdef \dim \left( \mathrm{hull}\left( (U+V, U+W, U+V+W) \right)\right)
$$}
{
\begin{eqnarray}
X_{1} & \eqdef & \dim \left( \mathrm{hull}\left( (U+V, U, U+W) \right)\right),\\
X_{2} & \eqdef & \dim \left( \mathrm{hull}\left( (U+V, U, U+V+W) \right)\right),\\
X_{3} & \eqdef & \dim \left( \mathrm{hull}\left( (U+V, U+W, U+V+W) \right)\right),
\end{eqnarray}}
where $U$, $V$, and $W$ are sampled independently and uniformly among binary $[n/3,k_U]$, $[n/3,k_V]$, and $[n/3,k_W]$ linear codes, respectively.

\begin{figure}[h]
    \centering

	\includegraphics[scale=0.8]{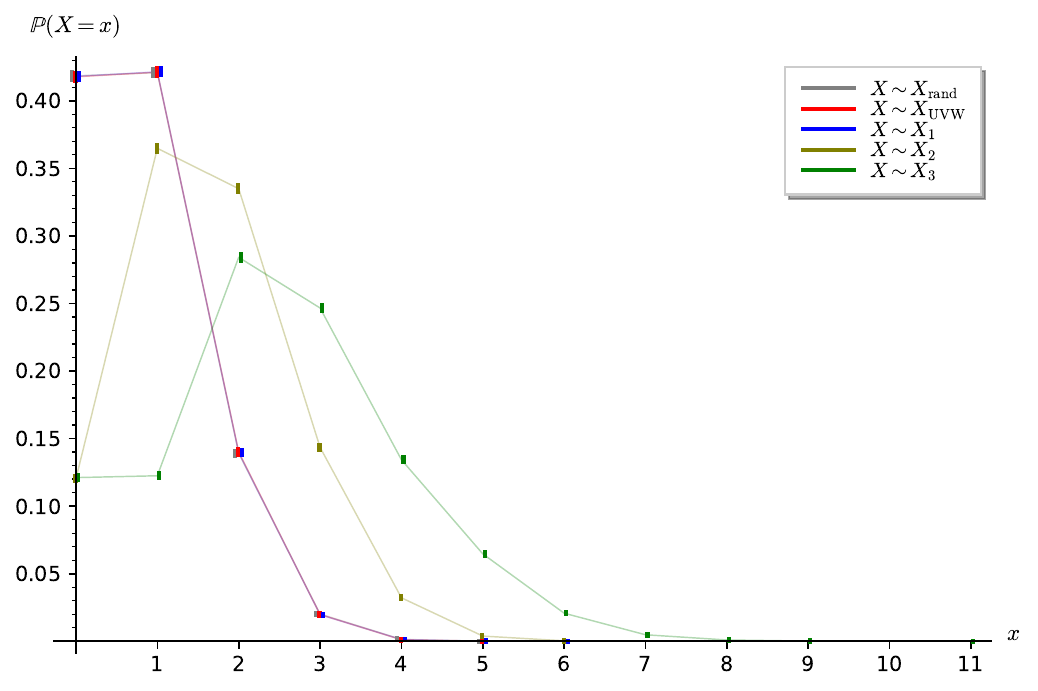}
	
	\caption{Empirical distributions of the hull dimension, estimated from $N=130\,000$ independent samples of each of $X_{\mathrm{rand}}$, $X_{\mathrm{UVW}}$, $X_1$, $X_2$, and $X_3$, for our first parameter set $(n,k,t,k_U,k_V,k_W)=(15498,6617,3739,3723,1447,1447)$. For $X_{\mathrm{UVW}}$, we fix $\vec a=(\vec 0_{2n/9},\vec 1_{n/9})$ and $\vec b=(\vec 0_{n/9},\vec 1_{2n/9})$. For $X_1$, $X_2$, and $X_3$, the pair $(\vec a,\vec b)$ is fixed to $(\vec 0_{n/3},\vec 0_{n/3})$, $(\vec 0_{n/3},\vec 1_{n/3})$, and $(\vec 1_{n/3},\vec 1_{n/3})$, respectively. Vertical error bars indicate pointwise $95\%$ Wilson score confidence intervals for the estimated probabilities.}
    \label{fig:hull-distribution}
\end{figure}

Figure~\ref{fig:hull-distribution} reports the resulting empirical distributions. For the parameters considered, we observe no significant difference between the empirical distributions of $X_{\mathrm{UVW}}$ and $X_{\mathrm{rand}}$. This provides evidence that the hull dimension alone does not yield an effective distinguisher for the generalized construction.

Figure~\ref{fig:hull-distribution} also motivates the use of a mixed construction. The distributions of $X_2$ and $X_3$ visibly differ from that of $X_{\mathrm{rand}}$, indicating that the constructions $(U+V,U,U+V+W)$ and $(U+V,U+W,U+V+W)$ may leak structural information through their hulls. By contrast, the distribution of $X_1$ is much closer to that of a random code. However, the corresponding construction contains lower-weight structural codewords in the code or its dual, making it more vulnerable to the decoding-based structural attacks. The random coordinate-wise choice of $\vec{a}$ and~$\vec{b}$ therefore balances these two effects: it mitigates hull-based distinguishers while making the weight distribution of structural codewords less favourable to decoding-based attacks.

\subsection{Hull after Puncturing}
\label{sec:punctured-hull}

Although the hull of the full code behaves similarly to that of a random code, carefully puncturing, {\it i.e.,} removing, some coordinates may reveal a simpler structured code.

By absorbing a suitable coordinate permutation into the secret permutation, we may assume without loss of generality that
\iftoggle{llncs}{$\vec{a}
	=
	(\vec{0}_{2n/9},\vec{1}_{n/9})$ and $\vec{b}
	=
	(\vec{0}_{n/9},\vec{1}_{2n/9})$}{
\begin{equation}
    \vec{a}
    =
    (\vec{0}_{2n/9},\vec{1}_{n/9})
    \qquad\text{and}\qquad
    \vec{b}
    =
    (\vec{0}_{n/9},\vec{1}_{2n/9}),
\end{equation}
}

After undoing the secret permutation, the generator matrix of the generalized $(U,U+V,U+V+W)$ code can therefore be viewed as the $3\times 9$ block matrix given below.

\[
\mathbf{G} =
\vcenter{\hbox{
\begin{tikzpicture}[baseline=(M.center), scale=0.8]
\small
\matrix (M) [
    matrix of math nodes,
    left delimiter={(},
    right delimiter={)},
    nodes={
        minimum width=1.1cm,
        minimum height=6mm,
        inner sep=1pt
    },
    row sep=0pt,
    column sep=0pt
] {
    \mathbf{G}_U^{(1)} &
    \mathbf{G}_U^{(2)} &
    \mathbf{G}_U^{(3)} &
    \mathbf{G}_U^{(1)} &
    \mathbf{G}_U^{(2)} &
    \mathbf{G}_U^{(3)} &
    \mathbf{G}_U^{(1)} &
    \mathbf{G}_U^{(2)} &
    \mathbf{G}_U^{(3)} \\
    \mathbf{G}_V^{(1)} &
    \mathbf{G}_V^{(2)} &
    \mathbf{G}_V^{(3)} &
    \mathbf{0} &
    \mathbf{0} &
    \mathbf{0} &
    \mathbf{0} &
    \mathbf{G}_V^{(2)} &
    \mathbf{G}_V^{(3)} \\
    \mathbf{0} &
    \mathbf{0} &
    \mathbf{0} &
    \mathbf{0} &
    \mathbf{0} &
    \mathbf{G}_W^{(3)} &
    \mathbf{G}_W^{(1)} &
    \mathbf{G}_W^{(2)} &
    \mathbf{G}_W^{(3)} \\
};

\foreach \j in {1,2,4,5,7,8}{
    \draw[line width=0.3pt]
        (M-1-\j.north east) -- (M-3-\j.south east);
}

\foreach \j in {3,6}{
    \draw[line width=1pt]
        (M-1-\j.north east) -- (M-3-\j.south east);
}

\draw[line width=0.3pt]
    (M-1-1.south west) -- (M-1-9.south east);

\draw[line width=0.3pt]
    (M-2-1.south west) -- (M-2-9.south east);

\draw[<->, >=stealth]
    ([xshift=22mm]M-1-9.north west)
    --
    ([xshift=22mm]M-1-9.south west)
    node[midway,right] {$k_U$};

\draw[<->, >=stealth]
    ([xshift=22mm]M-2-9.north west)
    --
    ([xshift=22mm]M-2-9.south west)
    node[midway,right] {$k_V$};

\draw[<->, >=stealth]
    ([xshift=22mm]M-3-9.north west)
    --
    ([xshift=22mm]M-3-9.south west)
    node[midway,right] {$k_W$};

\foreach \j in {1,...,9}{
    \draw[<->, >=stealth]
        ([yshift=-2mm]M-3-\j.south west)
        --
        ([yshift=-2mm]M-3-\j.south east)
        node[midway,below] {$\frac{n}{9}$};
}

\end{tikzpicture}
}}
\]

Puncturing the coordinates corresponding either to blocks $\{1,5,6\}$ or to blocks $\{3,7,8\}$ reveals, up to a coordinate permutation and elementary row operations, a code of the form $(U',U'+V')$ of length $2n/3$ and for which
\begin{equation}
    \dim(U') = \min( k_U + k_V, n/3) = n/3
    \qquad\text{and}\qquad
    \dim(V')= k_W = k_V.
\end{equation}
We can then apply the same argument as in~\cite{DST17b} to show that the punctured code has hull dimension
\begin{equation}
    \dim\left(\mathrm{hull}(\mathcal{C}_{\mathrm{punc}})\right)
    =
    \frac{n}{3}-k_V.
\end{equation}
Therefore, once the appropriate coordinates have been identified, the resulting punctured code can be efficiently distinguished from a random code through its abnormally large hull. Identifying such coordinates is, however, precisely the main difficulty of the attack.

More generally, puncturing only $n/3-k_U$ coordinates belonging to either of the two puncturing patterns is already sufficient to make the hull of the punctured code distinguishable from that of a random (punctured) code. However, finding such a pattern by random search requires essentially
\begin{equation}
    \frac{\binom{n}{\,n/3-k_U\,}}
         {2\binom{n/3}{\,n/3-k_U\,}}
\end{equation}
trials on average, which remains well above the targeted security level for all our parameter sets.

	\iftoggle{llncs}{\bibliographystyle{src/splncs04}}{
	\section{Acknowledgements}

Thomas Debris-Alazard was supported by the European Union (ERC StG, \textsf{IQ-SCALe}, 101219089). Views and opinions expressed are however those of the author(s) only and do not necessarily reflect those of the European Union or the
European Research Council. Neither the European Union nor the granting authority can be held responsible for them. 	\bibliographystyle{alpha}}   
	\newcommand{\etalchar}[1]{$^{#1}$}

	\iftoggle{llncs}{
	\newpage
	\appendix
	\section{Proofs of Section~\ref{sec:unif}}\label{app:unif}

Our aim in this appendix is to prove all the statements of Section~\ref{sec:unif} whose goal is to show that distribution of signatures of \scheme{} are independent of the used trapdoor. 

Recall that we use the following notation 
\begin{equation}
\left(\vec{e}_U^{\mathrm{unif}}, \vec{e}_V^{\mathrm{unif}}, \vec{e}_W^{\mathrm{unif}}\right) = \varphi_{\vec{a},\vec{b},\vec{P}}^{-1}\left( \vec{e}^{\mathrm{unif}} \right) \quad \mbox{ where } \; \vec{e}^{\mathrm{unif}} \Unif \mathcal{S}_{t}^{n} \ .
\end{equation}
where $\vec{a} \star \vec{b} = \vec{a}$ and, 
\begin{equation}
	\begin{array}{lccc}
	\varphi_{\vec{a},\vec{b}, \vec{P}}: & \mathbb{F}_{2}^{n/3} \times \mathbb{F}_{2}^{n/3} \times \mathbb{F}_{2}^{n/3} & \longrightarrow &  \mathbb{F}_{2}^{n/3} \times \mathbb{F}_{2}^{n/3} \times \mathbb{F}_{2}^{n/3} \\
	& (\vec{u},\vec{v},\vec{w}) & \longmapsto &  \left( \vec{u} + \vec{v}, \vec{u}  + \vec{a}\star \vec{w}, \vec{u} + \vec{b}\star \vec{v}+ \vec{w} \right)\vec{P} \ .
\end{array}
\end{equation}
Recall that by Lemma~\ref{lemma:varphiInverse},
\begin{multline}
\varphi_{\vec{a},\vec{b},\vec{P}}^{-1}: (\vec{x},\vec{y},\vec{z})\vec{P} \longmapsto \bigl( \vec{a} \star \vec{x} + \vec{y} + \vec{a} \star \vec{z}, \\ (\vec{1}_{n/3}+\vec{a})\star \vec{x} + \vec{y} + \vec{a} \star \vec{z}, \\ \vec{b} \star \vec{x} + (\vec{1}_{n/3}+\vec{b})\star\vec{y} + \vec{z} \bigr)
\end{multline}
with $\vec{x},\vec{y},\vec{z} \in \mathbb{F}_{2}^{n/3}$.

\subsection{Uniformity Properties of $\vec{e}^{\textup{unif}}$}

\propoStat*

\begin{proof}
    Fix $i\in[1,n/3]$, and write
    $a\eqdef\vec{a}(i)$, $b\eqdef\vec{b}(i)$, and
    $(\alpha,\beta)\eqdef(\vec{x}_V(i),\vec{x}_W(i))$.
    For $\vec{x}_U(i)=u$, the three-bit vector
    whose Hamming weight is $\Omega_i$ is
    \[
        (u+\alpha,\ u+a\beta,\ u+b\alpha+\beta)\ .
    \]
    Replacing $u$ by $1+u$ complements all three bits,
    so the corresponding Hamming weights sum to three.

    When $(\alpha,\beta)=(0,0)$, the choices $u=0$
    and $u=1$ yield $(0,0,0)$ and $(1,1,1)$,
    respectively. For the remaining cases, substituting
    the minimizing bit $c$ given in the table yields
    \[
    \begin{array}{c|c}
        (\alpha,\beta) & \text{Three-bit vector for }u=c\\
        \hline
        (1,0) & (1+b,b,0)\\
        (0,1) & (a,0,1+a)\\
        (1,1) & (a+b,1+b,a)
    \end{array}
    \]
    The first two vectors have Hamming weight one.
    In the last case, the condition $ab=a$ restricts
    $(a,b)$ to $(0,0)$, $(0,1)$, or $(1,1)$,
    yielding $(0,1,0)$, $(1,0,0)$, and $(0,0,1)$,
    respectively. Thus, the minimizing bit always
    yields weight one when $(\alpha,\beta)\neq(0,0)$,
    while the complementary bit yields weight two.
    This establishes the table.

    We now establish the weight identity~\eqref{eq:weight}.
    For positions in $\mathcal M_{00}$, exactly
   $h_{00}$ three-bit vectors have weight three,
    while the remaining ones have weight zero.
    For each $(\alpha,\beta)\neq(0,0)$, exactly~$h_{\alpha\beta}$ positions contribute weight two,
    while the remaining
    $m_{\alpha\beta}-h_{\alpha\beta}$ positions
    contribute weight one. Consequently,
    \begin{align*}
        \big|\varphi_{\vec{a},\vec{b},\vec{P}}
        (\vec{x}_U,\vec{x}_V,\vec{x}_W)\big|
        &=3h_{00}
        +\sum_{(\alpha,\beta)\neq(0,0)}
        \big(2h_{\alpha\beta}
        +m_{\alpha\beta}-h_{\alpha\beta}\big)\\
        &=3h_{00}
        +m_{10}+m_{01}+m_{11}
        +h_{10}+h_{01}+h_{11}\\
        &=\nu+3h_{00}+h_{10}+h_{01}+h_{11},
    \end{align*}
    where the last equality follows from
    $\nu=n/3-m_{00}=m_{10}+m_{01}+m_{11}$.

    Finally, for fixed $\vec{x}_V$ and $\vec{x}_W$,
    prescribing $\vec{h}$ amounts to choosing,
    within each $\mathcal M_{\alpha\beta}$,
    exactly~$h_{\alpha\beta}$ positions at which
    $\vec{x}_U(i)$ takes the larger-weight value.
    Since each choice uniquely determines $\vec{x}_U$,
    the number of such vectors is
    \[
        N_{\vec{m}}(\vec{h})
        =\prod_{\alpha,\beta\in\{0,1\}}
        \binom{m_{\alpha\beta}}{h_{\alpha\beta}},
    \]
    proving~\eqref{eq:Nm}. \qed
\end{proof}

This proposition enables to show the following lemma proving that probabilities associated to $\vec{e}^{\mathrm{unif}}$ are only functions of statistics from Definition~\ref{def:statistics}. 

\lemmaRefUnif*

\begin{proof}
    Write $\vec{e}^{\mathrm{unif}}=(\vec{x},\vec{y},\vec{z})\vec{P}$.
    Since $\vec{P}$ is a permutation matrix,
    $(\vec{x},\vec{y},\vec{z})$ is uniformly distributed
    over vectors of Hamming weight $t$.

    We first prove (1). By Lemma~\ref{lemma:varphiInverse}, for $i\in[1,n/3]$,
    \[
        \vec{e}_W^{\mathrm{unif}}(i)=
        \begin{cases}
            \vec{x}(i)+\vec{z}(i)
                & \text{if }\vec{b}(i)=1,\\
            \vec{y}(i)+\vec{z}(i)
                & \text{otherwise}.
        \end{cases}
    \]
    Define $\vec{p},\vec{r}\in\mathbb{F}_2^{n/3}$ by
    \[
        \vec{p}(i)\eqdef
        \begin{cases}
            \vec{x}(i) & \text{if }\vec{b}(i)=1\\
            \vec{y}(i) & \text{otherwise}
        \end{cases}
        \qquad
        \vec{r}(i)\eqdef
        \begin{cases}
            \vec{y}(i) & \text{if }\vec{b}(i)=1,\\
            \vec{x}(i) & \text{otherwise}.
        \end{cases}
    \]
    Thus, $\vec{e}_W^{\mathrm{unif}}=\vec{p}+\vec{z}$.
    The map
    $(\vec{x},\vec{y},\vec{z})\mapsto
    (\vec{p},\vec{r},\vec{z})$
    only exchanges $\vec{x}(i)$ and $\vec{y}(i)$
    at positions where~$\vec{b}(i)=0$, and hence it keeps the distribution invariant.

    Fix $\vec{x}_W\in\mathbb{F}_2^{n/3}$ and let
    $\omega\eqdef\omega(\vec{x}_W)$.
    The condition $\vec{e}_W^{\mathrm{unif}}=\vec{x}_W$
    imposes odd parity on exactly~$\omega$ pairs
    $(\vec{p}(i),\vec{z}(i))$ and even parity on
    the remaining $n/3-\omega$ pairs.
    Each odd-parity pair has two possible values,
    both of weight one.
    Among the even-parity pairs, choose $j$ positions
    to take the value $(1,1)$, while the others
    take the value $(0,0)$.
    The vector $\vec{r}$ must then have weight
    $t-\omega-2j$.

    Consequently, the number of vectors
    $(\vec{p},\vec{r},\vec{z})$ of weight $t$
    satisfying $\vec{p}+\vec{z}=\vec{x}_W$ is
    \[
        2^\omega\sum_j
        \binom{n/3-\omega}{j}
        \binom{n/3}{t-\omega-2j}
        =A_\omega.
    \]
    Since there are $\binom{n}{t}$ vectors of weight $t$,
    we obtain (1).

    Now, let us prove (2).
    Fix $\vec{x}_V,\vec{x}_W\in\mathbb{F}_2^{n/3}$
    and let $\nu\eqdef\nu(\vec{x}_V,\vec{x}_W)$.
    By Proposition~\ref{prop:counts}, each position
    in $\mathcal{M}_{00}$ contributes either weight
    zero or three, while each position in
    $\mathcal{M}_{\alpha\beta}$ with~$(\alpha,\beta)\neq(0,0)$ contributes either
    weight one or two.
    Moreover, each choice of weight uniquely
    determines the corresponding bit of $\vec{x}_U$.

    Suppose that $j$ of the $n/3-\nu$ positions
    in $\mathcal{M}_{00}$ contribute weight three,
    and that $s$ of the remaining $\nu$ positions
    contribute weight two.
    The total Hamming weight is then
    $$
        3j+2s+(\nu-s)=\nu+3j+s \ .
    $$
    Hence, the condition that the total weight
    equals $t$ imposes $s=t-\nu-3j$.
    For each $j$, there are~$\binom{n/3-\nu}{j}$ ways to choose
    the positions contributing weight three,
    and~$\binom{\nu}{t-\nu-3j}$ ways to choose
    those contributing weight two.
    Since these choices uniquely determine $\vec{x}_U$,
    the number of vectors~$\vec{x}_U$ compatible
    with the fixed pair $(\vec{x}_V,\vec{x}_W)$ is
    \[
        B_\nu
        =\sum_{j=0}^{n/3-\nu}
        \binom{n/3-\nu}{j}
        \binom{\nu}{t-\nu-3j}\ .
    \]
    Therefore,
    \[
        \mathbb{P}\left(
        \vec{e}_V^{\mathrm{unif}}=\vec{x}_V,
        \vec{e}_W^{\mathrm{unif}}=\vec{x}_W
        \right)
        =\frac{B_\nu}{\binom{n}{t}}\ .
    \]
    Combining this with (1), we obtain
    \[
        \mathbb{P}\left(
        \vec{e}_V^{\mathrm{unif}}=\vec{x}_V
        \mid
        \vec{e}_W^{\mathrm{unif}}=\vec{x}_W
        \right)
        =\frac{B_\nu}{A_\omega},
    \]
    which proves (2).

    Finally, we prove (3).
    Fix $\vec{x}_V,\vec{x}_W$ such that
    $B_\nu>0$.
    Conditioned on
    $(\vec{e}_V^{\mathrm{unif}},
      \vec{e}_W^{\mathrm{unif}})
    =(\vec{x}_V,\vec{x}_W)$,
    the vector $\vec{e}_U^{\mathrm{unif}}$
    is uniformly distributed over the $B_\nu$
    compatible vectors.
    Thus, for any $\vec{x}_U$,
    \[
        \mathbb{P}\left(
        \vec{e}_U^{\mathrm{unif}}=\vec{x}_U
        \mid
        \vec{e}_V^{\mathrm{unif}}=\vec{x}_V,
        \vec{e}_W^{\mathrm{unif}}=\vec{x}_W
        \right)
        =
        \begin{cases}
            \dfrac{1}{B_\nu}
            & \text{if }
            \big|\varphi_{\vec{a},\vec{b},\vec{P}}
            (\vec{x}_U,\vec{x}_V,\vec{x}_W)\big|=t,\\[1.2ex]
            0 & \text{otherwise}.
        \end{cases}
    \]
    By Equation~\eqref{eq:weight},
    \[
        \big|\varphi_{\vec{a},\vec{b},\vec{P}}
        (\vec{x}_U,\vec{x}_V,\vec{x}_W)\big|
        =\nu+3h_{00}+h_{10}+h_{01}+h_{11}\ .
    \]
    Consequently,
    \[
        \mathbb{P}\left(
        \vec{e}_U^{\mathrm{unif}}=\vec{x}_U
        \mid
        \vec{e}_V^{\mathrm{unif}}=\vec{x}_V,
        \vec{e}_W^{\mathrm{unif}}=\vec{x}_W
        \right)
        =
        \begin{cases}
            \dfrac{1}{B_\nu}
            & \text{if }
            \nu+3h_{00}+h_{10}+h_{01}+h_{11}=t,\\[1.2ex]
            0 & \text{otherwise}.
        \end{cases}
    \]
    Since $\nu=n/3-m_{00}$, both the denominator
    and the condition depend only on
    $(\vec{m},\vec{h})$.
    This proves~(3) and completes the proof. \qed
\end{proof}

\subsection{Reaching $\vec{e}^{\mathrm{unif}}$ via Statistics of Definition~\ref{def:statistics}}

We are now ready to show that if statistics of Definition~\ref{def:statistics} for $\vec{e}_W$, $\vec{e}_V$ knowing $\vec{e}_W$ and~$\vec{e}_U$ knowing $(\vec{e}_W,\vec{e}_V)$ and those of~$\vec{e}^{\mathrm{unif}}_W$, $\vec{e}^{\mathrm{unif}}_V$ knowing $\vec{e}^{\mathrm{unif}}_W$ and $\vec{e}^{\mathrm{unif}}_U$ knowing $(\vec{e}^{\mathrm{unif}}_W,\vec{e}^{\mathrm{unif}}_V)$ coincide, then $\varphi_{\vec{a},\vec{b},\vec{P}}(\vec{e}_W,\vec{e}_V,\vec{e}_U)$ and $\vec{e}^{\mathrm{unif}}$ are equidistributed. 

\lemmaUnifSampl*

\begin{proof}
    Fix $\vec{x}\in\mathcal{S}_t^n$ and write $(\vec{x}_U,\vec{x}_V,\vec{x}_W)
        =\varphi_{\vec{a},\vec{b},\vec{P}}^{-1}(\vec{x})$, with corresponding statistics $\widetilde{\omega}\eqdef\omega(\vec{x}_W)$, $\widetilde{\nu}\eqdef\nu(\vec{x}_V,\vec{x}_W)$, $\widetilde{\vec{m}}
        \eqdef\vec{m}(\vec{x}_V,\vec{x}_W)$ and $\widetilde{\vec{h}} \eqdef\vec{h}(\vec{x}_U;\vec{x}_V,\vec{x}_W)$.

There are $\binom{n/3}{\widetilde{\omega}}$ vectors $\vec{x}_W$ of weight $\widetilde{\omega}$. For a fixed $\vec{x}_W$, a vector $\vec{x}_V$ with statistic $\widetilde{\nu}$ must have exactly $\widetilde{\nu}-\widetilde{\omega}$ non-zero coordinates among the $n/3-\widetilde{\omega}$ positions where $\vec{x}_W$ is zero, while its coordinates on the support of $\vec{x}_W$ are arbitrary. There are therefore $2^{\widetilde{\omega}}
        \binom{n/3-\widetilde{\omega}}
              {\widetilde{\nu}-\widetilde{\omega}}$ such vectors. Finally, for a fixed pair $(\vec{x}_V,\vec{x}_W)$, Equation~\eqref{eq:Nm} counts $N_{\widetilde{\vec{m}}}(\widetilde{\vec{h}})$ possible vectors $\vec{x}_U$ with statistic~$\widetilde{\vec{h}}$.

By the assumptions of the lemma, all vectors sharing the same statistics have equal (conditional) probabilities. Therefore, each probability can be expressed as the probability of the corresponding statistic divided by the number of compatible vectors.

Using the chain rule for conditional probabilities, we obtain
    \begin{align*}
        \mathbb{P}(\vec{e}^{\mathrm{out}}=\vec{x})
        &=
        \mathbb{P}\big(
            \vec{e}_U=\vec{x}_U
            \mid
            \vec{e}_V=\vec{x}_V,
            \vec{e}_W=\vec{x}_W
        \big) \cdot
        \mathbb{P}\big(
            \vec{e}_V=\vec{x}_V
            \mid
            \vec{e}_W=\vec{x}_W
        \big)
        \cdot
        \mathbb{P}\big(
            \vec{e}_W=\vec{x}_W
        \big)\\[1ex]
        &=
        \frac{
            \mathbb{P}\big(
                \vec{h}(\vec{e}_U;\vec{e}_V,\vec{e}_W)
                =\widetilde{\vec{h}}
                \mid
                \vec{m}(\vec{e}_V,\vec{e}_W)
                =\widetilde{\vec{m}}
            \big)
        }{
            N_{\widetilde{\vec{m}}}
              (\widetilde{\vec{h}})
        }\\
        &\qquad\qquad \cdot
        \frac{
            \mathbb{P}\big(
                \nu(\vec{e}_V,\vec{e}_W)
                =\widetilde{\nu}
                \mid
                \omega(\vec{e}_W)
                =\widetilde{\omega}
            \big)
        }{
            2^{\widetilde{\omega}}
            \binom{n/3-\widetilde{\omega}}
                  {\widetilde{\nu}-\widetilde{\omega}}
        }\cdot
        \frac{
            \mathbb{P}\big(
                \omega(\vec{e}_W)=\widetilde{\omega}
            \big)
        }{
            \binom{n/3}{\widetilde{\omega}}
        }.
    \end{align*}

By the assumptions of the lemma, the last expression equals
		\begin{align*}
			&\frac{\mathbb{P}(\vec{h}(\vec{e}_U^{\mathrm{unif}};\vec{e}_V^{\mathrm{unif}},\vec{e}_W^{\mathrm{unif}})=\widetilde{\vec{h}}\mid
				\vec{m}(\vec{e}_V^{\mathrm{unif}},\vec{e}_W^{\mathrm{unif}})=\widetilde{\vec{m}})}{N_{\widetilde{\vec{m}}}(\widetilde{\vec{h}})}\\
				&\qquad\qquad \cdot
			\frac{\mathbb{P}(\nu(\vec{e}_V^{\mathrm{unif}},\vec{e}_W^{\mathrm{unif}})=\widetilde{\nu}\mid
				\omega(\vec{e}_W^{\mathrm{unif})}=\widetilde{\omega})}
			{2^{\widetilde{\omega}}\binom{n/3-\widetilde{\omega}}{\widetilde{\nu} - \widetilde{\omega}}}\cdot
			\frac{\mathbb{P}(\omega(\vec{e}_W^{\mathrm{unif}})=\widetilde{\omega})}{\binom{n/3}{\widetilde{\omega}}}\\
			&\qquad=\mathbb{P}(\vec{e}_U^{\mathrm{unif}}=\vec{x}_U\mid
			\vec{e}_V^{\mathrm{unif}}=\vec{x}_V,\vec{e}_W^{\mathrm{unif}}=\vec{x}_W)\cdot\mathbb{P}(\vec{e}_V^{\mathrm{unif}}=\vec{x}_V\mid\vec{e}_W^{\mathrm{unif}}=\vec{x}_W)
			\mathbb{P}(\vec{e}_W^{\mathrm{unif}}=\vec{x}_W)\notag\\
			&\qquad=\mathbb{P}(\vec{e}^{\mathrm{unif}}=\vec{x})
		\end{align*}
		where we used Lemma~\ref{lem:reference-uniformity} in the first
		equality and the bijectivity of $\varphi_{\vec{a},\vec{b},\vec{P}}$ in the last one.
		We have therefore proved that
		$\mathbb{P}(\vec{e}^{\mathrm{out}}=\vec{x})=\mathbb{P}(\vec{e}^{\mathrm{unif}}=\vec{x})$ for every
		$\vec{x}\in \mathcal{S}_t^n$ which concludes the proof.  \qed
\end{proof}

Our algorithm\ref{algo:trap} instantiated with rejection sampling steps enables to output $\vec{e}^{\mathrm{out}}$ being almost distributed as $\vec{e}^{\mathrm{unif}}$. 

\propoRejetUnif* 

\begin{proof}
Throughout the proof, the matrices $\vec{H}_W$, $\vec{H}_V$, $\vec{H}_U$ and vectors $\vec{s}_W$, $\vec{s}_V$, $\vec{s}_U$ are fixed, and~$(\vec e_U,\vec e_V,\vec e_W)$ denotes the random tuple produced by Algorithm~\ref{algo:inv2} at Step~\ref{algomark:inv2}, once all three loops have been accepted.

We use the following elementary rejection-sampling fact. Suppose that the probability of proposing and accepting a value $x$ in a single trial is
\[
    c\,\mathbb P(X=x)\,\mathbf 1_{\{x\in S\}},
\]
where $c>0$ does not depend on $x$ and
$\mathbb P(X\in S)>0$.
Then independent repetitions until acceptance produce the conditional distribution of $X$ given $X\in S$. Indeed, the acceptance probability in one trial is $c\,\mathbb P(X\in S)$, and so the probability of proposing and accepting $x$ after the trials is 
\[
\begin{array}{lcl}
\sum_{j\geq 0} \Big( 1 - c\,\mathbb P(X\in S) \Big)^j c\,\mathbb P(X=x)\,\mathbf 1_{\{x\in S\}} & = & \dfrac{c\,\mathbb P(X=x)\,\mathbf 1_{\{x\in S\}}}
         {c\,\mathbb P(X\in S)} \\
& = & \mathbb P(X=x\mid X\in S)
\end{array}
\]
where $j$ is the number of rejected trials before acceptance.

We will apply this fact to each loop, keeping the outputs
of the preceding loops fixed.

By definition of $M_W$, $M_V$, and $M_U$, $\vec{r}_W$, $\vec{r}_V$ and $\vec{r}_U$ always lie in $[0,1]$.
Thus, comparing them with an independent uniform sample from $[0,1]$ implements the corresponding acceptance decision.

\medskip
\noindent\emph{The $W$-loop.}
Fix $\vec x_W$ and, for brevity, write $\omega=\omega (\vec x_W) \eqdef |\vec x_W|$. By Definition~\ref{def:uniformity},
\[
    \mathbb{P}\big(
        \mathsf{DecodeW}(\vec H_W,\vec s_W)=\vec x_W
    \big)
    =
    \begin{cases}
        2^{n/3-k_W}p_W(\omega)
        &\text{if }\vec x_W\vec H_W^\top=\vec s_W,\\
        0&\text{otherwise}.
    \end{cases}
\]
Therefore, the probability of
outputting and accepting a valid $\vec x_W$ in one trial is given by
\begin{align*}
    2^{n/3-k_W}p_W(\omega)\,\vec r_W(\omega)
    &=
    \frac{2^{n/3-k_W}}{M_W}
    \frac{A_\omega}{\binom nt}\\
    &=
    \frac{2^{n/3-k_W}}{M_W}
    \mathbb P(\vec e_W^{\mathrm{unif}}=\vec x_W)
\end{align*}
where the last equality follows from
Lemma~\ref{lem:reference-uniformity}. On the other hand, if $\vec{x}_W\vec{H}_W^{\top} \neq \vec{s}_W$, it cannot be output and thus accepted.
Using the rejection-sampling fact proves
Equation~\eqref{eq:rej_W}.

\medskip
\noindent\emph{The $V$-loop.}
Fix an output $\vec x_W$ of the $W$-loop 
and write~$\omega=\omega(\vec x_W)$. For an output $\vec x_V$ of the decoder, we denote $\nu$ as the statistic $\nu (\vec x_V,\vec x_W)$. By Definition~\ref{def:uniformity}, the definition of $\vec r_V$, and Lemma~\ref{lem:reference-uniformity}, the probability of outputting and accepting $\vec{x}_V$ 
satisfying
$\vec x_V\vec H_V^\top=\vec s_V$ in one trial is
$$
\frac{2^{n/3-k_V}}{M_V(\omega)}
    \mathbb P\big(
        \vec e_V^{\mathrm{unif}}=\vec x_V
        \mid \vec e_W^{\mathrm{unif}}=\vec x_W
    \big) \ .
$$
Since $\omega$ is fixed throughout the loop, the prefactor does not depend on $\vec x_V$. The same rejection-sampling fact therefore proves Equation~\eqref{eq:rej_V}.

\medskip
\noindent\emph{The $U$-loop.}
Fix outputs $(\vec x_V,\vec x_W)$ of the preceding loops occurring with positive probability, and write
\[
    \vec m=\vec m(\vec x_V,\vec x_W),
    \qquad \nu=n/3 - m_{00}\ .
\]
For a decoder output $\vec x_U$, write
$\vec h=\vec h(\vec x_U;\vec x_V,\vec x_W)$. By Definition~\ref{def:uniformity},  Lemma~\ref{lem:reference-uniformity}  and the definition of
$\vec r_U$, the probability of outputting and accepting $\vec{x}_U$ satisfying $\vec{x}_U\vec{H}_U^{\top}$ if  $\nu+3h_{00}+h_{10}+h_{01}+h_{11}=t$ 
in one trial is
$$
\frac{2^{n/3-k_U}}{M_U(\vec m)}
    \mathbb P\big(
        \vec e_U^{\mathrm{unif}}=\vec x_U
        \mid
        (\vec e_V^{\mathrm{unif}},\vec e_W^{\mathrm{unif}})
        =(\vec x_V,\vec x_W)
    \big) \ .
$$
Since $\vec m$ is fixed throughout the loop, the prefactor
does not depend on $\vec x_U$.
Applying the same rejection-sampling fact proves
Equation~\eqref{eq:rej_U} and concludes the proof. \qed
	\end{proof}

We are now ready to prove one of our main theorem showing that signatures are not leaking information (under the condition of perfect decoders from Definition~\ref{def:uniformity}).

\theoremNoLeakage*

\begin{proof}
		Let us recall the following standard decomposition of the statistical distance: given two pairs of random variables $(X,Y)$ and $(X',Y')$,
		\begin{equation}\label{eq:statDistPairs}
			\Delta\Big( (X,Y),(X',Y') \Big) \leq \Delta(X,X') +\sum_x\mathbb P(X'=x)\,
			\Delta\bigl(Y\mid X=x,\;Y'\mid X'=x\bigr)\ .
		\end{equation}

		Fix for now a trapdoor, in particular $\vec{H}_{U}$, $\vec{H}_{V}$ and $\vec{H}_{W}$. Notice that since $\varphi_{\vec{a},\vec{b},\vec{P}}$ is a bijection, we have 
		$$
		\Delta\big(\vec{e}^{\mathrm{out}},\vec{e}^{\mathrm{unif}}\big) = \Delta\big( (\vec{e}_W,\vec{e}_V,\vec{e}_U), (\vec{e}_{W}^{\mathrm{unif}},\vec{e}_V^{\mathrm{unif}},\vec{e}_U^{\mathrm{unif}} ) \big) \ . 
		$$
Applying now Equation~\eqref{eq:statDistPairs} twice gives,
\begin{multline}\label{eq:upStatDist}
			\Delta\big(\vec e^{\mathrm{out}},\vec e^{\mathrm{unif}}\big) \leq \\ \Delta\big(\vec e_W,\vec e_W^{\mathrm{unif}}\big) +\sum_{\vec x_W}\mathbb{P}(\vec e_W^{\mathrm{unif}}=\vec x_W)\cdot 
				\Delta\bigl(\vec e_V\mid\vec e_W=\vec x_W,\;
				\vec e_V^{\mathrm{unif}}\mid\vec e_W^{\mathrm{unif}}=\vec x_W\bigr)
		\\ +\sum_{\vec x_V,\vec x_W}
				\mathbb P(\vec e_V^{\mathrm{unif}}=\vec x_V,
				\vec e_W^{\mathrm{unif}}=\vec x_W)\cdot\\
		  \Delta\bigl(
			\vec e_U\mid(\vec e_V,\vec e_W)=(\vec x_V,\vec x_W),\;
			\vec e_U^{\mathrm{unif}}\mid
			(\vec e_V^{\mathrm{unif}},\vec e_W^{\mathrm{unif}})
			=(\vec x_V,\vec x_W)\bigr) \ .
		\end{multline}

		Let us now compute the above statistical distance. First, since $\vec{s}$ is uniform and $\vec{S}$ is invertible,~$(\vec{s}_U,\vec{s}_V,\vec{s}_W)=\vec{s}\vec S^{\top}$ is a uniform. 
		Recall now that we assume that decoders $\mathsf{DecodeW}$, $\mathsf{DecodeV}$ and $\mathsf{DecodeU}$ of Algorithm \ref{algo:inv2} are perfect (see Definition~\ref{def:uniformity}). Therefore Proposition \ref{propo:rejetUnif} applies. In particular, for a fixed $\vec{H}_W$,
		\begin{align*} 
		\mathbb{P}(\vec{e}_W = \vec{x}_W) &= \mathbb{P}\big(\vec{e}_W = \vec{x}_W \mid \vec{s}_W = \vec{x}_W \vec{H}_{W}^{\top}\big)\mathbb{P}\big( \vec{s}_W = \vec{x}_W \vec{H}_{W}^{\top} \big) \\
		&= \mathbb{P}\big(\vec{e}_W = \vec{x}_W \mid \vec{s}_W = \vec{x}_W\vec{H}_{W}^{\top}\big)  \;\frac{1}{2^{n/3-k_W}} \quad \left(\mbox{$\vec{s}_W$ is uniform}\right)  \\
		&= \mathbb{P}\big(\vec{e}^{\mathrm{unif}}_W = \vec{x}_W \mid \vec{e}_W^{\mathrm{unif}}\vec{H}_W^{\top} = \vec{x}_{W}\vec{H}_{W}^{\top} \big)    \;\frac{1}{2^{n/3-k_W}} \quad \left(\mbox{By Proposition \ref{propo:rejetUnif} } \right) \\
		&= \frac{\mathbb{P}\big(\vec{e}^{\mathrm{unif}}_W = \vec{x}_W\big)}{\mathbb{P}\big( \vec{e}_W^{\mathrm{unif}}\vec{H}_W^{\top} = \vec{x}_{W}\vec{H}_{W}^{\top} \big)} \;\frac{1}{2^{n/3-k_W}}
		\end{align*} 
		Therefore, 
		\begin{align*}
			\Delta\big(\vec{e}_W,\vec{e}^{\mathrm{unif}}_W\big) &= \frac{1}{2}\sum_{\vec{x}_W} \left| \mathbb{P} \big( \vec{e}_W = \vec{x}_W \big) - \mathbb{P}\big( \vec{e}_W^{\mathrm{unif}} = \vec{x}_W \big) \right|   \\
			&= \frac{1}{2}\sum_{\vec{y}} \sum_{\vec{x}_W: \; \vec{x}_W\vec{H}_W^{\top}=\vec{y}}  \left| \frac{\mathbb{P}\big(\vec{e}^{\mathrm{unif}}_W = \vec{x}_W\big)}{\mathbb{P}\big( \vec{e}_W^{\mathrm{unif}}\vec{H}_W^{\top} = \vec{x}_{W}\vec{H}_W^{\top} \big)} \;\frac{1}{2^{n/3-k_W}}-\mathbb{P}(\vec{e}_W^{\mathrm{unif}}=\vec{x}_W) \right| \\
			&=\frac{1}{2} \sum_{\vec{y}} \sum_{\vec{x}_W: \; \vec{x}_W\vec{H}_W^{\top}=\vec{y}}  \left| \frac{1}{\mathbb{P}\big( \vec{e}_W^{\mathrm{unif}}\vec{H}_W^{\top} = \vec{y} \big)} \;\frac{1}{2^{n/3-k_W}}-1 \right| \cdot \mathbb{P}(\vec{e}_W^{\mathrm{unif}}=\vec{x}_W) \\
			 &= \frac{1}{2}\sum_{\vec{y}}  \left| \frac{1}{\mathbb{P}\big( \vec{e}_W^{\mathrm{unif}}\vec{H}_W^{\top} = \vec{y} \big)} \;\frac{1}{2^{n/3-k_W}}-1 \right| \cdot \mathbb{P}(\vec{e}_W^{\mathrm{unif}}\vec{H}_W^{\top}=\vec{y}) \\
			 &=\Delta\big(\vec{s}_W, \vec{e}_W^{\mathrm{unif}}\vec{H}_W^{\top}\big)
		\end{align*}
		Now by averaging over $\vec{H}_W$ and using Lemma \ref{lemma:binaryLHL}, we easily deduce (using the notation of Lemma~\ref{lem:reference-uniformity}) that,
		\begin{equation*}
			\mathbb{E}_{\vec{H}_W}\left( \Delta\big(\vec{e}_W,\vec{e}_W^{\mathrm{unif}}\big)\right) \leq \frac{1}{2} \sqrt{2^{n/3-k_W} C_W} \ . 
		\end{equation*}
	The same computations identify the statistical distances $\Delta\bigl(\vec{e}_V^{\mathrm{unif}}\vec{H}_V^{\top},\vec{s}_V
	\mid\vec{e}_W^{\mathrm{unif}}=\vec{x}_W\bigr)$ and $\Delta(\vec e_U^{\mathrm{unif}}\vec H_U^{\top},\vec{s}_U
	\mid\vec e_V^{\mathrm{unif}}=\vec x_V,
	\vec e_W^{\mathrm{unif}}=\vec x_W)$ from which we conclude the proof by using Lemma~\ref{lemma:binaryLHL} and plugging this bound into Equation~\eqref{eq:upStatDist}.  \qed
	\end{proof}

\subsection{No Leakage Proof of the Instantiation via Prange Decoders}

Our aim now is to show that our decoders instantiated in Subsection \S~\ref{subsec:instPrange} are close enough of perfect decoders. 

\propoPrangeProb*

\begin{proof}
At the first step, the weight is the sum of $\ell$ and the number of ones among $n/3-k_W'$ uniform bits. Thus, its probability of being $\omega$ is $2^{-(n/3-k_W')}\binom{n/3-k_W'}{\omega-\ell}$ for a fixed $\ell$. All vectors of weight $\omega$ have the same probability. Dividing by $\binom{n/3}{\omega}$ and averaging over $\ell$ proves the first equality.

At the second step, the number of positions in $\mathcal{M}_{10}$ (that is where $\vec x_V$ and $\vec x_W$ are $1$ and~$0$, respectively) is $\ell$ plus the number of ones among the remaining $n/3-\omega-\min(k_V',n/3-\omega)$ independent uniform bits at positions where $\vec x_W$ is zero. Hence, its probability of being $\nu - \omega$ is~$2^{-(n/3-\omega-\min(k_V',n/3-\omega))}\binom{n/3-\omega-\min(k_V',n/3-\omega)}{\nu - \omega-\ell}$ for a fixed $\ell$. The other $\omega$ bits where $\vec x_W$ is nonzero are uniform and independent of these choices. Therefore, there are $2^\omega\binom{n/3-\omega}{\nu - \omega}$ vectors with this statistic, all with the same probability. This proves the second equality.

Finally, fix $(\vec j,\boldsymbol\ell)$. For each $(\alpha,\beta)\in\{0,1\}^2$, the statistic $h_{\alpha\beta}$ is $\ell_{\alpha\beta}$ plus the number of larger local weights among~$m_{\alpha\beta}-j_{\alpha\beta}$ uniform bits. Thus,
\[
    h_{\alpha\beta}-\ell_{\alpha\beta}
    \Unif \operatorname{Binomial}(m_{\alpha\beta}-j_{\alpha\beta},1/2)\ .
\]
These four random variables are independent, so
\[
    \mathbb P(\vec h\mid\vec j,\boldsymbol\ell)
    =
    2^{-(n/3-k_U')}
    \prod_{(\alpha,\beta)\in\{0,1\}^2}
    \binom{m_{\alpha\beta}-j_{\alpha\beta}}
          {h_{\alpha\beta}-\ell_{\alpha\beta}},
\]
where we used $\sum_{\alpha,\beta}(m_{\alpha\beta}-j_{\alpha\beta}) =n/3-k_U'$. By symmetry under permutations within each~$\mathcal{M}_{\alpha\beta}$, the $N_{\vec m}(\vec h)$ vectors counted in Equation~\eqref{eq:Nm} are equally likely. Dividing by $N_{\vec m}(\vec h)$ and averaging over $(\vec j,\boldsymbol\ell)$ concludes the proof. \qed
\end{proof}

\lemmaPrangeAlmostPerfect* 

\begin{proof}
For any fixed set $\mathcal F$ selected at step $X\in\{W,V,U\}$, we have $|\mathcal{F}|\le k_X-d$.
	If $\vec H_{X,\overline{\mathcal F}}$ is not of full rank, then there exists some non-zero
	$\vec{y}\in\mathbb{F}_2^{n/3-k_X}$ satisfying
	$\vec{y}\vec{H}_{X,\overline{\mathcal{F}}}=\vec{0}$.
	For a uniform~$\vec{H}_X$, this event, for each fixed non-zero $\vec{y}$, has probability
	$2^{-(n/3-|\mathcal{F}|)}$. A union bound therefore gives
	\begin{equation}\label{eq:prange-rank-bound}
		\mathbb P_{\vec H_X}\big(\mathsf{rank}(\vec H_{X,\overline{\mathcal F}})<n/3-k_X\mid\mathcal{F}\big)
		\le(2^{n/3-k_X}-1)\; 2^{-(n/3-|\mathcal{F}|)}<2^{-d}\ .
	\end{equation}
	Notice now that $\vec e_X^{(0)}$ is independent of
	$\vec H_X$. Thus, even after fixing $\vec e_X^{(0)}=\vec x_X$, the
	conditional choice of $\mathcal{F}$ is independent of $\vec H_{X}$.
	Averaging~\eqref{eq:prange-rank-bound} over this conditional choice proves
	\eqref{eq:prange-pointwise-rank}.
	
	Now fix $\vec H_X$ and a $(\mathcal{F},\vec{x'})$ passing the rank test of Instruction \ref{inst:rankTest} in Algorithm \ref{algo:prange-components}. A uniform input $\vec s_X$, together with the $k_X-|\mathcal F|$ uniform free entries, gives uniform entries on all of $\overline{\mathcal{F}}$: the map between these two descriptions is bijective
, since $\vec H_{X,\overline{\mathcal J}}$ is invertible. Algorithm~\ref{algo:prange-components} redraws $(\mathcal{F},\vec{x'})$ until $\vec H_{X,\overline{\mathcal F}}$ has rank $n/3-k_X$. Its output for a uniform input therefore has the distribution of~$\vec e_X^{(0)}$ conditioned on that rank condition. Let $b_X\eqdef\mathbb{P}\big( \mathsf{rank}(\vec H_{X,\overline{\mathcal F}})<n/3-k_X \big)$ for a fixed $\vec H_X$. We deduce that, 
	$$
	\Delta\big(\vec e_X^{(0)}\mid \mathsf{rank}(\vec H_{X,\overline{\mathcal F}})=n/3-k_X,\;\vec e_X^{(0)}\big)\le b_X \ .
	$$
	Finally, using Equation~\eqref{eq:prange-rank-bound} gives $\mathbb E_{\vec H_X}(b_X)<2^{-d}$ which concludes the proof. \qed
\end{proof}

\theoremPrangeNoLeakge*

\begin{proof}
	We first consider the $W$ decoding step, {\it i.e.,} $X=W$. Let us fix $\vec{H}_W$. Use $g_W$ from Lemma~\ref{lem:prange-almost-perfect} and $b_W \eqdef \sum_{\vec{x}_W} p_{W}(\omega(\vec{x}_W))\left( 1-g_W(\vec{x}_W) \right)$. First, suppose $b_W<1$. For a uniform input, the probability of returning $\vec x_W$ in one call is $p_W(\omega(\vec{x}_W))g_W(\vec{x}_W)/(1-b_W)$. Consequently, for a fixed input $\vec{s}_W$ and $\vec x_W\vec H_W^{\top}=\vec{s}_W$, we obtain
	$$
	\mathbb{P}\big(\mathsf{DecodeW}(\vec H_W,\vec{s}_W)=\vec{x}_W\big)
	=\frac{2^{n/3-k_W}p_W(\omega(\vec{x}_W))g_W(\vec{x}_W)}{1-b_W} \ .
	$$
	Multiplying by $r_W(\omega(\vec{x}_W))$ from Proposition~\ref{propo:rejetUnif} removes $p_W(\omega(\vec{x}_W))$. After repeating the rejection with this same input, the output probabilities are therefore proportional to
	$$
	\mathbb{P}\big(\vec{e}_W^{\mathrm{unif}}=\vec{x}_W\big)g_W(\vec{x}_W)
	\quad\hbox{on}\quad
	\left\{\vec{x}_W:\vec{x}_W\vec{H}_W^{\top}=\vec{s}_W\right\} \ .
	$$
	Therefore, for a fixed $\vec{y}$, the output is obtained from $\vec{e}_W^{\mathrm{unif}}$ conditioned on $\vec{e}_W^{\mathrm{unif}}\vec{H}_W^{\top}=\vec y$ by additionally accepting $\vec{x}_W$ with probability $g_W(\vec{x}_W)$. Hence
	\begin{equation}\label{eq:statCdtse}
		\Delta\Big(\vec{e}_W\mid\vec{s}_W=\vec{y},\;
		\vec{e}_W^{\mathrm{unif}}\mid
		\vec{e}_W^{\mathrm{unif}}\vec{H}_W^{\top}=\vec{y}\Big)
		\leq
		\frac{\mathop{\sum}\limits_{\vec x_W:\vec x_W\vec H_W^{\top}=\vec y}
			\mathbb P(\vec{e}_W^{\mathrm{unif}}=\vec x_W)
			\big(1-g_W(\vec{x}_W)\big)}
		{\mathbb{P}(\vec{e}_W^{\mathrm{unif}}\vec{H}_W^{\top}=\vec{y})} \ .
	\end{equation} 
	If no acceptance is possible, the right-hand side is one and the inequality still holds. The case~$b_W=1$ is covered in the same way, since then $g_W=0$ on the support of $p_W$.
	
	We now take $\vec{s}_W$ uniform. Apply Equation~\eqref{eq:statDistPairs} to $(\vec{e}_W, \vec{s}_W)$ and $(\vec{e}_W^{\mathrm{unif}}, \vec{e}_W^{\mathrm{unif}}\vec{H}_W^{\top})$, using the input vectors as the first components in that equation. The conditional distances are weighted by~$\mathbb{P}(\vec{e}_W^{\mathrm{unif}}\vec{H}_W^{\top}=\vec y)$, which cancels the denominator above. We obtain
	\begin{align*}
		\Delta\Big(\vec e_W,\vec e_W^{\mathrm{unif}}\Big) &= \Delta\Big((\vec{e}_W, \vec{s}_W), (\vec{e}_W^{\mathrm{unif}}, \vec{e}_W^{\mathrm{unif}}\vec{H}_W^{\top}) \Big)\\ 
		&\leq\Delta(\vec{s}_W,\vec e_W^{\mathrm{unif}}\vec H_W^{\top})+ \sum_{\vec{y}} \mathbb{P}(\vec{e}_W^{\mathrm{unif}}\vec{H}_W^{\top} = \vec{y})\Delta\Big(\vec{e}_W\mid\vec{s}_W=\vec{y},\;
		\vec{e}_W^{\mathrm{unif}}\mid
		\vec{e}_W^{\mathrm{unif}}\vec{H}_W^{\top}=\vec{y}\Big)  \\
		&\leq\Delta(\vec{s}_W,\vec e_W^{\mathrm{unif}}\vec H_W^{\top})+\sum_{\vec x_W}
		\mathbb{P}(\vec{e}_W^{\mathrm{unif}}=\vec x_W)
		\big(1-g_W(\vec{x}_W)\big)
	\end{align*}
	where in the first inequality we used Equation~\eqref{eq:statDistPairs} and in the second one Equation~\eqref{eq:statCdtse}.

	Averaging over $\vec{H}_W$, using Equation~\eqref{eq:prange-pointwise-rank} and Lemma~\ref{lemma:binaryLHL} (as we did in the proof of Theorem~\ref{thm:perfect-no-leakage}) therefore gives
	\begin{equation}\label{eq:Wcase} 
	\mathbb{E}_{\vec H_W}\big(\Delta(\vec{e}_W,\vec{e}_W^{\mathrm{unif}})\big)
	\le\frac{1}{2}\sqrt{2^{n/3-k_W}C_W}+2^{-d}\ .
	\end{equation} 
	
	For the case of the $V$ decoding step, \ie\, $X=V$, fix $\vec{x}_W$ and use the target $\vec{e}_V^{\mathrm{unif}}$ conditioned on $\vec{e}_W^{\mathrm{unif}}=\vec x_W$. Considering now the $U$ decoding step, we fix this time $(\vec{x}_V,\vec{x}_W)$ and use the corresponding conditional target.
	The same computation and reasoning apply with $p_V,\vec r_V,g_V$ and $p_U,\vec r_U,g_U$, respectively. The next input remains uniform after fixing the preceding outputs, and each conditional target is independent of the matrix being averaged. To conclude the proof we proceed as in the proof of Theorem~\ref{thm:perfect-no-leakage} by considering Equation~\eqref{eq:upStatDist} and substituting the three bounds for $W$, $V$ and $U$ (as in Equation~\eqref{eq:Wcase}). \qed 
\end{proof} }{}

\end{document}